\documentclass{aa}  

\usepackage[colorlinks=true, linkcolor = blue,
            urlcolor  = blue,
            citecolor = blue,
            anchorcolor = blue, bookmarks=false]{hyperref}
\usepackage{graphicx}
\usepackage{txfonts}
\usepackage{soul}
\usepackage{natbib}
\usepackage[section]{placeins}
\usepackage{enumitem} 
\usepackage{color}
\usepackage{threeparttable}
\usepackage{tikz,amsmath}
\usetikzlibrary{shapes.geometric, arrows}
\usepackage{ulem}
\usepackage{wrapfig}
\usepackage{rotating}
\usepackage{MnSymbol}
\usepackage{lscape} 
\tikzstyle{process} = [rectangle, minimum width=1.5em, minimum height=3.5em, text centered, draw=blue, fill=gray!10]
\tikzstyle{process2} = [rectangle, minimum width=1.5em, minimum height=3.5em, text centered, draw=white, fill=white]
\tikzstyle{arrow} = [thick,->,>=stealth]
\usepackage{cancel}
\usepackage{ulem}
\bibpunct{(}{)}{;}{f}{}{,} 

\newcommand{\teffa}{$T_{\mathrm{eff}}$(H$\alpha$)}

\newcommand{\teff}{$T_{\mathrm{eff}}$}
\newcommand{\teffI}{$T_{\mathrm{eff}}^{\mathrm{IRFM}}$}
\newcommand{\teffP}{$ T_{\text{eff}}^{\text{phot}} $}

\newcommand{\logg}{\mbox{log \textit{g}}}
 
\newcommand{\vsini}{$v\mathrm{sin}\,i$}
\newcommand{\titan}{\textsc{Titans}}

\begin{document}

   \title{The metal-poorest tail of the Galactic halo: hypothesis on its origin from precise spectral analysis
   \thanks{XXX}
   }


   \author{Riano E. Giribaldi\inst{1}
            \and
           Laura Magrini\inst{1}
           \and
           Martina Rossi \inst{2,3}
           \and
           Anish M. Amarsi\inst{4}
           \and 
           Donatella Romano \inst{3}
           \and
           Davide Massari\inst{3}
          }
   \institute{INAF – Osservatorio Astrofisico di Arcetri, Largo E. Fermi 5, 50125
Firenze, Italy 
             \email{riano.escategirbaldi@inaf.it; rianoesc@gmail.com}
             \and 
   Dipartimento di Fisica e Astronomia, Alma Mater Studiorum, Università di Bologna, Via Gobetti 93/2, 40129 Bologna, Italy 
   \and 
   INAF, Osservatorio di Astrofisica e Scienza dello Spazio, Via Gobetti 93/3, 40129 Bologna, Italy
   \and
   Theoretical Astrophysics, Department of Physics and Astronomy, Uppsala University, Box 516 SE-751 20 Uppsala, Sweden}

    \date{Received December, 2024; Accepted 24 March of 2025}

 
  \abstract
{The origin of the Galactic halo is one of the fundamental topics linking the study of galaxy formation and evolution to cosmology.}  
{We aim at deriving precise and accurate stellar parameters, Mg abundances, and ages for a sample of metal-poor stars with [Fe/H]$<-2$ dex from high signal-to-noise and high resolution spectra.  }
{We derive effective temperatures from H$\alpha$ profiles using three-dimensional non local thermodynamic equilibrium (3D NLTE) models, and surface gravities and ages from isochrone fitting based on Gaia data. Iron abundances were derived in one-dimensional (1D) NLTE, while Mg abundances were derived in 1D LTE, 1D NLTE, 3D LTE, and 3D NLTE to show the increasing level of accuracy.  }
{The stars show a tight trend in the [Mg/Fe] vs [Fe/H] plane with a knee at [Fe/H]$\approx -2.8$ dex,  which indicates a low level of stochasticity. 
Their location in the Lindblad diagram confirms their belonging to the Galactic halo, but does not show a distinct clustering that might be expected for a merger with a single low-mass galaxy. 
Comparison with chemical evolution models is also not definitive on whether the sample stars were born in-situ or in accreted low-mass galaxy mergers.}
{We find two plausible explanations for the chemical sequence traced by the stars in the [Mg/Fe] vs [Fe/H] plane.
One is that the sample stars originated in the already formed Milky Way, which at that time (12.5 Gyr ago) was already the main galaxy of its Local Group surroundings. Another one is that the sample stars originated in several small galaxies with similar properties, which later merged with the Galaxy. 
Only accurate spectroscopic analysis such as that done here can reveal trustworthy chemical diagrams required to observe the traces of the Galaxy evolution.
Other elements are required to discern between the two hypotheses.
}
   
\titlerunning{The metal-poorest Galactic halo}
   \maketitle
%

\section{Introduction}

The evolutionary path of the Milky Way has been strongly influenced by several mergers that happened in the past \citep[see, e.g.][]{kruijssen20, Malhan2022ApJ...926..107M}, with  at least one major accretion event that occurred about 9.5 Gyr ago  
with the so-called Gaia-Enceladus or Gaia-Sausage progenitor \citep{Belokurov2018,helmi2018, gallart2019NatAs...3..932G, massari19, belokurov2020MNRAS.494.3880B,borre2022MNRAS.514.2527B, giribaldi2023A&A...673A..18G}.
This scenario  is supported by plenty of evidence of the observational  properties of possible members belonging to the remnants of merging galaxies, such as chemical abundances from stellar spectra,  and kinematic properties derived from both radial velocities and proper motions \citep[e.g.][]{nissen2010A&A...511L..10N, myeong2019MNRAS.488.1235M, Monty2020MNRAS.497.1236M, bonaca2020ApJ...897L..18B, feuillet2021MNRAS.508.1489F,da_silva2023A&A...677A..74D, ceccarelli24}.  
Precise and accurate chemical and kinematic data are required as constraints to develop and improve cosmological and numerical chemo-dynamical simulations of galaxy formation \citep[e.g.][]{brook2003ApJ...585L.125B,grand2020MNRAS.497.1603G,rey_10.1093/mnras/stad513}, built in the frameworks of the two compelling scenarios:   "monolithic collapse" \citep[e.g.][]{eggen1962ApJ...136..748E} and "hierarchical accretion" \citep{searle1978ApJ...225..357S}.
In particular, the presence of the remnant of the Gaia-Enceladus merger in the nearby halo is characterised by a different pattern with respect to the Milky Way in-situ populations in several chemical planes, such as [Mg\footnote{Mg or other sensitive $\alpha$ elements such as O, Ca, and Si, for example.}/Fe]\footnote{[A/B] = $ \log{\left( \frac{N(\text{A})}{N(\text{B})} \right )_\text{Star}} - \log{\left( \frac{N(\text{A})}{N(\text{B})} \right )_\text{Sun}} $, where $N$ denotes the number abundance of a given element.} vs [Fe/H], [Ni/Fe] vs [(C+N)/O], and [Mg/Mn] vs [Al/Fe] \citep[e.g.][]{feuillet2021MNRAS.508.1489F,montalban2021NatAs...5..640M,Horta21,giribaldi2023A&A...673A..18G}; these differences  have been exhaustively explored thanks to the observations carried out with the large spectroscopic surveys, such as the Apache Point Observatory Galactic Evolution Experiment  \citep[APOGEE,][]{APOGEE} and the Galactic Archaeology with HERMES survey \citep[GALAH, ][]{GALAH}. 
However, although these surveys are reasonably sampled down to [Fe/H] $\sim -1.5$~dex, they are
moderately limited for [Fe/H] $\lesssim -1.5$~dex, 
and severely limited for [Fe/H] $\lesssim -2$~dex.
Nevertheless, a dedicated search for the oldest most metal-poor stars by the Pristine survey \citep{Starkenburg2017MNRAS.471.2587S} is progressively pushing these limits.
Also, it is worth to mention the Measuring at Intermediate Metallicity Neutron Capture Element (MINCE) survey \citep{cescutti2022A&A...668A.168C,francois2024A&A...686A.295F}, which is currently acquiring high resolution spectra of metal-poor stars with several facilities in diverse observatories, with the aim of studying the nucleosynthesis processes at early stages of the Milky Way assembling.

In fact, this is precisely the most metal-poor stars that allow us to investigate the most distant past of our Galaxy, probably dominated by a large number of mergers \citep[see, e.g.][]{Witten2024NatAs...8..384W}. While some orbital properties  are likely to be lost when tracing back in time \citep[e.g.][]{JeanBaptiste2017}, chemical properties, such as photospheric abundance ratios, remain largely preserved.
Detailed chemical analysis of a few hundred stars with $-4 \lesssim$ [Fe/H] $\lesssim -3$~dex and a few dozen with  [Fe/H] $ \lesssim -4 $ have been done by moderate resolution spectroscopy \citep[e.g.][among other works]{cayrel2004A&A...416.1117C,norris2013ApJ...762...28N,yong2013ApJ...762...26Y,aoki2013AJ....145...13A,cohen2013ApJ...778...56C,jacobson2015ApJ...807..171J, norris2007ApJ...670..774N, caffau2012A&A...542A..51C, spite2013A&A...552A.107S, hansen2013A&A...551A..57H, frebel2005Natur.434..871F,   aoki2006ApJ...639..897A, christlieb2002Natur.419..904C,keller2014Natur.506..463K, aguado2019ApJ...874L..21A,aguado2022A&A...668A..86A,arentsen2023MNRAS.519.5554A, 
sestito2023MNRAS.518.4557S}.
The vast majority of these stars are red giants, as exploring distant environments, such as the Galaxy halo and bulge, requires the observation of the brightest objects.
However, there are also observations of several dwarf stars in the sample of the nearest stars.

As for the main stellar populations of the Milky Way, such as thin and thick discs, the halo, and the bulge, the [Mg/Fe] vs. [Fe/H]  (or alternatively vs. [Be/H]) diagram proved to be a key tool for identifying the different origins of the various populations and  their star formation history and  efficiency, also in the metal-poor regime \citep[e.g.][]{smiljanic2009A&A...499..103S,nissen2010A&A...511L..10N}. 
Most of the literature spectroscopic works adopted the standard spectroscopic analysis: one dimensional (1D) hydrostatic model atmospheres and the assumption of local thermodynamic equilibrium (LTE). Their results usually display a substantial scatter in the [Mg/Fe] vs [Fe/H] diagram. 
Only a few works analysing  metal-poor stars ([Fe/H] $<-2$~dex) indicate a flat trend with  dispersions compatible with the errors of individual [Mg/Fe] measurements \citep[see, e.g.][]{francois2004A&A...421..613F,arnone2005A&A...430..507A, andrievsky2010A&A...509A..88A, aoki2013AJ....145...13A,jacobson2015ApJ...807..171J}.
\cite{arnone2005A&A...430..507A} analysed 23 dwarfs and obtained a scatter of 0.06~dex consistent with star-to-star errors. This suggests that the typical [Mg/Fe] scatter observed in the literature with 1D-LTE analysis is mainly dominated by data and modelling errors.
In this respect, \cite{andrievsky2010A&A...509A..88A} showed that 1D with non-LTE (NLTE) line modelling might alleviate the [Mg/Fe] offset of $\sim$0.2~dex between giants and dwarfs displayed by their 1D LTE modelling, thereby it can significantly decrease the scatter and dispersion. 
Therefore, it is evident that in the analysis of metal-poor stars,  NLTE effects must be considered \citep[see, e.g.][]{Sitnova2019MNRAS.485.3527S}. 

In addition, the implementation of three-dimensional (3D) model atmospheres for line synthesis (under NLTE) in the  most metal-poor regime is becoming even more necessary so as to obtain accurate effective temperatures of FGK type stars; and also to obtain accurate individual element abundances. 
For the former,  efforts along the years testing diverse model recipes \citep[e.g.][]{vidal1970,stehle1999,BPO2000,BPO2002,barklem2007,allard2008,ludwig2009,pereira2013,amarsi2018}  indicated that the spectral synthesis of Balmer line profiles under 3D NLTE \citep{amarsi2018} can accurately reproduce the observational profiles of dwarfs and giants from their {\it actual}\footnote{It refers to inferences from the most fundamental or least model-dependent methods: eclipsing binaries, interferometry, and InfraRed Flux Method (IRFM).} \teff\ values \citep{giribaldi2021A&A...650A.194G,giribaldi2023A&A...679A.110G}. Once stars are benchmarked by those methods, the differential approach can be used with confidence for stars with similar parameters, as done by \cite{reggiani2016A&A...586A..67R} and \cite{reggiani2018MNRAS.475.3502R}, for instance.
The latter is related to the line spectral  synthesis of some elements: the abundances measured under the 3D NLTE synthesis may drastically change patterns of element ratios \citep[e.g.][]{amarsi2019} or even emphasise the pattern separation of stellar populations \citep[e.g.][]{matsuno2024A&A...688A..72M}.

Recent findings \citep{Horta2023ApJ...943..158H} highlighted that the time of accretion and the stellar mass of an accreted satellite are key factors shaping the spatial distribution, orbital energy, and chemical compositions of stellar debris. In particular, in the inner halo (R$_{GC}<$ 30 kpc), where phase mixing dominates, the debris from both lower- and higher-mass satellites are expected to be fully spatially blended with stars formed in situ. Despite this extensive mixing, the persistence of distinct chemical signatures might provide insight into the assembly history of the Milky Way.
In this work, we therefore propose to re-analyse a sample of metal-poor stars with state-of-the-art spectral analysis tools, including whenever possible 3D and NLTE, in order to reveal their pattern in the [Mg/Fe] vs [Fe/H] plane and, from it, to infer their origin. We compare our results with chemical evolution models, which are also discussed in light of recent chemo-dynamical simulations.

The paper is structured as follows: in Section~\ref{sec:data} we present our three samples, whereas in Section~\ref{sec:parameters} we describe how we derived stellar parameters. 
In Section~\ref{sec:abundances} we report how the abundances of Mg are computed in the various approximations. In Section~\ref{sec:results} we provide our results on the chemical and kinematic properties of the samples. 
We discuss our results in Section~\ref{sec:discussion}, comparing with two chemical evolution models, and we provide our summary and conclusion in Section~\ref{sec:conclusions}.

\section{Data and sample}
\label{sec:data}

Our sample includes halo stars in the  metal-poor regime ([Fe/H] $< -2$~dex), belonging to three different samples: Precision samples I and II, and the {\it Gaia}-ESO sample. Our targets are selected to have high-resolution and high-S/N spectra so that it is possible to derive 
precise abundances and to constrain the  [Mg/Fe]--[Fe/H] pattern.
In what follows, we briefly describe our three samples.

\paragraph{Precision sample I:} It is composed of stars selected from \citet[][Paper~I henceforth]{giribaldi2023A&A...673A..18G}. It consists of dwarf FGK stars: 15 over 48 stars from \citet{giribaldi2021A&A...650A.194G} named \titan~I. 
Their parameters were derived from Ultraviolet and Visual Echelle Spectrograph  \citep[UVES, ][]{2000SPIE.4008..534D} spectra of the European Southern Observatory (ESO), with resolution higher than $R= \lambda/\Delta \lambda = 40\,000$.
In most cases, for the element abundance analyses, we combined spectra of the same resolution to optimise the S/N. When available, also spectra taken by the High Accuracy Radial velocity Planet Searcher  \citep[HARPS of $R = 115\,000$, ][]{mayor2003Msngr.114...20M}  and the Echelle SPectrograph for Rocky Exoplanets and Stable Spectroscopic Observation \citep[ESPRESSO of $R = 190\,000$, ][]{pepe2021A&A...645A..96P} instruments were used. 
Specific resolution and S/N values are listed in Table~\ref{tab:new_teff}.
The stars in the \titan~I sample   are accurate templates, since their spectroscopic \teff\ and  parallax-based \logg\ are compatible with values determined by fundamental methods: interferometry and InfraRed Flux Method (IRFM).
Paper~I also includes 212 stars from the GALAH DR3 catalogue \citep{buder2021MNRAS.506..150B}, but none has [Fe/H] $< -2$, therefore GALAH stars are absent in this sample; they are used only as reference to display the concentration of the stellar populations in chemical and kinematic diagrams.
We include five additional new dwarf stars selected from the sample of \cite{melendez2010A&A...515L...3M}. These stars allow us  to cover the [Fe/H] range between $-3$ and $-2.5$~dex, improving the \titan~I sub-sample. These are BPS BS16968$-0061$, BPS CS22177$-0009$, BPS CS22953$-0037$, and BPS CS29518$-0043$, which have UVES archival spectra, obtained  with a resolving power $R\sim42000$ and with $100 <$ S/N $< 170$.
Their spectra  fulfil our request to embrace the H$\alpha$ line profiles without  {\it distortions} of artificial or physical origin. 
Respectively, imperfect blaze subtraction and artefacts such as ripple-like patterns due to imperfect order merging \citep[][]{skoda2004ASPC..310..571S,skoda2008SPIE.7014E..5XS}, or flux emission at any of the two wings (e.g. Fig.~\ref{fig:Ha_emission}).
We also analysed the star BPS BS16023$-0046$ from \cite{melendez2010A&A...515L...3M}, which has a spectrum with S/N $\sim 50$. Since its spectrum has a very weak magnesium line  (at 5528~\AA), which leads to an uncertain abundance, we included this star in the lower precision sample, that is together with the  {\it Gaia}-ESO sample.  
Stars with spectra with such distortions in all samples are indicated in Table~\ref{tab:new_teff}.

\paragraph{Precision sample~II:} It contains 13 red giants from the \titan~II \citep{giribaldi2023A&A...679A.110G} sample with [Fe/H] $< -2$~dex and  spectra of S/N > 140.
Since red giants are cooler than main sequence FGK dwarf stars, their metal lines are more prominent and permit the determination of the abundances in very metal-poor stars with higher precision.
For a more homogeneous sample, we excluded the handful of Carbon-Enhanced Metal-Poor (CEMP) giants present in the \titan~II set.
The stars in Precision sample~II are also accurate templates because their atmospheric parameters are compatible with those from fundamental methods, as described in the analysis of \citet{giribaldi2023A&A...679A.110G}.

\paragraph{{\it Gaia}-ESO sample:}   The {\it Gaia}-ESO survey \citep{Randich2022A&A...666A.121R, Gilmore2022A&A...666A.120G}  is a large public spectroscopic survey that  observed for 340 nights at the Very Large Telescope (VLT) from the end of 2011 to 2018. It gathered  $\sim$190000 spectra, for nearly 115000 targets belonging to all the main Galactic populations. It is still the only stellar survey using 8 m class telescopes, with the explicit aim of being complementary to the data obtained by the {\it Gaia} satellite \citep[see, e.g.][for a thorough comparison with {
\it Gaia} spectroscopic results]{Van2024arXiv240704204V}.
{\it Gaia}-ESO observed its targets at two different $R$: the medium-resolution sample was observed with GIRAFFE at  $R \sim 20\,000$ and the high-resolution sample with UVES at 
 $R \sim 47\,000$. From this latter sample, we selected  very metal-poor field stars ([Fe/H] $ < -2$ and with GES$\_$FLD=GE$\_$MW) whose stellar parameters and abundances are available in the final data release {\sc dr 5.1} \citep{Hourihane2023A&A...676A.129H}. 
This sample is rather limited in number because the {\it Gaia}-ESO target selection did not favour halo populations \citep{Stonkut2016MNRAS.460.1131S}. 
We obtained their reduced spectra from the ESO archive, and we consistently re-analysed them  as our Precision samples. 
Our {\it Gaia}-ESO sample contains 27 stars with [Fe/H] $ < -2$~dex, from which we  discarded three stars  because 
our method failed to retrieve reliable parameters from their spectra.
One of these stars (GES~J13301360-4346323) has a very narrow H$\alpha$ line with a small flux emission in one of its wings, we suspect it may be a giant with \logg $\,< 1$~dex. 
Another star (GES~J08192562-1405040) is located at the base of the red giant branch, and although it has a spectrum with S/N = 50, its Fe lines appear at the limit of detection.
We also excluded GES~J14205624-3701526 from our sample because its Mg line at 5528~\AA\ is affected by an artifact.
We added to the {\it Gaia}-ESO sample one star from the ESO archive, as described above (see Precision sample I).

\paragraph{General properties}

\begin{figure}[t]
    \centering
    \includegraphics[width=0.97\linewidth]{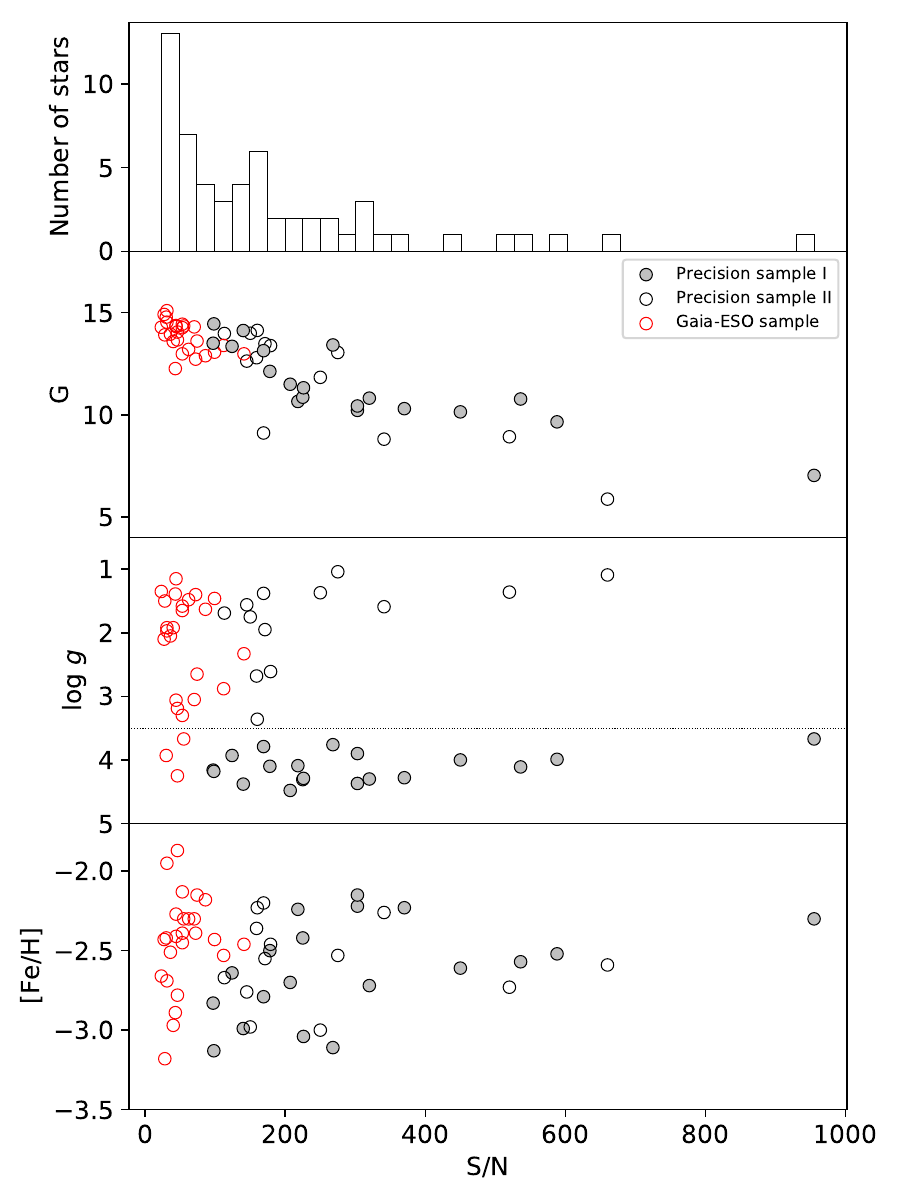}
    \caption{\tiny Observational properties of the full sample. Top panel displays a histogram of the S/N. From second panel downwards, G magnitude, \logg, and [Fe/H] are plotted as function of the S/N. 
    Precision samples I and II, and the {\it Gaia}-ESO sample are represented by different symbols.
    In third panel, the dotted line displays the division between dwarfs and giants.
    Surface gravity and [Fe/H] values are those listed in Table~\ref{tab:parameters}.
    }
    \label{fig:SNR}
\end{figure}

\begin{figure}[btp]
    \centering
    \includegraphics[width=0.99\linewidth]{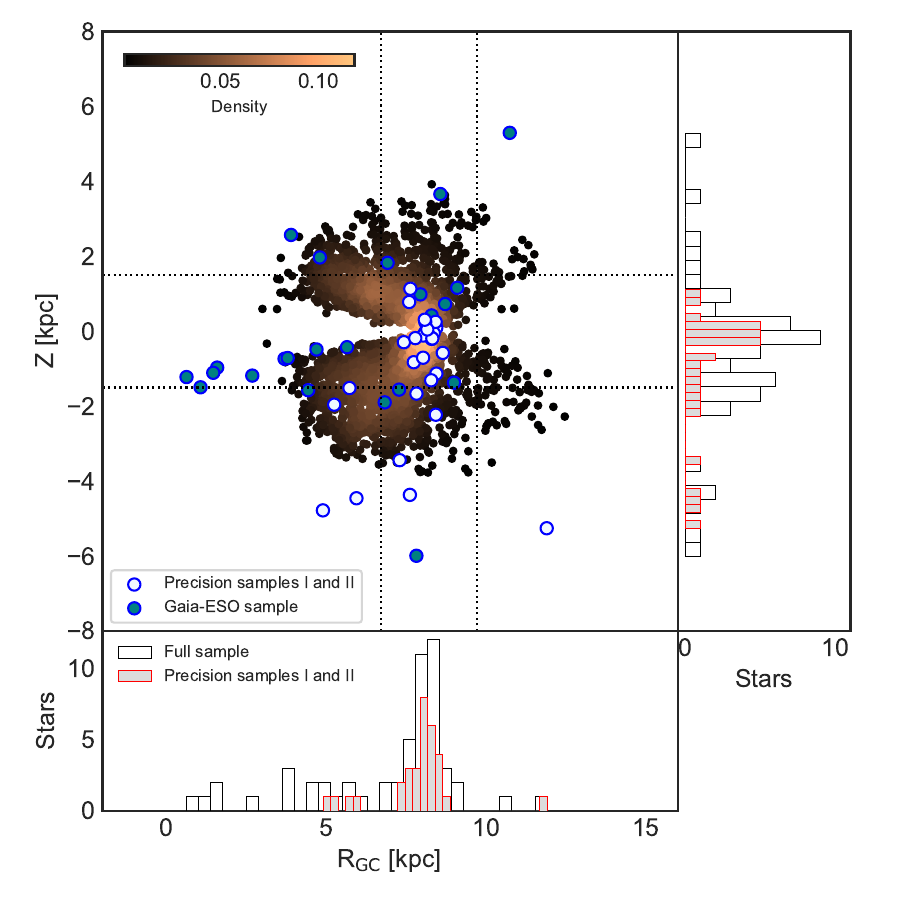}
    \caption{\tiny Spatial distribution of our sample stars with respect to the Galaxy centre.  The colours of the symbols distinguish the Precision samples (empty circles) and  the {\it Gaia}-ESO sample (filled blue circles).  
    A sample of 6250 halo stars of the GALAH survey are plotted as reference background, colour-code by density. 
    Dotted lines indicate $Z_{\odot} \pm1.5$ and $R_{\mathrm{GC\,\odot}} \pm 1.5$~kpc.
    Histograms of the distribution in R$_{\rm GC}$ and $Z$ of  the full and the Precision samples I and II are displayed.}
    \label{fig:space}
\end{figure}

Figures~\ref{fig:SNR} and ~\ref{fig:space} show some general properties of the overall sample. 
Figure~\ref{fig:SNR} displays the distribution of the S/N of our sample spectra. The spectra in the
Precision samples I and II, which together contain 32 stars, are those  with the highest S/N $\gtrsim 100$, whereas the spectra in the {\it Gaia}-ESO sample have S/N $\sim 50$. The stars in this latter sample are usually fainter, with $12.5 \lesssim G \lesssim 15$, whereas the Precision samples mostly have brighter stars.

Figure~\ref{fig:space} shows the distribution of the stars in the frame Galactocentric distance (R$_{\rm GC}$) vs heights on the Galactic plane ($Z$).
It shows that 72\% (23 of 32) of the stars in the Precision samples I and II are concentrated in the solar neighbourhood, defined here as a squared box of 3~kpc with the Sun at its centre.
Only five of 23 stars of the {\it Gaia}-ESO sample are located in the solar neighbourhood, while 10 of them lie in the direction of the Galactic centre. 
As a reference background, a colour-coded density dispersion of the 6250 stars selected by \cite{giribaldi2023A&A...673A..18G} is displayed in the figure.

\section{Atmospheric parameters and ages}
\label{sec:parameters}

\begin{table*}[h]
\caption{Stellar parameters of the stars}
\label{tab:parameters}
\centering
\tiny 
\begin{threeparttable}
\begin{tabular}{lcccccccccc}
\hline\hline
Star & \teff\ &  
\logg & [Fe/H] & $v_{mic}$ & \multicolumn{4}{c}{[Mg/Fe]} & age \\
&&&&&1D LTE &1D NLTE & 3D LTE & 3D NLTE &\\
(SIMBAD identifier) & [K] & [dex] & [dex] & [km/s] & [dex] & [dex] & [dex] & [dex] & [Gyr] \\
\hline
\multicolumn{10}{c}{Precision Sample I} \\
\hline
BD+24 1676 & $6433 \pm 23$ & $4.09 \pm 0.04$  & $-2.24 \pm 0.05$ & $1.35$ & 0.26 &$0.24 \pm 0.05$ & 0.29 & 0.40 & $12.9 \pm 0.5$\\
BD$-10$ 388 & $6285 \pm 19$ & $3.90 \pm 0.04$ & $-2.22 \pm 0.05$ & 1.33 & 0.27 & $0.26 \pm 0.05$ & 0.30 & 0.41 & $12.1 \pm 0.5$\\
BPS CS22166-030 & $6203 \pm 130$ & $3.76 \pm 0.04$ & $-3.16 \pm 0.05$ & 0 & 0.40 & $0.39 \pm 0.11$ & 0.44 & 0.52 & $13.2 \pm 0.6$\\
CD-35 14849 & $6380 \pm 10$ & $4.37 \pm 0.04$ & $-2.15 \pm 0.05$ & 1.14 & 0.28 & $0.23 \pm 0.06$ & 0.30 & 0.40 & $10.9 \pm _{1.4}^{1.3}$\\
BD+03 740 & $6388 \pm 13$ & $3.99 \pm 0.04$ & $-2.52 \pm 0.05$ & 1.22 & 0.19 & $0.17 \pm 0.06$ & 0.23 & 0.31 & $13.9 \pm 0.6$\\
BD$-13$ 3442 & $6444 \pm 6$ & $4.00 \pm 0.04$ & $-2.61 \pm 0.05$ & 1.34 & 0.36 & $0.35 \pm .0.06$ & 0.40 & 0.49 & $13.0 \pm 0.5$\\
BD+26 2621 & $6436 \pm 65$ & $4.31 \pm 0.04$ & $-2.42 \pm 0.05$ & 1.08 & 0.22 & $0.21 \pm 0.07$ & 0.25 & 0.33 & $11.9 \pm _{1.0}^{0.9}$\\
CD$-33$ 1173 & $6617 \pm 46$ & $4.30 \pm 0.04$ & $-2.72 \pm 0.05$ & 0.76 & 0.27 & $0.26 \pm 0.07$ & 0.31 & 0.38 & $11.2 \pm _{0.8}^{0.7}$\\ 
CD$-71$ 1234 & $6427 \pm 13$ & $4.28 \pm 0.04$ & $-2.23 \pm 0.05$ & 1.20 & 0.18 & $0.17 \pm 0.05$ & 0.21 & 0.30 & $12.3 \pm _{1.0}^{0.8}$\\
HE 0926$-0508$ & $6457 \pm 3$ & $4.10 \pm 0.04$ & $-2.50 \pm 0.05$ & 1.09 & 0.43 & $0.30 \pm 0.10$ & 0.35 & 0.38 & $13.2 \pm 0.5$\\
LP 815$-43$ & $6538 \pm 2$ & $4.11 \pm 0.04$ & $-2.85 \pm 0.05$ & 1.10 & 0.41 & $0.40 \pm 0.06$ & 0.46 & 0.52 & $11.1 \pm 0.4$\\
LP 831$-70$ & $6389 \pm 16$ & $4.48 \pm 0.04$ & $-2.84 \pm 0.05$ & 0.50 & 0.52 & $0.51 \pm 0.06$ & 0.56 & 0.62 & $8.7 \pm _{1.8}^{2.0}$\\
UCAC2 20056019 & $6401 \pm 25$ & $3.93 \pm 0.03$ & $-2.64 \pm 0.05$ & 0.50 & 0.25 & $0.24 \pm 0.16$ & 0.28 & 0.36 & $12.1 \pm _{0.8}^{0.7}$\\
Wolf 1492 & $6572 \pm 44$ & $4.29 \pm 0.04$ & $-3.04 \pm 0.05$ & 0.50 & 0.44 & $0.43 \pm 0.16$ & 0.48 & 0.54 & $13.6 \pm 0.7$\\
HD 140283 & $5805 \pm 11$ & $3.67 \pm 0.02$ & $-2.30 \pm 0.05$ & 1.17 & 0.19 & $0.14 \pm 0.05$ & 0.21 & 0.31 & $13.8 \pm 0.55$\\
BPS BS 16023$-046$ & $6502 \pm 7$ & $4.25 \pm 0.05$ & $-2.78 \pm 0.09$ & 0.80 & 0.15 & $0.14 \pm 0.22$ & 0.20 & 0.26 & $13.1 _{1.3}^{1.0}$\\
BPS BS 16968$-061$ & $6252 \pm 12$ & $3.79 \pm 0.03$ & $-2.79 \pm 0.11$ & 0.80 & 0.36 & $0.34 \pm 0.15$ & 0.40 & 0.48 & $12.4 \pm 0.6$\\
BPS CS 22177$-009$ & $6353 \pm 74$ & $4.38 \pm 0.04$ & $-2.99 \pm 0.15$ & 0.80 &  0.39 & $0.38 \pm 0.20$ & 0.44 & 0.50 & $12.4 \pm _{1.8}^{1.2}$\\
BPS CS 22953$-037$ & $6510 \pm 9$ & $4.16 \pm 0.03$ & $-2.83 \pm 0.11$ & 0.80 & 0.51 & $0.50 \pm 0.13$ & 0.55 & 0.63 & $13.7 \pm 0.5$\\
BPS CS 29518$-043$ & $6519 \pm 15$ & $4.18 \pm 0.05$ & $-3.15 \pm 0.13$  & 0.80 & 0.54 & $0.53 \pm 0.24$ & 0.59 & 0.65 & $13.9 \pm 0.5$\\
\hline
\multicolumn{10}{c}{Precision Sample II} \\
\hline
BPS BS 17569$-049$ & $4794 \pm 6$ & $1.04 \pm 0.15$ & $-2.53 \pm 0.13$ & 1.70 & 0.50 & $0.34 \pm 0.13$ & 0.54 & --- & ---\\
BPS CS 22953$-003$ & $5160 \pm 12$ & $1.95 \pm 0.15$ & $-2.55 \pm 0.13$ & 0.90 & 0.17 & $0.12 \pm 0.14$ & 0.10 & --- & ---\\
HD 186478 & $4793 \pm 22$ & $1.59 \pm 0.15$ & $-2.25 \pm 0.09$ & 1.40 & 0.64 & $0.35 \pm 0.10^{\spadesuit}$ & 0.63 & --- & ---\\
BD+09 2870 & $4724 \pm 1$ & $1.38 \pm 0.15$ & $-2.22 \pm 0.08$ & 1.10 & 0.57 & $0.19 \pm 0.10^{\spadesuit}$ &  0.45 & --- & ---\\
BD$-18$ 5550 & $4981 \pm 85$ & $1.83 \pm 0.15^\lozenge$ & $-2.63 \pm 0.10$ & 1.30 & 0.26 & $0.20 \pm 0.11$ & 0.22 & --- & ---\\
BPS CS 22186$-025$ & $5042 \pm 14$ & $1.69 \pm 0.15$ & $-2.67 \pm 0.10$ & 1.70 & 0.36 & $0.31 \pm 0.12$ & 0.38 & --- & ---\\
BPS CS 22891$-209$ & $4835 \pm 43$ & $0.93 \pm 0.15^\lozenge$ & $-3.01 \pm 0.08$ & 1.40 & 0.37 & $0.34 \pm 0.09$ & 0.32 & --- & ---\\
BPS CS 22896$-154$ & $5288 \pm 25$ & $2.61 \pm 0.15$ & $-2.46 \pm 0.08$ & 0.80 & 0.34 & $0.27 \pm 0.10$ & 0.28 & --- & ---\\
BPS CS 22956$-050$ & $5028 \pm 32$ & $1.75 \pm 0.15$ & $-3.07 \pm 0.07$ & $1.20$ & 0.46 & $0.43 \pm 0.09$ & 0.42 & --- & ---\\
BPS CS 22966$-057$ & $5536 \pm 40$ & $3.36 \pm 0.15$ & $-2.23 \pm 0.08$  & 0.90 & 0.33 & $0.27 \pm 0.09$ & 0.30 & --- & ---\\
BPS CS 29491$-053$  & $4834 \pm 38$ & $1.56 \pm 0.15$ & $-2.76 \pm 0.08$ & 1.30 & 0.46 & $0.37 \pm 0.09$ & 0.40 & --- & ---\\
BPS CS 29518$-051$ & $5339 \pm 6$ & $2.68 \pm 0.15^\lozenge$ & $-2.36 \pm 0.07$ & 1.10 & 0.31 & $0.23 \pm 0.08$ & 0.28 & --- & ---\\
HD 122563 & $4627 \pm 13$ & $1.09 \pm 0.15$ & $-2.59 \pm 0.09$ & 1.20 & 0.43 & $0.20 \pm 0.09$ & 0.34 & --- & ---\\
\hline
\multicolumn{10}{c}{{\it Gaia}-ESO sample} \\
\hline
GESJ17543319-4110487 & $5108 \pm 21$ & $1.90  \pm 0.25$ & $-2.97 \pm 0.11$ & $0.90$ & 0.43 & $0.40 \pm 0.32$ & 0.39 & --- & ---
\\
GESJ17545552-3803393 & $4748 \pm 20$ & $1.10 \pm 0.25$ & $-3.16 \pm 0.24$ & $1.20$ & 0.26 & $0.24 \pm 0.37$ & 0.21 & --- & ---\\
GESJ17574658-3847500$^{\filledstar}$ & $5251 \pm 306$ & $2.10 \pm 0.25$ & $-2.43 \pm 0.15$ & 1.50 &  0.39 & $0.29 \pm 0.36$ & 0.36 & --- & ---\\
GESJ18185545-2738118$^{\filledstar}$ & $4921\pm197$ & $1.90 \pm 0.25$ & $-2.68 \pm 0.18$ & 1.20 & 0.35 & $0.28 \pm 0.42$ & 0.31 & --- & --- \\
GESJ18215385-3410188 & $4925 \pm 10$ & $1.10 \pm 0.25$ & $-2.34 \pm 0.04$ & 1.35 & 0.66 &  $0.45 \pm 0.09$ & 0.62 & --- & ---\\
GESJ18261627-3154272 & $4924 \pm 132$ & $1.50 \pm 0.25$ & $-2.05 \pm 0.11$ & 1.20 & 0.74 & $0.36 \pm 0.15$ & 0.70 & --- & --- \\
GESJ18265160-3159404$^{\filledstar}$ & $4586\pm102$ & $1.00 \pm 0.25$ & $-2.82 \pm 0.02$ & 1.80 & 0.46 &  $0.31 \pm 0.09$ & 0.48 & --- & ---\\
GESJ18361733-2700053 & $4818 \pm 65$ & $2.05 \pm 0.25$ & $-2.52 \pm 0.06$ & $1.10$ & 0.35 & $0.27 \pm 0.08$ & 0.29 & --- & ---\\
GESJ18370372-2806385 & $4878 \pm 122$ & $1.50 \pm 0.25$ & $-2.20 \pm 0.04$ & 1.10 & 0.60 &$0.22 \pm 0.14^{\spadesuit}$ & 0.50 & --- & ---\\
GESJ18374490-2808311 & $4627 \pm 99$ & $1.00 \pm 0.25$ & $-2.00 \pm 0.16$ & 1.40 & 0.23 & $0.12 \pm 0.58$ & 0.20 & --- & ---\\
GESJ01293652-5020327$^{\ddagger}$ & $6368 \pm 35$ & $3.00 \pm 0.25$ & $-2.18 \pm 0.20$ & 0.70 & 0.17 & $0.14 \pm 0.50$ & 0.14 & --- & ---\\
GESJ03372526-2724127$^{\filledstar}$ & $5816 \pm 61$ &  $3.70 \pm 0.25$ & $-2.18 \pm 0.07$ &1.20& 0.37 & $0.32 \pm 0.16$ & 0.37 & 0.50 & $13.5 \pm _{1.2}^{1.0}$\\
GESJ09475504-1027247$^{\filledstar}$ & $5403 \pm 64$ & $3.20 \pm 0.25$ & $-2.53 \pm 0.05$ & 1.20 & 0.36 & $0.29 \pm 0.09$ & 0.38 & --- & ---\\
GESJ10091577-4127155$^{\filledstar}$ & $5361 \pm 100$ & $3.00 \pm 0.25$ & $-2.30 \pm 0.10$ & 1.30 & 0.35 & $0.26 \pm 0.12$ & 0.35 & --- & ---\\
GESJ10142785-4052503$^{\filledstar}$ & $5201 \pm 102$ & $2.40 \pm 0.25$ & $-2.46 \pm 0.06$ &  1.60 & 0.79 & $0.64 \pm 0.09$ & 0.78 & --- & ---\\
GESJ10580638-1542390 & $4838 \pm 53$ & $1.60 \pm 0.25$ & $-2.27 \pm 0.08$ & 1.50 & 0.23 & $0.09 \pm 0.09$ & 0.24 & --- & ---\\
GESJ12555146-4507342$^{\filledstar}$ & $5057 \pm 73$ & $1.50 \pm 0.25$ & $-2.45 \pm 0.07$ & 1.50 & 0.36 & $0.27 \pm 0.19$ & 0.35 & --- & ---\\
GESJ14214860-4408399$^{\filledstar}$ & $4674 \pm 69$ & $1.10 \pm 0.25$ & $-2.41 \pm 0.10$ & 1.30 & 0.46 & $0.25 \pm 0.12$ & 0.40 & --- & ---\\
GESJ14410090-4007413 & $4926 \pm 43$ & $1.50 \pm 0.25$ & $-2.41 \pm 0.03$ & 1.50 & 0.49 & $0.27 \pm 0.09$ & 0.47 & --- & ---\\
GESJ15300154-2005148 & $5811 \pm 65$ & $2.20 \pm 0.25$ & $-1.94 \pm 0.07$ & 1.60 & 0.39 & $0.31 \pm 0.09$ & 0.33 & --- & ---\\
GESJ22125747-4539203 & $5500 \pm 62$ & $3.40 \pm 0.25$ & $-2.12 \pm 0.06$ & 1.30 & 0.32 & $0.24 \pm 0.16$ & 0.31 & --- & ---\\
GESJ22494718-5001048 & $6302 \pm 25$ & $3.90 \pm 0.25$ & $-2.42 \pm 0.17$ & 1.50 & 0.40 & $0.37 \pm 0.40$ & 0.45 & 0.57 & $13.5 \pm _{1.0}^{0.9}$\\
\hline
\end{tabular}
\begin{tablenotes}
\item{} \textbf{Notes.} {The second part of the table lists the atmospheric parameters of the \titan~II giants \cite{giribaldi2023A&A...679A.110G}. Unlike the original source, here [Fe/H] is derived fully under NLTE, therefore our values are slightly different.
Surface gravity values accompanied by the symbol ($\lozenge$) are slightly different to those in \cite{giribaldi2023A&A...679A.110G}, see main text in Sect.~\ref{sec:parameters}.
Stars accompanied by with the symbol ($\filledstar$) indicate \teff\ determined only with Gaia colour calibrations.
The star accompanied by the symbol ($\ddagger$) has unreliable atmospheric parameters, see main text in Sect.~\ref{sec:parameters}.
Magnesium to iron ratios indicated with the symbol ($\spadesuit$) indicate uncertain values according to the analysis in Sect.~\ref{sec:LTE_NLTE}. 
} 
\end{tablenotes}
\end{threeparttable}
\end{table*}

\begin{figure}
    \centering
    \includegraphics[width=0.99\linewidth]{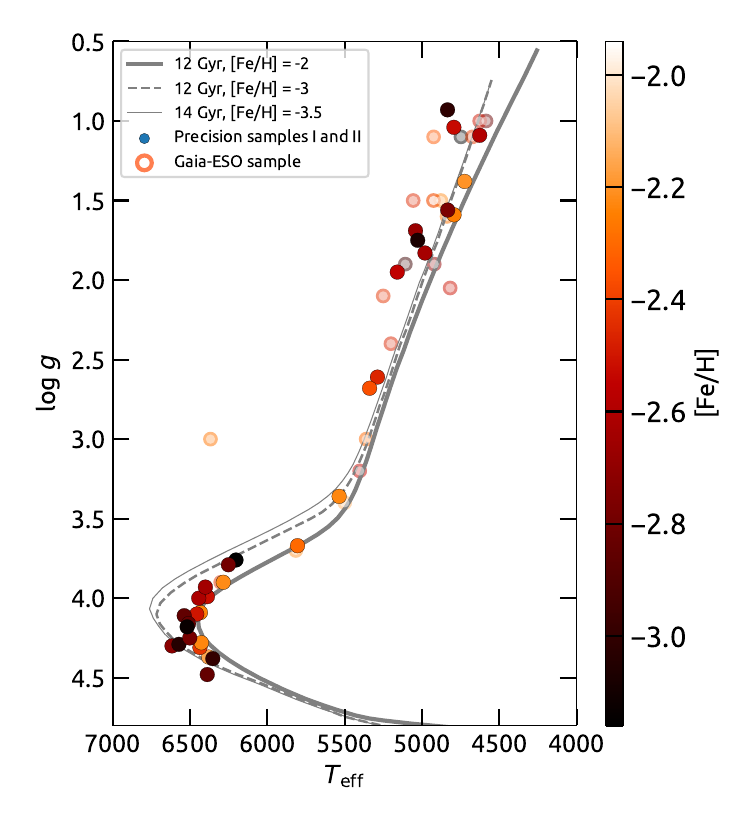}
    \caption{\tiny Kiel diagram (\teff\, vs \logg):  Precision samples I and II, and the {\it Gaia}-ESO sample are displayed as indicated in the legends. Symbols are colour-coded according to the metallicity. Yonsey-Yale isochrones \citep{kim2002,yi2003} with ages and metallicity values given \textcolor{magenta}{in the legend} are overplotted as reference.}
    \label{fig:kiel}
\end{figure}

In this section, we describe the methodology used for the determination of atmospheric parameters and ages. The method changes slightly depending on the S/N of the spectrum and whether dwarf or giant stars are analysed.
Figure~\ref{fig:kiel} shows our sample stars in the Kiel Diagram.

\begin{figure}
    \centering
    \includegraphics[width=0.99\linewidth]{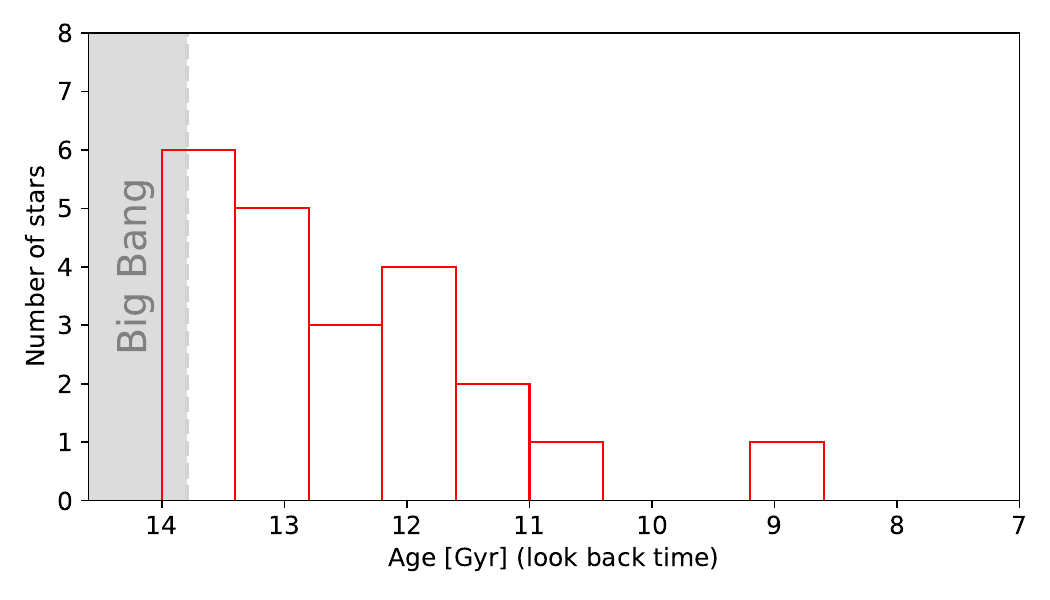}
    \caption{\tiny Histogram of isochronal ages of dwarf stars in Table~\ref{tab:parameters}, binned to 0.6~Gyr, which is the median of the individual errors.}
    \label{fig:hist}
\end{figure}

\paragraph{Precision sample I} The determination of the atmospheric parameters of dwarf stars is described in \cite{giribaldi2021A&A...650A.194G}, and details of the determination of their ages are given in \cite{giribaldi2023A&A...673A..18G}.
In the present work, we adopted the values obtained in such papers.
Here we briefly recall the used method: the sample has two sets of accurate and compatible temperatures: one derived by IRFM \citep[\teffI]{Casagrande2010} and another derived by fitting observational H$\alpha$ profiles (\teffa) with the 3D NLTE synthetic profiles of \cite{amarsi2018}. 
We combine the two sets of temperatures as described below to improve the precision to a few Kelvin. 
Surface gravity and age are based on isochrone fitting \citep[using Yonsey-Yale, ][]{kim2002,yi2003} using the frequentist approach implemented in the $q^2$ code \citep{ramirez2013ApJ...764...78R}, the formalism of which is exposed in \cite{melendez2012A&A...543A..29M}. We used as inputs the Gaia parallax, [Fe/H], extinction-corrected $V$, and \teff; obtaining outputs of typical precisions of 0.05~dex and 0.6~Gyr in \logg\ and age, respectively.
Figure~\ref{fig:hist} shows a histogram of the ages of all dwarf stars in Table~\ref{tab:parameters}.
Metallicity is based on spectral synthesis of weak lines with reduced equivalent width, REW $< -5$. The line list used is given in Table~\ref{tab:Fe_linelist}.

We include in this sample  five very metal-poor dwarf stars studied by 
\cite{melendez2010A&A...515L...3M}, as described above. 
To improve the \teffa\ precision, we degraded the spectral resolution from $R \sim 42\,000$ to $R \sim 25\,000$, this way we duplicate the S/N. Figure~\ref{fig:Teffa} shows the fits done as  in \cite{giribaldi2019A&A...624A..10G} with the synthetic 3D NLTE line profiles. 

Some stars in the sample do not have \teffI\footnote{This is, direct determinations from the IRFM method, not by colour-\teff\ relations.}, for them we derive photometric temperatures (\teffP) using Gaia colour-\teff\ relations
calibrated with IRFM determinations \citep{casagrande2021}. 
We only use $BP - RP$ and $G - BP$ colours, as $G - BR$ was observed to produce  outcomes slightly biased to cooler values \citep[Fig.~10 in][]{giribaldi2023A&A...679A.110G}. 
\teffP\ is given by the weighted average of the outcomes from each colour-\teff\ relation, where the weights ($\omega_i$)
are the squared inverse of their internal errors.
These errors are obtained by randomising [Fe/H] and \logg\ within Gaussian distributions of $\sigma$ equal to typical uncertainties; for this sample $\sigma = 0.05$~dex. 
The internal error of \teffP\ is obtained by the formula:

\begin{equation}
    \label{eq:error}
    \delta T = \sqrt{ \frac{ \sum_i \left( T - T_i \right)^2 \omega_i}{ \frac{\left(n-1\right)}{n} \sum_i \omega_i }  }
\end{equation}

\noindent
where $T$ is the averaged value, $T_i$ represents the temperature from each colour-\teff\ relation, $\omega_i$ is the corresponding weight, and $n$ is the number of quantities averaged.
The total error of \teffP\ is obtained by adding in quadrature the internal error with the error induced by the extinction uncertainty ($\delta T_{\mathrm{eff}}^{E(B-V)}$) and the precision of the colour calibrations \citep[$\sim$60~K, see Table~1 in][]{casagrande2021}.

For Precision samples I and II, the final \teff\ is given by the weighted average of \teffa\ and either interferometric temperatures ($T_{\mathrm{eff}}^{\mathrm{int}}$), \teffI, or \teffP; 
the two first ($T_{\mathrm{eff}}^{\mathrm{int}}$ and \teffI) are preferred over the last because they are more fundamental.
The internal error of the final \teff\ is obtained by Eq.~\ref{eq:error}.
Table~\ref{tab:parameters} 
lists the atmospheric parameters of all stars, 
whereas Table~\ref{tab:new_teff}
lists the temperatures derived by each method.

\paragraph{Precision sample II} The determination of the atmospheric parameters of giants is described in detail and tested in \cite{giribaldi2023A&A...679A.110G}.
Their method consists to derive \teff\ by H$\alpha$ profiles with the same family of 3D NLTE models used for Precision Sample I.
We derive \teffP\ for these stars as done with Precision Sample I.
Therefore, here we assume \teff\ equals the weighted average of both quantities.
The star HD~122563 is an exception, for which \teffa\ and $T_{\mathrm{eff}}^{\mathrm{int}}$ were averaged, as the latter is available in \cite{karovicova2020A&A...640A..25K}.
The method in \cite{giribaldi2023A&A...679A.110G} relies on the Mg~{\sc i}~b triplet lines to derive \logg\ compatible with asteroseismologic measurements, and it derives [Fe/H] by spectral synthesis from weak \ion{Fe}{ii} lines with REW $< -5$ considering NLTE corrections. 
We avoid calculating the ages of the giants because of the high degeneracy between isochrones in the region where they are located in the Kiel diagram. 

In addition, in the present work  we refined the determination of  [Fe/H]  by performing line synthesis under NLTE using the radiative transfer code Turbospectrum~2020\footnote{\url{https://github.com/bertrandplez/Turbospectrum_NLTE}}
\citep{gerber2023} with MARCS model atmospheres \citep{gustafson2008}. We considered the atomic parameters from  \cite{heiter2021A&A...645A.106H} and the iron NLTE departure coefficients based on the model atom developed in \cite{bergemann2012} and \cite{semenova2020}.
We compiled a Fe line list initially selecting lines from \cite{jofre2014A&A...564A.133J}, \cite{2016A&A...585A..75D}, and \cite{melendez2009A&A...497..611M}.
Then, we made preliminary selections of new lines by visual inspection by comparing observational with synthetic spectra. 
Our criterion consists on selecting all lines away from regions seriously contaminated by other spectral characteristics. 
We finally cleaned this selection by removing lines with excitation potential ($\chi_{ex}$) lower than 1.4~eV to avoid sources of overestimation \citep[e.g.][]{sitnova2015ApJ...808..148S,sitnova2024A&A...690A.331S}. The line list is presented in Table~\ref{tab:Fe_linelist}.

Since for several stars, Mg abundances will be derived from saturated lines, it is important to fine tune $v_{\mathrm{mic}}$  with similarly strong lines.
We did it under NLTE assuming the [Fe/H] balance of lines with REW $<-4.8$; [Fe/H] and associated microturbulent velocity ($v_{mic}$) values are listed in Table~\ref{tab:parameters}.
From the analysis of the spectra of giant stars we found that the excitation equilibrium of iron is systematically retrieved under NLTE.
See Fig.~\ref{fig:FeH}, where the excitation equilibrium is recovered from [Fe/H] and $v_{mic}$ values in \cite{giribaldi2023A&A...679A.110G}.
Exceptions found are the stars BD$-18\,5550$, BPS~CS 22891$-209$, and BPS~CS 29518$-051$, whose former \logg\ values (derived from Mg~{\sc i}~b triplet lines) were biased likely because of the interdependence with the Mg abundance. 
For these stars, we derive \logg\ assuming the excitation equilibrium of iron under NLTE. Later, after Mg abundance is derived, we tune \logg\ from the Mg~{\sc i}~b triplet; this way, all stars in the sample share the same \logg\ accuracy and precision.
We show the determination of \logg\ of the star BPS~CS 29518$-051$ in Fig.~\ref{fig:logg_BPS_CS_29518-051} as an example.

\begin{figure}
    \centering
    \includegraphics[width=0.8\linewidth]{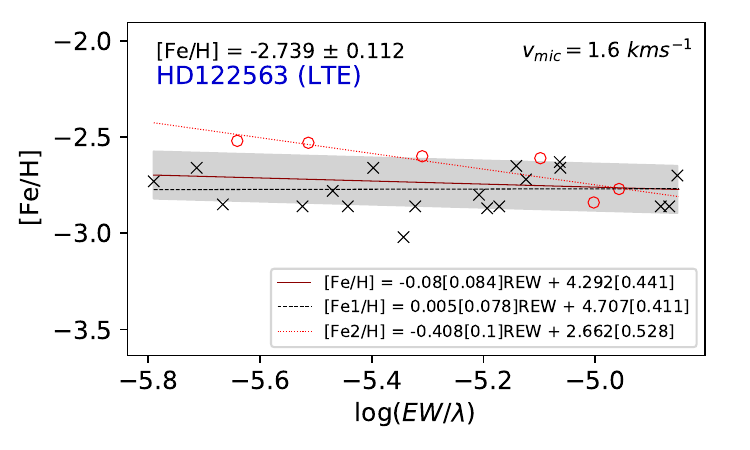}
    \includegraphics[width=0.8\linewidth]{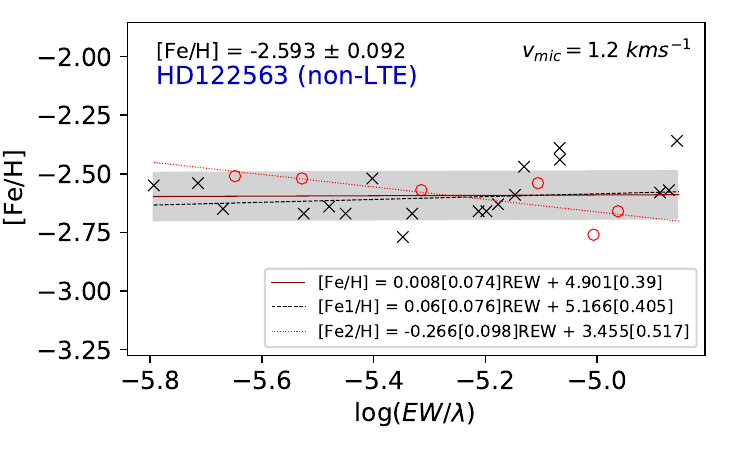}
    \caption{\tiny {\bf Ionisation equilibrium of the benchmark star HD~122563.} Top and bottom panels display [Fe/H] and $v_{mic}$ determination under LTE and NLTE respectively. Cross and circles indicate neutral and ionised species, respectively. Linear regressions corresponding to each species are displayed by different line-styles, their equations are given in the legends, where the values in brackets are the errors of the coefficients. The red line represents the regression of both neutral and ionised species together, the shade is the 1$\sigma$ standard deviation.  
    Determined [Fe/H] and $v_{mic}$ values are given in the plots.}
    \label{fig:Fe_H_determination}
\end{figure}

The ionisation equilibrium under NLTE differs from what was found under LTE conditions: a general offset of 0.1-0.2~dex between \ion{Fe}{i} and \ion{Fe}{ii} exists in LTE that is removed in NLTE.
As example, we show the case of the star HD~122563 in Figure~\ref{fig:Fe_H_determination}, where the offset between both Fe species under LTE is visible.  

This again shows the importance of accurate \teff\ determination and NLTE corrections for metal poor giants.
In addition, we found that our sample of giants have $v_{mic}\sim1.2\,\pm 0.3$~km~s$^{-1}$ (median and standard deviation, respectively), which may be somewhat lower than typical values in the literature derived under LTE for giant stars \citep[e.g.][]{francois2007A&A...476..935F,jofre2014A&A...564A.133J}. A different micro-turbulence value may have important effects in the determination of abundances \citep[see. e.g.][]{Baratella2020A&A...634A..34B, Baratella2021A&A...653A..67B}. 
In the case of HD~122563 shown in the figure, the LTE analysis provides a $v_{mic}$ value of 1.6~km~s$^{-1}$, which is 0.4~km~s$^{-1}$ higher than in NLTE. This higher value of  $v_{mic}$ produces a [Fe/H] value biased by $-0.15$~dex.

\paragraph{{\it Gaia}-ESO sample} 
Most {\it Gaia}-ESO stars are  giants, with typical S/N $\sim50$. 
We derive their \teff\  
same way as done for Precision Sample II.
Several of their spectra present H$\alpha$ profiles with {\it distortions} (e.g. Fig.~\ref{fig:Ha_emission}); the last column in Table~\ref{tab:new_teff} indicates the stars with such problems. For the corresponding stars,  we only derive \teffP\ from $BP - RP$ and $G - BP$ colours.
This temperature scale
depends mildly on \logg\ and [Fe/H], and
strongly on the extinction\footnote{The routine provided by \cite{casagrande2021} at \url{https://github.com/casaluca/colte} uses and convert extinction $E(B-V)$ into the Gaia photometric system.
Extinction was retrieved with the online tool at \url{https://irsa.ipac.caltech.edu/applications/DUST/}, assuming the average of the values of \cite{schlegel1998ApJ...500..525S} and \cite{Schlafly2011ApJ...737..103S}.} correction applied on the colour.
For instance, an error of $E(B-V)\pm0.05$~mag leads to a \teff\ variation of $\sim$120~K.
Therefore, 
to estimate the total uncertainty of \teffP, we add in quadrature 
the errors related to extinction, \logg, and [Fe/H], and the standard deviation of the colour-\teff\ scale.
We estimate the error of the extinction ($\delta_{\mathrm{ext}}$), by simply accounting  the difference between the values given by \cite{Schlafly2011ApJ...737..103S} and \cite{schlegel1998ApJ...500..525S}.
We estimate the errors 
related to \logg\ and [Fe/H] by randomising their values within Gaussian distributions of $\sigma$(\logg) $= 0.50$~dex and $\sigma$([Fe/H]) $= 0.20$~dex.

We obtain first estimates of the full set of atmospheric parameters by iteratively deriving \teffP, \logg\ from isochrones, and [Fe/H] by line synthesis until \teff\ does not vary by more than its internal precision.
Then, we refine \teff\ by H$\alpha$ fitting, when possible. 
We derive [Fe/H] and \logg\
accepting the ionization equilibrium under NLTE, as this is the outcome observed from the giants in the Precision sample II.
We did not fit Mg~I~b triplet lines to derive their \logg\ because the spectral noise leads to imprecise determinations.
The error is considered to be the value that makes the median of \ion{Fe}{ii} out of the 1$\sigma$ dispersion of \ion{Fe}{i} abundances.
We obtain a typical value of 0.25~dex, which we associate to the sample.
The star GESJ01293652-5020327 exhibits atypical atmospheric  parameters incompatible with evolutionary tracks, see Fig.~\ref{fig:kiel}. We suspect our \teff(H$\alpha$) and \teffP\ determinations are biased to a high value due to the potential presence of a binary component. The parameters of this star must be considered unreliable.

Fig.~\ref{fig:GESO_comparison} shows a comparison between the atmospheric parameters derived by our method and those listed in the {\it Gaia}-ESO survey.
Overestimations of \teff\ and \logg\ in the {\it Gaia}-ESO survey, based on 1D LTE analysis,  are visibly larger for the coolest stars with low surface gravities (i.e. \logg\ $\lesssim 2$~dex).
For giant stars in the red clump (\logg $\sim 2.5$~dex), \teff\ and \logg\ of both sets of parameters are in agreement. 
We find that, in general, [Fe/H] in the {\it Gaia}-ESO survey is underestimated by $\sim$0.3~dex with respect to our determination. 
We conclude that the negative bias in [Fe/H] is mainly due to the relatively higher \teff\ scale of the {\it Gaia}-ESO survey.

\section{Magnesium abundance}
\label{sec:abundances}
While magnesium abundances for Precision samples I and II were first derived in \cite{giribaldi2023A&A...673A..18G} and \cite{giribaldi2023A&A...679A.110G} considering  generic NLTE corrections \citep{mashonkina2013A&A...550A..28M}, we re-derived Mg abundances with updated NLTE corrections  for both samples in this study and determined Mg abundances in the {\it Gaia}-ESO sample for the first time.
We examined [Mg/Fe] vs [Fe/H] diagrams under different assumptions: 1D LTE, 1D NLTE, 3D LTE, and (only for dwarfs) 3D NLTE. 
For dwarf stars, we assumed a constant offset of $+$0.05~dex (in the [Mg/Fe] scale) to correct for diffusion, following the values in Table~4 of \cite{korn2007ApJ...671..402K}.

The reference solar Mg abundance in all cases  (1D LTE, 1D NLTE, 3D LTE, and 3D NLTE) is taken from the 3D NLTE analysis of \cite{asplund2021A&A...653A.141A}, namely $7.55 \pm 0.03$~dex. To verify that this is consistent with our analysis, we performed line synthesis to fit a solar spectrum (light reflected in Ganymede) from the ESO archive. 
Using the standard parameters \teff$_\odot$ = 5771~K \citep{2016AJ....152...41P}, \logg\ = 4.44~dex, [Fe/H] = 0~dex, and \vsini~= 2.04 km s$^{-1}$  \citep{2016A&A...592A.156D}, we obtain A(Mg) = 7.40 and 7.50~dex from the lines at 5528 and 5711~\AA, respectively; corresponding line fits are shown in Fig.~\ref{fig:Mg_fits}. The 1D LTE to 3D NLTE corrections described below for each line result in $+$0.11 and +0.08~dex, respectively. The 3D NLTE corrected abundances averaged over the two lines are 7.55 dex, which is compatible with the result of \cite{asplund2021A&A...653A.141A} that is based on several weaker lines of \ion{Mg}{i} as well as of \ion{Mg}{ii}.
We remark that the line 5528~\AA\ in the solar spectrum is strong and therefore a fully solar-differential analysis with it is prone to systematic biases. On the other hand, the line at 5711~\AA\ is only saturated with REW = $-4.71$, and therefore may be more suitable to that method.

\subsection{1D LTE vs 1D NLTE (Turbospectrum code)}
\label{sec:LTE_NLTE}

In the spectra of our three samples, Mg abundance is obtained from two lines, the Mg~{\sc I} lines at 5528~\AA\  and 5711~\AA.
For the atomic parameters of the lines, we used the Gaia-ESO linelist \citep{heiter2021A&A...645A.106H}.  In particular, the oscillator strength values (log $gf$)
of the two lines were from the theoretical study of \citet{1990JQSRT..43..207C}, $-0.498$ and $-1.724$ respectively.
These compare well against the values adopted by \citet{matsuno2024A&A...688A..72M}, $-0.547$ and $-1.742$
(experimental and theoretical values, respectively, from \citealt{2017A&A...598A.102P}).
For the 5528~\AA, broadening by hydrogen collisions was described using the so called Anstee-Barklem-O\textquotesingle Mara (ABO) theory \citep{1997MNRAS.290..102B},
with parameters $\alpha=1461$ and $\sigma=0.312$.
Both lines are not available in all spectra, so it is necessary to establish in which regime they give consistent A(Mg) values, so as to ensure that we always have abundances on the same scale for giant and dwarf stars.

\begin{figure*}
    \centering
    \includegraphics[width=1\linewidth]{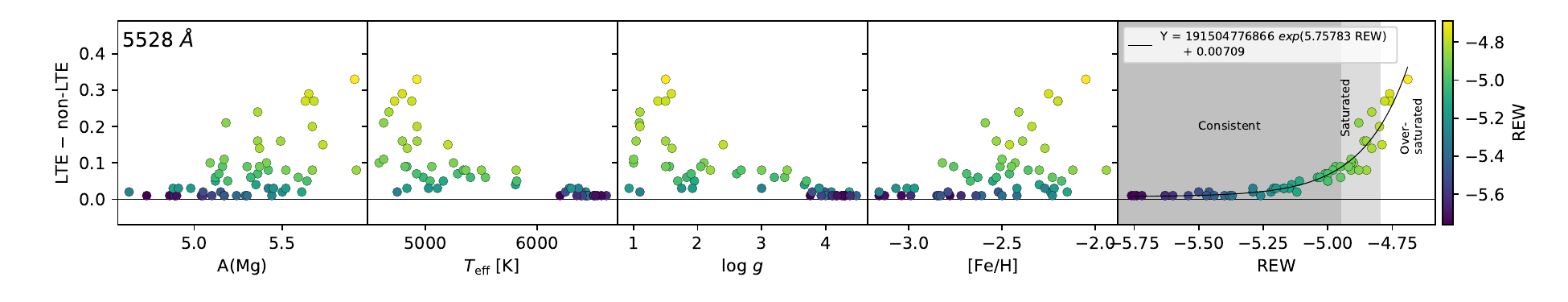}\\
    \includegraphics[width=1\linewidth]{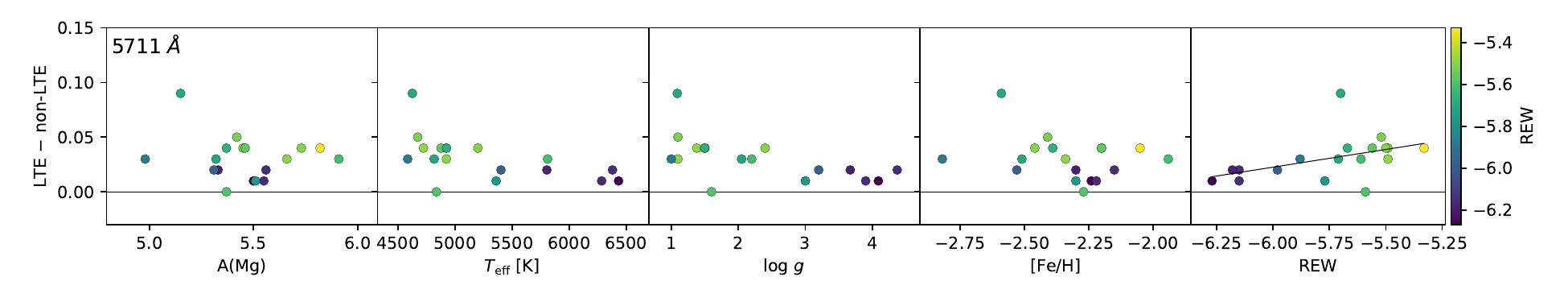}\\
    \caption{\tiny Comparison between 1D LTE and 1D NLTE Mg determination.  
    {\it Top panels:}
    Difference between 1D LTE and 1D NLTE  abundance as function of atmospheric parameters, abundance, and REW for the line at 5528~\AA. 
    The rightmost panel displays an exponential function fitted to the dispersion, its coefficients are given in the plot.
    Shades cover three regions of REW: Consistent ($REW < -4.95$), saturated ($-4.95 \leq REW < -4.80$), and over-saturated ($REW > -4.8)$); the explanation is given in the main text.
    Dispersions of the points around the fitted function are 0.0072 (consistent region), 0.0399 (saturated region), and 0.0397~dex (over-saturated region).
    {\it Bottom  panel:} Similar to top panels for the line at 5511~\AA.}
    \label{fig:LTE_NLTE}
\end{figure*}

We applied line synthesis with Turbospectrum using NLTE departure coefficients based on the model atom used in \cite{bergemann2017ApJ...847...15B}.
Figure~\ref{fig:LTE_NLTE} shows the difference between LTE and NLTE abundances retrieved from the Mg~{\sc I} lines at 5528~\AA\ (top panel) and 5711~\AA\ (bottom panel) as a function of atmospheric (A(Mg), \teff, \logg, [Fe/H]) and spectroscopic parameters (reduced equivalent widths, REW).
Considering NLTE abundances as our reference, the differences displayed can be deemed as LTE biases. 
The maximum extent of the LTE biases of the line at 5711~\AA\ are significantly smaller than those of line 5528~\AA, as already reported in previous works \citep[e.g.][]{mashonkina2013A&A...550A..28M,bergemann2017ApJ...847...15B}. Namely, except for the coolest stars with lower surface gravity, its biases remain within the error bars of our determinations. 
For dwarf stars, LTE biases are observed to be lower than 0.04 and 0.025~dex, respectively, and therefore can be considered negligible.
That is to say, 1D LTE and 1D NLTE Mg abundances of our metal-poor dwarfs are almost equivalent.
These results differ from those of \cite{andrievsky2010A&A...509A..88A}, where the abundances of the dwarfs are more affected than those of giants by NLTE corrections.  Namely, in average, NLTE Mg of dwarfs appears larger than Mg LTE by $\sim$0.3 dex  in that work.

For giants, the LTE biases of the line at 5528~\AA\ become larger as \logg\ and \teff\ decrease.
Since the absolute Mg abundance is naturally correlated with [Fe/H], the largest LTE biases remain at [Fe/H] $>-2.5$~dex.
The bias of the line at 5528~\AA\ responds more precisely to the REW variation, following the exponential function displayed in the plot of Fig. \ref{fig:LTE_NLTE}.
However, the standard deviation around the exponential function  increases with REW.
The "consistent" region (REW $< -4.95$) has a negligible standard deviation value of 0.007~dex, whereas the "saturated" and over-saturated regions have a much larger  standard deviation of $\sim$0.039~dex, which is comparable to the [Mg/Fe] average uncertainty of Precision Samples I and II.
The plots in the figure show that out of the consistent region\footnote{As a generic rule, stars stars out of the consistent region (REW~$\geq -4.95$) have \teff~$< 5000$~K and \logg~$< 2$~dex.}, the LTE bias is influenced by the combination of several stellar parameters, and therefore it is more susceptible to the accuracy of all of them.
Section~\ref{sec:growth} 
shows how the NLTE corrections of the Mg abundance evolve as A(Mg)\footnote{$A(X)$ is the logarithmic abundance of element X in the scale $\log N(\rm{H}) = 12$.} increases. Large values are observed out of the consistent region for the line at 5528~\AA, unlike for the line at 5711~\AA, which remains unsaturated even for very high Mg abundances.

\begin{figure*}
    \centering
    \includegraphics[width=1\linewidth]{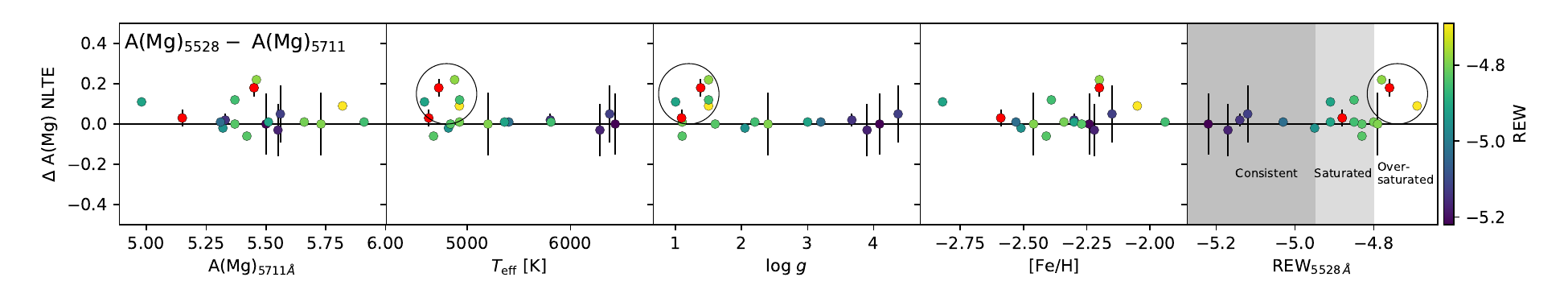}\\
    
    \caption{\tiny 
    NLTE abundance differences from each line. Error bars are fitting errors related to the noise added in quadrature; only those lower than 0.2~dex are shown.
    Large circles point the parameter regions with the largest offsets. Red points indicate the stars HD~122563 and BD+09~2870.}
    \label{fig:Mg_NLTE_dif}
\end{figure*}

Most stars in our sample (37 of 55) have abundances only from the line at 5528~\AA; either because the spectra do not include the weaker line at 5711~\AA, or because it is not detected. 
We perform the following analysis to verify that our abundance scale is independent of the line choice and invariant  within the parameter range of our stars. 
With 18 stars, we compare the consistency of the NLTE outcomes of both lines.
In Figure~\ref{fig:Mg_NLTE_dif} we show the abundance difference, where only error bars lower than 0.2~dex are plotted. Only the giants HD~122563 and BD+09~2870 have very precise measurements, these are indicated by the red points.
Values are compatible for dwarfs and for giants with \teff~$\gtrsim 5000$~K and \logg~$\gtrsim 1.8$~dex.

\subsection{3D LTE vs 1D LTE (Scate code)}

We used the 3D LTE code \texttt{Scate} \citep{2011A&A...529A.158H} to
calculate 3D LTE versus 1D LTE abundance corrections for the $5528$ \AA\ line.
The calculations were
performed for $20$ snapshots across $113$ available models from the original
\texttt{Stagger}-grid of 3D radiative-hydrodynamical model atmospheres
\citep{2013A&A...557A..26M}; see also the updated grid of
\cite{2024A&A...688A.212R}, with [Fe/H] of $0$, $-1$, $-2$, and
$-3\,\mathrm{dex}$, and spanning dwarfs and giants.  The 1D calculations were performed on the
1D version of the \texttt{Stagger}-grid 
\cite[the \texttt{ATMO}-grid; Appendix A of][]{2013A&A...557A..26M}.  The 1D calculations were done with four
different values of microturbulence: $0$, $1$, $1.5$, and
$2\,\mathrm{km\,s^{-1}}$.
The model atmospheres were constructed assuming the 
solar chemical composition of 
\citet{2009ARA&A..47..481A}, scaled by [Fe/H]
and with $\alpha$-enhancement of 
$+0.4\,\mathrm{dex}$ for the metal-poor models.

The magnesium abundance was always set to match that with which the model
atmosphere was constructed.  The abundance corrections were calculated by
varying $\log gf$ such that the equivalent width in the 3D synthesis matched
that of the 1D synthesis, for a given model atmosphere and given 1D $\Delta \log
gf$. The abundance corrections
are thus a function of \teff, \logg, 
[Fe/H], $v_{mic}$, and 1D ``abundance'' (as inferred from the solar-scaled
and $\alpha$-enhanced model atmosphere composition and $\Delta \log gf$). 
These corrections were then interpolated onto the stellar parameters
and 1D LTE abundance measured from the 5582 \AA\ line.

\subsection{3D NLTE vs 1D LTE (Balder code; dwarfs)}

We took 3D NLTE vs 1D LTE abundance corrections for the 5582 \AA\, from \citet{matsuno2024A&A...688A..72M}.  These were calculated with
the code \texttt{Balder} \citep{amarsi2018}, which is based on
\texttt{Multi3D} \citep{2009ASPC..415...87L}.  As with the 3D LTE abundance
corrections described above, these 3D non-LTE abundance corrections are based on
post-processing of the \texttt{Stagger}-grid of 3D radiative-hydrodynamical
model atmospheres \citep{2013A&A...557A..26M}.  The corrections are a function
of \teff, \logg, $v_{mic}$, and 1D LTE abundance; the abundances of other elements
were scaled with the magnesium abundance.

\subsection{Determination of the uncertainties on [Mg/Fe]}
The errors of A(Mg) are computed as explained in  
\cite[][Sect. 6.1]{giribaldi2023A&A...679A.110G}. We briefly recall the method used. We prepared grids of 1D LTE theoretical line profiles at 5528~\AA\ using the following parameters: for dwarfs, \teff\ from 5500 to 6500~K with steps of 200~K, [Fe/H] from $-1.5$ to $-3.0$~dex with steps of 0.5 dex, log $g$ = 4, and A(Mg) from 4.8 to 7.2~dex with steps of 0.4~dex; for giants, \teff\ from 4500 to 6500~K with steps of 500~K, [Fe/H] from $-1.0$ to $-3.0$~dex with steps of 1~dex, log $g$ from 1.0 to 3.0~dex with steps of 1~dex, and A(Mg) from 4.8 to 7.2~dex with steps of 0.4~dex. Then, we used those grids as they were observational spectra to derive Mg abundances from biased \teff, [Fe/H], \logg, and $v_{mic}$ by +40~K, +0.10~dex, +0.10~dex, and +0.10~dex, respectively.
This procedure gives biased A(Mg) values,  whose difference with the actual A(Mg) values provides the offsets $\Delta$(A(Mg))$_{param}$ related to the parameter biases above, where ${param}$ represents \teff,\logg, [Fe/H], or $v_{mic}$.
The errors related to the uncertainties of the parameters in Table~\ref{tab:parameters} are obtained by multiplying the required factor. 
For example, if the \teff\ error of the star is 80~K, $\Delta$(A(Mg))$_{T_{\mathrm{eff}}}$ is multiplied by 2; or if the [Fe/H] error of the star is 0.05, $\Delta$(A(Mg))$\mathrm{_{[Fe/H]}}$ is multiplied by 0.5.

For both dwarfs and giants, the dominant source of error is due to the uncertainty in \teff, while uncertainties in $v_{mic}$ and [Fe/H] errors produce negligible $\Delta$(A(Mg)) of the order of 0.001~dex. For dwarfs, also \logg\ errors produce negligible $\Delta$(A(Mg)), while it can be important for giant stars.
Total A(Mg) errors are computed by adding all $\Delta$(A(Mg))$_{param}$ in quadrature, and to compute the [Mg/Fe] total error we added in quadrature those of [Fe/H] in Table~\ref{tab:parameters}.
The table lists the [Mg/Fe] errors in the column of 1D NLTE; however; they are also ascribed to all 1D LTE, 1D NLTE, 3D LTE and 3D NLTE determinations. 
Since the temperatures are listed along with their internal errors (they do not include the uncertainty of the model 48 K), [Mg/Fe] errors of many stars are dominated by the errors of [Fe/H]. The stars with the largest [Mg/Fe] errors, e.g. those from BPS BS 16968-061 to BPS CS 29518-043 and those of the {\it Gaia}-ESO sample, own their errors to weak lines and low S/N.

\section{Results}
\label{sec:results}

In this section, we present the results of the chemical and kinematic analysis of our sample stars, and the two chemical evolution models used to compare with them.

\subsection{Compatibility of [Mg/Fe] scales between samples}
\label{sec:compat}

The main results of our analysis are shown 
in Fig.~\ref{fig:pop}.
There, we emphasise the Gaia-Enceladus and Milky Way (MW) populations separated in Paper~I as reference in background.
The latter embraces the populations called Splash or heated thick disc \citep[][and references in the paper]{belokurov2020MNRAS.494.3880B} and Erebus, which is an old (10-12~Gyr) retrograde population of low binding Energy with [Fe/H] extending up to $-0.5$~dex.
We expand on the characteristics of these populations in Sect.~\ref{sec:kinematics}.

In the four panels, we show the abundances of magnesium derived with different assumptions (1D LTE, 3D LTE, 1D NLTE and 3D NLTE) in the [Mg/Fe] vs. [Fe/H] plane; [Fe/H] is instead derived under 1D NLTE assumptions. 
Magnesium abundances of the Milky Way (Splash and Erebus) and Enceladus sequences in the figure were derived by means of the 1D MARCS models \citep{gustafson2008} under LTE assumption. 
This stands for the two sub-samples from which these populations were identified: the \titan~I \citep{giribaldi2021A&A...650A.194G} and GALAH  \citep{buder2021MNRAS.506..150B}, as stated in their respective papers.
The Mg abundance scale of GALAH considers only the line at 5711~\AA, whereas that of the \titan~I includes both lines at 5528 and 5711~\AA, which, in any case, were confirmed to be equivalent, as shown in Fig.~3 of Paper~I.
Both the GALAH and \titan~I subsamples apply nearly identical 1D NLTE corrections. For the former, values are in between 0 and +0.05 \citep[Fig.~7 in][]{amarsi2020A&A...642A..62A}, whereas for the latter, a constant value of +0.05~dex  \citep{mashonkina2013A&A...550A..28M} was used.
The Mg scale of the GALAH sub-sample was originally derived from a \teff\ scale distinct to that of the \titan~I.
Therefore, they were adjusted for compatibility by subtracting the relative offsets of the former, which were determined to be a constant of +0.07dex; see Sect.~5.3 of Paper~I for details. 
Regarding the [Fe/H] scales, metallicity values of the GALAH sub-sample were not adjusted to those of the \titan~I. This is because, according to the tests in \cite{buder2021MNRAS.506..150B}, [Fe/H] of metal-poor GALAH stars appear to be compatible with the scale of the Gaia benchmark stars \citep{jofre2014A&A...564A.133J}; 
whereas the TITANS~I were proven to be compatible to that scale in \cite{giribaldi2021A&A...650A.194G}.
Consequently, considering all the above, we can assert that the [Mg/Fe] scales of the GALAH and \titan~I subsamples from Paper~I (which compose the Splash, Erebus, and Enceladus sequences in Fig.~\ref{fig:pop}) are compatible and apply equivalent 1D LTE to 1D NLTE corrections.

We remark that, in Paper~I, the [Mg/Fe] of the Splash, Erebus, and Enceladus populations considered a diffusion correction of +0.05~dex only for stars with [Fe/H] $< -1.65$~dex following the findings of \cite{korn2007ApJ...671..402K} and \cite{gruyters2013A&A...555A..31G}.
In line with the findings of \cite{souto2018ApJ...857...14S,souto2019ApJ...874...97S}, we instead apply a constant diffusion correction of +0.05~dex for all dwarfs across the entire metallicity range, including the very metal-poor tail.
However, see the work of \cite{nordlander2024MNRAS.52712120N}, which suggests that the effect of diffusion on [Mg/Fe] in stars with [Fe/H] $\sim -1$~dex is even closer to zero.
In any case, the [Mg/Fe] sequences of the stars we compare our very metal-poor sample with are only marginally affected by these corrections.

We interpolated the 1D LTE to 3D NLTE Mg abundances of \cite{matsuno2024A&A...688A..72M}, in order to obtain a 3D NLTE [Mg/Fe] scale of the Splash, Erebus, and Enceladus stars in the right bottom panel of Fig.~\ref{fig:pop}.
We obtain 
the 1D LTE scale of these stars in the top left panel by subtracting their 
1D NLTE correction of 0.05.
The 1D NLTE abundances in the top right panel remain as described above.
Thus, we can effectively assess which chemical sequence best connects with the most metal-poor tail of the Galactic halo.

\subsection{The [Mg/Fe] vs [Fe/H] patterns of our samples}
\label{sec:results2}

In each panel of Fig.~\ref{fig:pop}, robust LOWESS\footnote{Locally Weighted Scatterplot Smoothing (LOWESS) regressions are applied by the Python {\textit{moepy}} package \cite{LOWESS} available at \url{https://ayrtonb.github.io/Merit-Order-Effect/}, using the parameter frac=0.55.} regressions are shown along with their dispersions characterised by 20 to 80\% quantiles. 
The first outcome to remark is the decrease of scatter when using NLTE corrections: at a given [Fe/H] the typical scatter in [Mg/Fe] changes from 0.07-0.27 to 0.04-0.08~dex (see shades in the plots). With 3D LTE models, the scatter remains between 0.06 and 0.12~dex, it is computed only from dwarfs; 
Figure~\ref{fig:dwarfs_dispersions} shows the [Mg/Fe] dispersion variation when only dwarfs are considered.
The difference between the panels on the right side in Fig.~\ref{fig:pop} lies in the possibility of applying the 3D models and NLTE corrections only to dwarf stars, i.e. to the Precision I sample and two stars of the {\it Gaia}-ESO sample. 
In the case of the 1D NLTE instead, abundances are calculated for both dwarf and giant stars. 
Using NLTE corrections (right panels), a clear sequence in the [Mg/Fe] vs [Fe/H] plane emerges for the  metal-poorer population: from [Mg/Fe] $\approx +0.6$~dex for [Fe/H]~$<-3$~dex to  [Mg/Fe]~$\approx+0.4$~dex for [Fe/H]~$\approx- 2$~dex.

In the case of 1D LTE synthesis, the scatter is enhanced by the giant stars at $-2.7 \lesssim$ [Fe/H] $\lesssim -2$~dex to high [Mg/Fe] values, which prevents distinguishing a clear pattern, as typically observed in the literature.
The case of 1D NLTE displays a narrow dispersion (i.e. 0.07-0.09~dex for $-2.5 < $ [Fe/H] < $-2.0$~dex),
where dwarfs and giants overlap. One giant of the {\it Gaia}-ESO sample (GESJ10142785-4052503) lies far out of the sequence  at about  [Mg/Fe] $= 0.64$~dex. Its H$\alpha$ profile displays a strong asymmetry, and instead we derived its \teff\ by photometric relations. Therefore, we do not rule out a possible [Mg/Fe] bias arising from atmospheric parameters.

The plot with [Mg/Fe] derived in 3D NLTE contains only dwarf stars; as mentioned above, for this sample we were able to calculate ages (see histogram in Fig.~\ref{fig:hist}). These are shown as a colour scale in the bottom panel on the right. As expected, the sample is composed of very old stars, with ages larger than 11 Gyr, and a median age of 12.5~Gyr. 
The dispersion has a similar shape to that of 1D NLTE, but it is offset upwards by +0.2~dex (see also Fig.~\ref{fig:dwarfs_dispersions}).
Thus, from a purely observational point of view, we can conclude that the selected sample of old and metal-poor halo stars (see Fig.~\ref{fig:space} for their spatial distribution) defines a  somewhat tight sequence in the [Mg/Fe] vs [Fe/H] plane with a trend in [Mg/Fe] that decreases up to [Fe/H]~$\sim-2.8$~dex, and then remains flat at [Mg/Fe] $\sim0.4$~dex for higher metallicities. This sequence becomes better defined only when NLTE corrections are properly taken into account.
This could be one of the reasons why in large surveys, which generally make use of 1D LTE analysis, this portion of the [Mg/Fe] vs. [Fe/H] plane looks generally more dispersed. 
In the next sections, we combine the chemical and kinematic properties of these stars to make some hypotheses on their origin.

\subsection{The kinematic properties of our samples}
\label{sec:kinematics}
We computed orbital parameters using the code GalPot\footnote{\url{https://github.com/PaulMcMillan-Astro/GalPot}} of \cite{McMillan2017MNRAS.465...76M} assuming the same Galactic potential and solar kinematic parameters in \cite{buder2021MNRAS.506..150B} to perform an analysis compatible with that in Paper~I.
Coordinates, proper motions, distances, and radial velocities are the required inputs.
We use Gaia DR3 coordinates and proper motions. Radial velocities were computed from Doppler shifts, coordinates, and observation dates in the spectral files for Samples I and II; catalogued velocities were assumed for Sample III.
Distances were extracted from the catalogue of \cite{bailer-jones2023AJ....166..269B}.

\begin{figure*}
    \centering
    \includegraphics[width=0.47\linewidth]{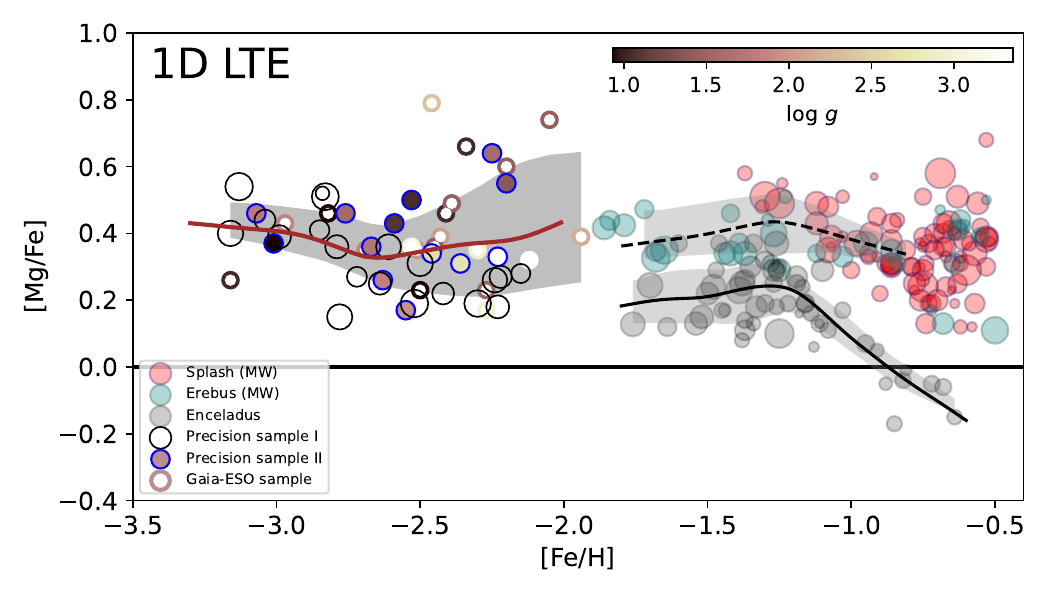}
    \includegraphics[width=0.47\linewidth]{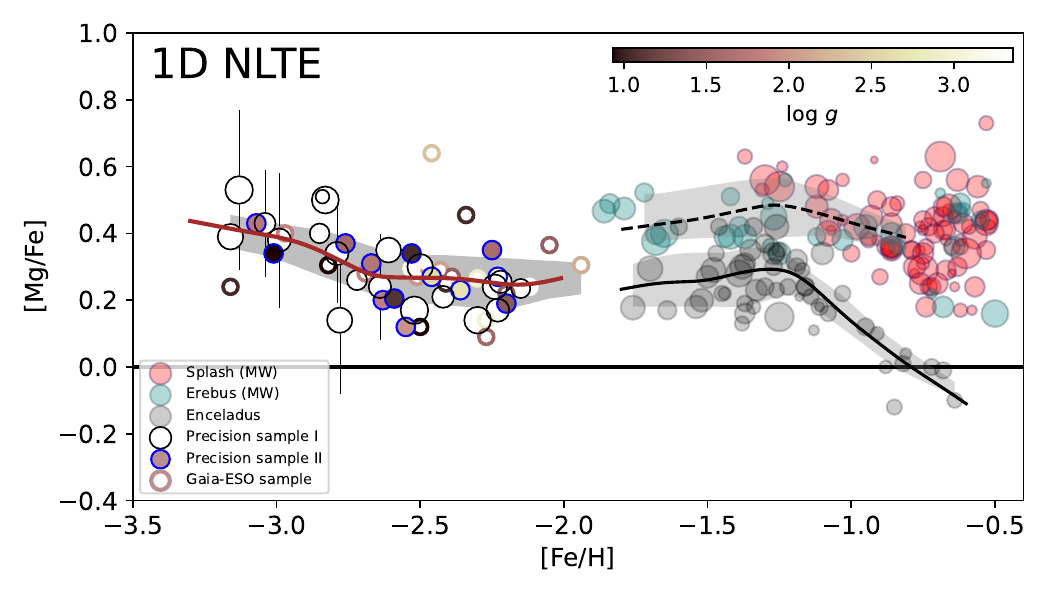}
    \includegraphics[width=0.47\linewidth]{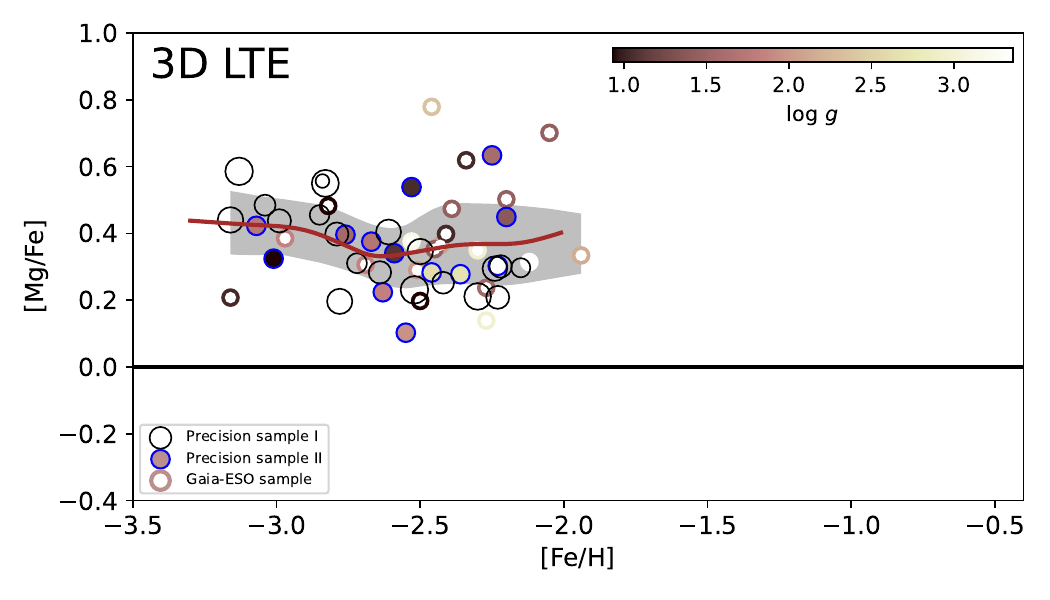}
    \includegraphics[width=0.47\linewidth]{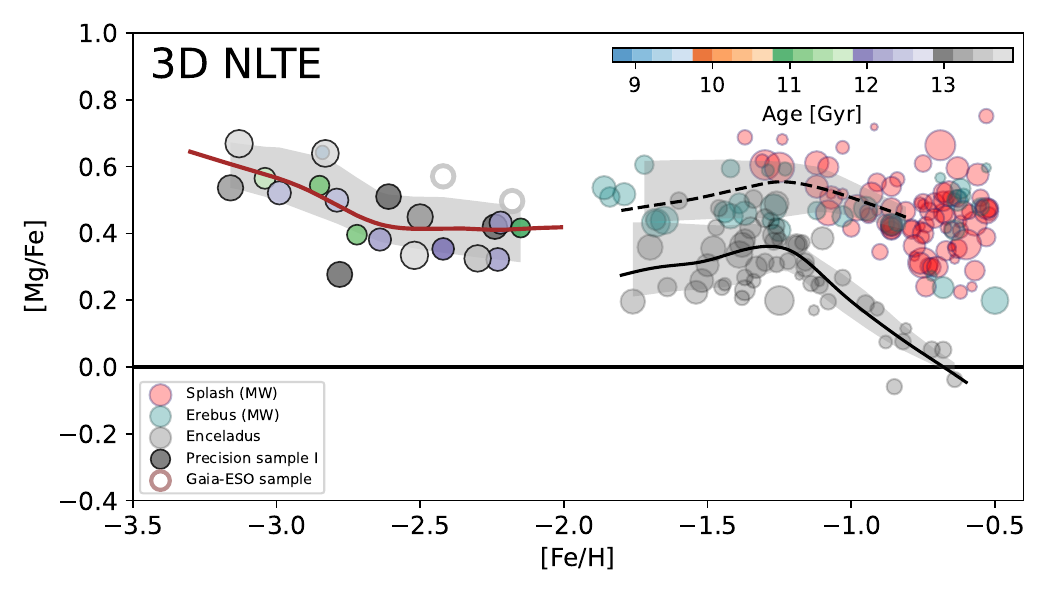}
    
    \caption{
    [Mg/Fe] vs. [Fe/H] for our sample stars:
    1D LTE, 1D NLTE, 3D LTE, and 3D NLTE abundances are displayed according to the legends; all panels are built with [Fe/H] under 1D NLTE. Symbols of dwarfs  are size-coded according to their ages.
    As a reference, stars classified as the in-situ populations Splash (red) and  Erebus (turquoise), and   the ex-situ population Gaia-Enceladus (gray) in \cite{giribaldi2023A&A...673A..18G} are plotted.
    Continuous red,  dashed black, and continuous black lines display 
    LOWESS regressions. 
    The shades display 20 to 80\% quantiles.
    {\it Bottom right panel: } Only dwarfs are displayed, as only these have 3D NLTE corrections.  
    They are colour-coded according to age.
    {\it Other panels: }
    Precision sample I (only dwarfs) has no colour coding.
    The Precision sample II and {\it Gaia}-ESO sample are colour codded according to \logg. Their symbols follow the legends in the plots.
    The panel 1D NLTE displays error bars of the stars in Precision samples I and II with errors larger than 0.14~dex.
    }
    \label{fig:pop}
\end{figure*}

\begin{figure*}
    \centering
    \includegraphics[width=0.8\linewidth]{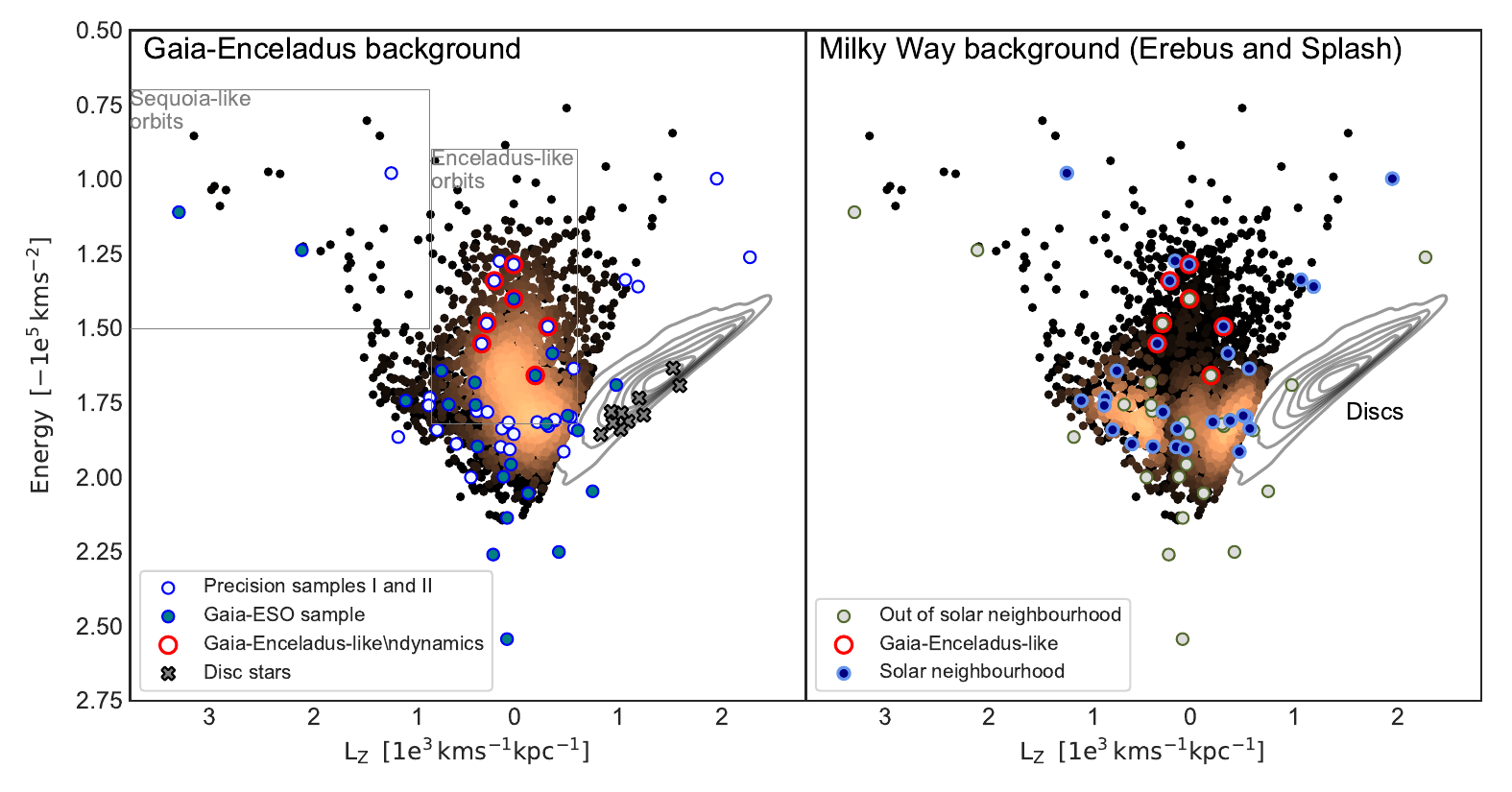}
    
    \caption{ Lindblad diagrams (Energy vs L$_{Z}$): 
    Halo stars of the GALAH survey are plotted as reference background, colour-code by density and in  each panel the colour-map emphasises the Gaia-Enceladus and Milky Way (Erebus and Splash) populations, respectively, using the separation  in \cite{giribaldi2023A&A...673A..18G}.
    \textit{Left panel:}
    The Precision samples and the {\it Gaia}-ESO sample are represented with the same symbols as in Fig.~\ref{fig:space}.
    The boxes enclose the areas with the highest probability of containing Gaia-Enceladus and Sequoia stars according to \cite{massari19}.
    Red circles point stars with Gaia-Enceladus-like dynamics in the current sample.
    {\it Right panel:}  
    Stars in and out of the solar neighbourhood are distinguished by different colour symbols.
    Also the stars with Gaia-Enceladus-like orbits are highlighted.
    }
    \label{fig:lindblad}
\end{figure*}

The Enceladus and the Milky Way populations (the Splash and Erebus) were identified in Paper~I as follows.
First, clustering was performed in the [Mg/Fe]–[Fe/H] space, considering only stars with halo-like orbits. These were identified by excluding regions occupied by the Milky Way discs in the Lindblad diagram. 
The distinctive knee-like shape of Enceladus, characterized by a relatively low [Mg/Fe] plateau (at $\sim$0.25~dex), was successfully recovered. Its stars were confirmed to cluster around a net $L_{Z}$ of approximately zero while being widely distributed along the Energy axis of the Lindblad diagram (see the background concentration in the left panel of Fig.~\ref{fig:lindblad}).
The Enceladus stars were removed from the initial sample to allow for clustering separation of the remaining stars, which exhibit a relatively high [Mg/Fe] plateau at $\sim$0.5~dex, in the Lindblad diagram.
There, a prograde (Splash) and a retrograde (Erebus) populations appeared well concentrated (see the background concentration in the right panel in Fig.~\ref{fig:lindblad}). 
When plotted in the [Mg/Fe]–[Fe/H] diagram, the stars of the Splash appear concentrated at high [Fe/H], whereas those of Erebus are more uniformly distributed along the [Fe/H] axis.
This feature is displayed in Fig.~\ref{fig:pop}, where red and turquoise circles represent each population, respectively. 
The age-[Fe/H] relations of these populations were observed to exhibit higher iron abundances at earlier times than Enceladus (Fig.~8 in Paper~I), leading to the conclusion that both formed within a more massive galaxy and therefore likely formed in situ. 
These findings show that kinematic clustering alone does not definitively determine the origin of stellar populations, as seen in the simulations of \cite{JeanBaptiste2017} and \cite{koppelman20}.

The statistics of our very metal-poor sample is not large enough to  apply machine learning algorithms for an  efficient separation of different populations in kinematic diagrams \citep[e.g.][]{giribaldi2023A&A...673A..18G,da_silva2023A&A...677A..74D, Berni2024arXiv240911429B}.  Therefore, we made a visual inspection in  Fig.~\ref{fig:lindblad}.
The left panel separates Precision samples I and II from the {\it Gaia}-ESO sample. 
The latter has several stars that seem to be associated to the bulge, as these have very high binding energy  (highly negative) and are close to the Galaxy centre $R_{\mathrm{GC}} \lesssim 5$~kpc (see Fig.~\ref{fig:space}).
For example, the stars with identification numbers 
GESJ17545552-3803393, 
GESJ18185545-2738118, 
GESJ18265160-3159404, 
GESJ18361733-2700053,
GESJ18374490-2808311,
and GESJ15300154-2005148 have binding energy  $< -2\times10^{5}$~km~s$^{-2}$ and $R_{\mathrm{GC}} < 3.9$~kpc, and so they may be the bulge members.

The right panel in Fig.~\ref{fig:lindblad} shows that an important fraction of the stars are positioned in correspondence with the populations of the MW, regardless if stars belong to the solar neighbourhood (blue circles) or not (gray circles).
The stars BD+24~1676 (dwarf), BPS~CS 22166-030 (dwarf), HD~140283 (dwarf), BPS CS 22953-003 (giant), BD+09 2870 (giant), and GESJ01293652-5020327 (giant) have Gaia-Enceladus-like orbits:  nearly null rotation and high eccentricity (stars  marked in red in Fig.~\ref{fig:lindblad}).
Figure~\ref{fig:GE_like} shows that these stars do not have particularly low [Mg/Fe] with respect to the trend of the sample.
Also, our sample has three stars that would be candidates of belonging to the Sequoia \citep{barba2019ApJ...870L..24B,myeong2019MNRAS.488.1235M} according to the selection box of \cite{massari19} in the Lindblad diagram (Fig.~\ref{fig:lindblad}).
As Fig.~\ref{fig:GE_like} shows, neither of these stars displays low [Mg/Fe] with respect to the main trend, as expected by \cite{matsuno2019ApJ...874L..35M} and \cite{koppelman19}, for instance.
These results exclude the possibility of discerning [Mg/Fe]-[Fe/H] features of populations of potential diverse origin at the range [Fe/H] $\lesssim -2$~dex, assuming that selection by grouping into boxes in the Lindblad diagram is efficient at that.
It is worth noting that \cite{Belokurov2018} already demonstrated that, within this metallicity range,  
stars from Gaia-Enceladus \citep[a $\sim$10$^{9} M_{\odot}$ merger; e.g.,][]{vincenzo10.1093/mnrasl/slz070, belokurov2020MNRAS.494.3880B} are scarce,  
making its kinematic imprint no longer detectable in the radial–azimuthal velocity diagram.
Consequently, we consider it most viable to regard the stars in this metallicity range as having formed either in the early Milky Way—by then already the dominant mass contributor of the Local Group alongside M~31—or in smaller progenitor galaxies of comparable stellar mass, which were later assimilated \citep[e.g.][]{malhan2024ApJ...964..104M}. Given their similar age and metallicity, distinguishing between the two scenarios remains challenging, as no quantitative evidence currently favours one over the other.

\subsection{Chemical evolution models}
To interpret the chemical properties of our metal-poor star sample (see Sect.~\ref{sec:results2}), we applied two distinct chemical evolution models: one specifically tailored for Boötes~I, one of the most extensively studied Ultra Faint Dwarf (UFD) galaxies both observationally and theoretically \citep[see, e.g.,][]{Koch2009ApJ...690..453K, Lai2011ApJ...738...51L}, and a Galactic Chemical Evolution (GCE) model \citep{Romano17, Romano19, Rossi24CNO} that traces the composition of stars formed in situ in the Milky Way.

By comparing the data with the predictions of these two models, we aim to analyse the differences between accreted stars, i.e. stars born in UFD progenitors, and stars born in situ, thereby shedding light in a quantitative way on the origin of the population under study.

\subsubsection{Ultra faint dwarf (UFD) model}
 The model has been extensively described in a series of papers \citep{Rossi+21, Rossi+23, Rossi2024arXiv}, to which we refer for further details. Here, we present a brief summary of the main assumptions underlying the model. The model tracks the formation and evolution of a Boötes~I-like galaxy from the epoch of its formation until redshift $z =0$. It assumes that the UFD galaxy evolves in isolation, without undergoing major merger events, in line with cosmological models \citep{SS09, SS15}. At the epoch of virialisation, the total amount of gas  is assumed  to have a pristine chemical composition. Star formation becomes possible when the gas begins to fall into the central region of the halo and cools down \citep{Rossi+21}. The total mass of stars formed within a timescale of 1 Myr is evaluated by assuming a star formation rate, $\Psi$, regulated by the free-fall time ($\rm t_{ff}$) and the mass of cold gas accreted via infall ($\rm M_{cold}$): $\Psi = \epsilon_{\star}\rm M_{cold}/t_{ff}$, where $\epsilon_{\star}$ is a free parameter of the model. No specific assumption regarding the star formation history is made.
 
The model assumes that according to the metallicity of the interstellar medium (ISM)   Population III (Pop~III) or Population II/I (Pop II/I) stars can form. Following  the critical metallicity scenario \citep{bromm01, Schneider2003, Omukai05}, Pop~III stars form only if the metallicity of the gas\footnote{$\rm Z_{gas} = M_{ISM}{Z}/M_{gas}$, where $\rm M_{ISM}{Z}$ and $\rm M_{gas}$ represent the mass of metals and gas in the ISM, respectively.}, $\rm Z_{gas}$, is below a critical threshold, $\rm Z_{cr} = 10^{-4.5 \pm 1}Z_{\odot}$, i.e., when $\rm Z_{gas} \leq Z_{cr}$. Once $\rm Z_{gas}$ exceeds this threshold,  Pop~II/I stars begin to form. The newly formed stellar mass is distributed according to the Initial Mass Function (IMF) of \cite{Larson1981MNRAS.194..809L}, with a characteristic mass of $m_{ch} = 10$~$\rm M_{\odot}$ for Pop~III stars and $m_{ch} = 0.35$~$\rm M_{\odot}$ for Pop~II/I stars, within the mass ranges 0.8--1000~$\rm M_{\odot}$ and 0.1--100 $\rm M_{\odot}$, respectively.

The model accounts for the incomplete sampling of the stellar IMF for both Pop~III and Pop~II stellar populations. Each time the conditions for star formation are reached, a discrete number of stars is formed, with their masses randomly drawn from the specified IMF. This feature is crucial for accurately simulating the evolution of UFDs with low star formation rates, where the IMF is never fully sampled over the galaxy’s lifetime (see \citealt{Rossi+21}). The chemical enrichment of the gas is tracked by considering the mass-dependent evolutionary timescales of individual stars. Specifically, the semi-analytical model follows the release of chemical elements (from carbon to zinc) by supernovae (SNe) and AGB stars. The contribution of supernovae Type Ia (SNIa) has been included adopting the yields of \citet{Iwamoto99} and the bimodal delay time distribution observationally derived by \citet{Mannucci06}. Ad each time-step the rate of SNIa is computed following the prescriptions of  \citet{Matteucci06}.

For Pop~III SNe in the mass range 10--100~$\rm M_{\odot}$, we adopt the stellar yields of \cite{Heger+woosley10}, while for Pair Instability SNe (PISN) in the mass range $m_\text{PISN}=$ 140--260~$\rm M_{\odot}$ we use the ones of \cite{heger02}. For less massive Pop~III stars in the mass range 2--8~$\rm M_{\odot}$ we account for the AGB yields of \cite{Meynet+02} (rotating model). For Pop~II/I stars, we employed stellar yields dependent on the initial mass and metallicity of the stars. Specifically, for Pop~II/I SNe  arising from progenitors in the mass range 8--40~$\rm M_{\odot}$ we employed the stellar yields of \cite{Limongi+18}, their set R, which assumes no rotation, while for AGB stars the ones of \cite{VanDenHoek+97}. The mechanical feedback from SNe is accounted. Indeed, SNe explosions can produce blast waves that, if sufficiently energetic, are capable of overcoming the gravitational potential of the host halo, leading to the ejection of gas and metals into the intergalactic medium (IGM). The ejected mass, $\rm M_{ej}$, is given by: $\rm M_{ej} = \frac{2 \epsilon_w N_{SN} E_{SN}}{v_{\rm esc}^{2}}$, where $\epsilon_w$ is the efficiency of the SNe-driven winds and it is a free parameter of the model, $\rm N_{SN}$ is the number of SNe, $\rm E_{SN}$ is the explosion energy, and $v_{\rm esc}$ is the escape velocity of the halo, which depends on its mass, $\rm M_{h}$, and virial radius.  The ejected gas fully mixes with the ISM, ensuring its metallicity matches that of the ISM. The free parameters ($\epsilon_{\star}$, $\epsilon_{w}$) have been fixed to reproduce the observed properties of Boötes~I UFD \citep[its total luminosity, star formation history, metallicity distribution function, and average stellar iron abundance; see, e.g.,][]{Brown14,  Simon19, Jenkins2021}, here taken as representative of the prototype UFD that interacted and merged with the Milky Way in its earlier evolutionary phases. 

\subsubsection{Galactic chemical evolution  model}
To simulate the chemical enrichment of stellar populations that formed in situ in the early Milky Way \citep[see, e.g.,][]{belokurov22} we use a comprehensive GCE model fully described in \citet{Romano17,Romano19,Romano2020} and \citet{Rossi24CNO}. The model incorporates both homogeneous and stochastic enrichment processes and integrates contributions from Pop~III stars, which play a pivotal role in the early phases of the Galaxy’s chemical evolution  \citep[see][]{Rossi24CNO}. The GCE model is constructed with a multizone approach, segmenting the Galactic disc into concentric annuli of 2 kpc width each. This allows us to account for radial variations in star formation, gas inflow, and enrichment processes across the disc.  For the comparison with the data sample analysed in this paper, we consider the model predictions relevant to an annulus centred on the Sun's position ($\rm R_{GC} =$~7--9~kpc). 
The model assumes an initial accretion phase of primordial material, which serves as the foundation for the formation of the in-situ halo and thick-disc components. During this early phase  \citep[lasting 3--4~Gyr, see][]{Spitoni2019, Spitoni2021}, the rapid gas accretion triggers an intense star formation activity, depleting the initial gas reservoir and producing the most ancient stellar populations. A second, prolonged accretion event leads to the formation of the thin disc, characterized by an “inside-out” growth pattern where the inner regions evolve more rapidly than the outer ones \citep{Romano20, Chiappini01}. The star formation rate is governed by the Schmidt-Kennicutt law \citep{Kennicutt98}, with a surface density power-law index of 1.5 and a radially dependent star formation efficiency to reproduce the observed gradients in the present-day MW disc \citep[see, e.g.,][]{Palla20}. The newly formed stellar mass is distributed in the mass range 0.1--100~M$_{\odot}$ according to the field IMF of \citet{Kroupa1993}. To follow the chemical enrichment, the model tracks the contributions from H to Eu from stars of different masses and lifetimes. The chemical enrichment from low- and intermediate-mass stars (LIMS; 1--6~M$_{\odot}$) is modelled using non-rotating yields provided in  the FRUITY database \citep{Cristallo2009,Cristallo2011,Cristallo2015}, which successfully reproduces MW abundances across various components \citep[][Romano et al., in prep]{Molero2023MNRAS.523.2974M}. For massive stars (13--120~M$_{\odot}$), exploding as core-collapse supernovae (CCSNe), we use the yields of \citep{Limongi+18} (set~R and rotating models with velocity 150 km/s). For SNe Type Ia, we adopt the yields of \citet{Iwamoto99}, which are critical for reproducing the [$\alpha$/Fe] decline at higher metallicities. The model also includes novae that became important after $\sim 1$~Gyr from the beginning of star formation, due to the time delay associated with their binary nature \citep[][and references therein]{Romano99}. To address the inhomogeneous enrichment characteristic of the MW’s early phases, we incorporate a stochastic star formation component. In this scheme, initial gas clumps form stars under conditions that mimic the localized, self-enriching star-forming regions that later merge into the Galactic halo. Each clump hosts a Pop~III star of randomly chosen mass (10–100~M$_{\odot}$) and explosion energy, selected from the grid of models computed by \citet{Heger+woosley10}. The wide range of SNe yields for Pop~III stars, together with clump-specific dilution factors, explains the significant scatter of abundances observed in MW halo stars \citep[see][for further details]{Rossi24CNO}.  In particular, the model is calibrated against CNO abundance data of metal-poor MW stars from recent spectroscopic surveys \citep{Mucciarelli22, amarsi2019}.

\begin{figure*}
\centering
\includegraphics[width=0.9\linewidth]{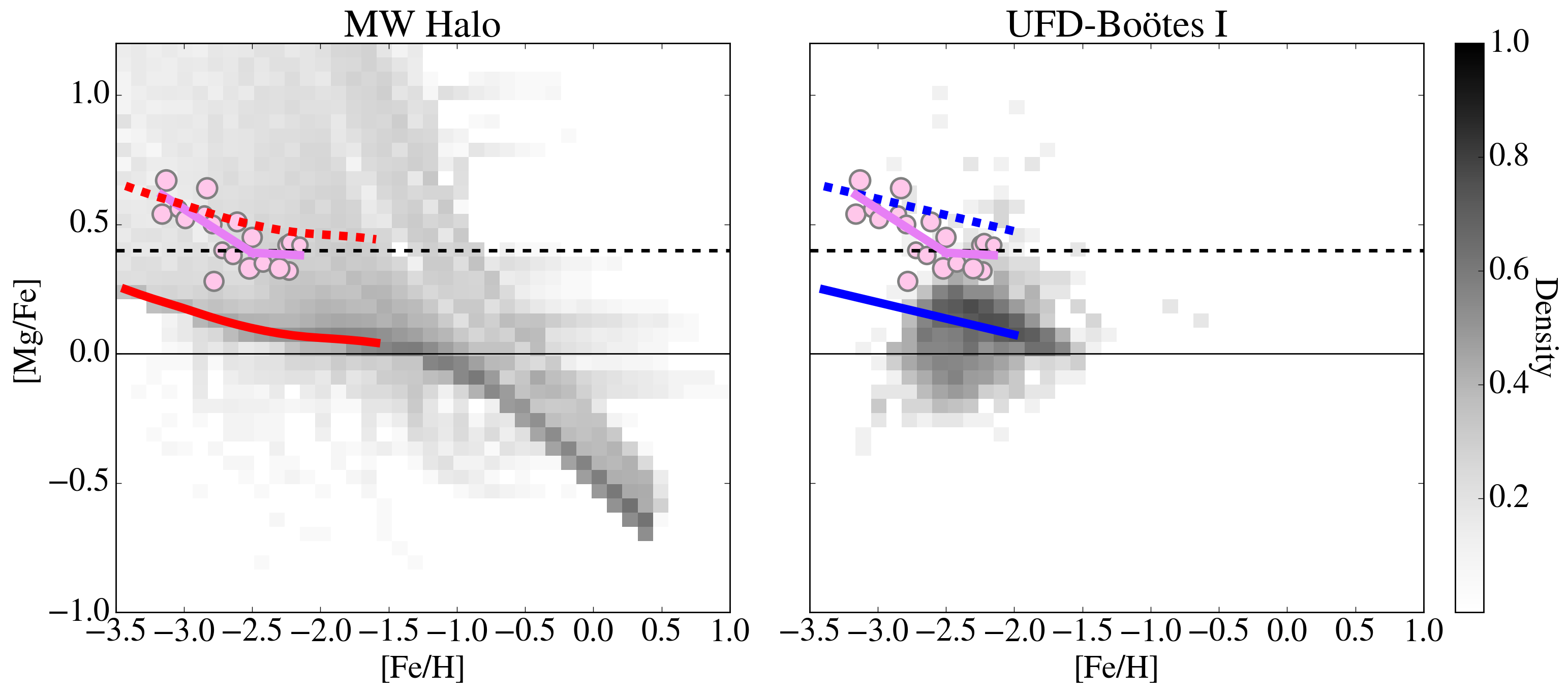}
    \caption{\tiny Predicted [Mg/Fe] versus [Fe/H] density maps. {\it Left panel:} Chemical evolution model for the Milky Way. {\it Right panel:} Chemical evolution model for UFD galaxy Boötes~I, taken as a prototype of the satellites accreted by the MW at early times. The highest density region of the GCE model and the average of UDF model are offset by +0.4~dex (dotted red line and dotted blue line, respectively) to allow a better comparison with the observed abundances (pink solid line is the LOWESS regression).  Circles represent Precision~I sample with Mg under 3D NLTE.} 
    \label{fig:model_martina}
\end{figure*}

\section{Discussion}
\label{sec:discussion}

By applying 1D NLTE and 3D NLTE spectral synthesis, we have demonstrated that at least part of the large [Mg/Fe] dispersion at the metal poorest tail ($-3.5 \lesssim$ [Fe/H] $\lesssim -2$ dex), typically observed in the literature, is due to the application of the LTE assumption on giant stars.
As shown in the right panels in Fig.~\ref{fig:pop}, our observational data composed of dwarfs and giants support a narrow [Mg/Fe] distribution, as previously proposed by \cite{arnone2005A&A...430..507A} and \cite{andrievsky2010A&A...509A..88A}, for instance.
Therefore, it indicates a low degree of stochastic enrichment $\sim$ 12.5~Gyr ago, which is the median age of our sample.  At the same time, a mildly decreasing trend with metallicity is observed, from [Mg/Fe]~$\simeq$~0.65 at [Fe/H]~$\simeq -$3.3 to [Mg/Fe]~$\simeq$~0.4 at [Fe/H]~$\simeq -$2.0.

We show that the sequence of the metal-poorest tail is compatible with that of the Milky Way (i.e. they join at [Mg/Fe] $\approx 0.4$~dex) only when 3D NLTE is applied (bottom right panel in Fig.~\ref{fig:pop}), whereas 
1D NLTE (and other assumptions) switch [Mg/Fe] down by $\sim0.2$~dex misleading continuity with the Gaia-Enceladus sequence.

By comparing the 3D NLTE chemical trend with the predictions of inhomogeneous chemical evolution models  that trace the chemical properties of stars formed in situ in the early MW and in a prototype UFD progenitor, taken as an example of a low-mass galaxy like the ones accreted by the Milky Way, we can try to place constraints on the origin of our metal-poor population. In Fig.\ref{fig:model_martina}, the predictions of the chemical evolution models for the early Milky Way (in situ component only) and the UFD galaxy Boötes~I are shown in the [Mg/Fe] versus [Fe/H] plane. The density maps represent stellar distributions predicted by the model. 
Individual data points correspond to our high-precision measurements  (3D NLTE abundances, shown also in the bottom left of Fig.~\ref{fig:pop})  allowing for a comparison between model predictions and observed abundances. 

The two density maps highlight distinct enrichment patterns for the  early MW and Boötes~I. The Milky Way  halo/early disc follows a more complex evolutionary path, with efficient and continuous star formation driving a rapid iron enrichment and a steady decline in [Mg/Fe] as [Fe/H] increases. In contrast, Boötes~I exhibits a simpler evolutionary path, marked by low star formation rates and more intermittent star formation episodes, shaped by stronger feedback effects and substantial mass loss \citep[see][]{Rossi2024arXiv}.  
The flat trend and the large scatter in [Mg/Fe] vs. [Fe/H] is largely due to the initial enrichment dominated by Pop~III stars, which contribute substantial magnesium in the early evolutionary phases. The Milky Way undergoes a more extended period of star formation, increasingly dominated by Pop~II stars, whose iron enrichment gradually reduces the initial chemical impact of the Pop~III contribution 
as the metallicity increases.
Both models display some scatter in [Mg/Fe] due to stochastic processes and  low star formation rate in  the earliest phases of their evolution,  although areas of higher stellar density show  more marked trends. In particular, in the model of the UDF, the regions of highest density, therefore most likely, are those with a high [Mg/Fe].  There is a good agreement -especially if a small offset is applied to the theoretical [Mg/Fe] ratio- with the values of [Mg/Fe] in our sample for  [Fe/H] around $-2.5$~dex, while stars at lower [Fe/H] with high [Mg/Fe] are less likely predicted in the UDF model. 
This might support the idea that some of the stars analysed in this work once belonged to dwarf galaxies. 
However, to confirm or not this result, other elements are needed to discriminate between in-situ and accreted origins; for instance, 
Zn could provide interesting hints \footnote{ At low metallicities, Zn is produced mostly by hypernovae  \citep[endpoints of the evolution of $m >$ 20~$\rm M_\odot$ stars with explosion energies much higher than those of normal CCSNe; see][]{kobayashi06,Romano10} that are expected much less numerous in low-mass systems undergoing inefficient star formation, such as UFDs.} \citep[see, e.g.,][]{Mucciarelli21}.

 Our sample could be composed of in-situ stars, as the comparison reveals an overall  consistency with the tail at high [Mg/Fe] for very metal-poor stars  in the MW model, once a positive offset is considered (as shown in the figure). This offset is indeed required due to the well-known problem of Mg under-production in stellar evolution and nucleosynthesis models, which leads to discrepancies between GCE model predictions and observations that are particularly pronounced at disc metallicities \citep[see, e.g.,][]{timmes95,Romano10,prantzos18,kobayashi20}. This sometimes prompted the usage of empirical yields to better fit the observations \citep[e.g.,][]{francois2004A&A...421..613F,Spitoni2019,Spitoni2021}. Here we deal with early inhomogeneous chemical enrichment utilizing different sets of yields/dilution factors \citep[see][]{Rossi24CNO} and do not apply any correction to the yields. However, we note that higher Mg yields from Pop~II stars would bring the high-density regions of the stellar maps shown in Fig.~\ref{fig:model_martina}  in better agreement with the loci occupied by the data. This is particularly true for the MW model, which is dominated by the contribution of Pop II stars. To guide the eye, a positive offset of +0.4~dex is shown (dashed line in both panels). 
 The trends in both the GCE and UDF models might provide both a plausible explanation for the origin of stars with [Fe/H] < -2.8~dex and high [Mg/Fe].   

Moreover, beyond chemical evolution models, it is also important to consider dynamical information to better constrain the origin of our sample stars.
As shown in the Lindblad diagram of Fig.~\ref{fig:lindblad}, most of the stars in our sample have orbital parameters consistent with those of halo populations,  like Gaia-Enceladus (accreted), Erebus or Splash (both in-situ). 
Most importantly, from a dynamical perspective, our sample lacks coherence, as would be instead expected for stars originating from a common progenitor galaxy, particularly if the merger occurred relatively recently \citep[see, e.g.,][]{Helmi2000MNRAS.319..657H}. However, clumps in dynamical spaces are typically neither complete nor pure \cite[][]{JeanBaptiste2017}. For instance, numerical models of galaxy interactions by \citet{Mori2024A&A...690A.136M} suggest that in high-mass mergers (mass ratios of $\sim$1:10), the stars of the accreted galaxy can more efficiently lose coherence in the E-Lz plane, whereas this is less likely for lower-mass mergers \citep{Amarante2022ApJ...937...12A, Khoperskov2023A&A...677A..90K}.   Additionally, recent models of the hierarchical assembly of galaxy halos indicate that stars accreted from progenitor fragments can be identified through their chemical signatures, enabling the distinction between in-situ, ex-situ, and endo-debris populations \citep{Gonzalez2025A&A...693A.282G}, but can be less evident in the kinematic space. 

Therefore, when analyzing stars like those in our sample, which are inhomogeneously distributed in dynamical space, a common origin cannot be definitively ruled out. At present, using only Mg and Fe, we are unable to discern subtle differences that could reveal their chemical origin. However, a more extensive analysis incorporating additional elements, carefully examined in 3D NLTE, may provide the necessary insights to disentangle their origin.


\section{Summary and conclusions}
\label{sec:conclusions}

Given the current understanding that orbital properties alone are likely insufficient to determine the membership of halo stars to a debris field or a disrupted satellite in the inner halo (R$_{GC} < $30~kpc) \citep[e.g.][]{Horta2023ApJ...943..158H, Gonzalez2025A&A...693A.282G}, we conducted a high-precision spectral analysis of low-metallicity halo stars to obtain optimal chemical abundances.  
Our results reveal a remarkably tight sequence in [Mg/Fe] vs. [Fe/H], suggesting a shared chemical evolutionary pathway for our sample. This sequence likely traces the primordial halo, particularly at very early epochs with low metallicity, as its knee-like shape closely aligns with predictions from our GCE model. Alternatively, the observed sequence may also be compatible with UFD models,  characterized by low star formation efficiency.  Both models require an empirical shift of +0.4 dex at most to match both the observational data and the solar [Mg/Fe] = 0 constraint. 

The combination of chemical homogeneity and kinematic incoherence complicates the unambiguous identification of these stars’ origins, making them consistent with both the early evolution of the Milky Way halo and that of UFDs. Additionally, the tightness of the sequence suggests that stochastic enrichment, at least for Mg, is not highly dominant in the metallicity range of our sample, as it is significantly more constrained than model predictions.
Increasing the sample of metal-poor stars, and the analysis of elements such as Al and Na, for example, with the kind of NLTE analysis presented here will better reveal if they originate in a common progenitor.
This is indeed one of the drivers  of forthcoming projects for new instrumentation, such as, e.g.,  HRMOS \citep{Magrini2023arXiv231208270M} and WST \citep{Mainieri2024arXiv240305398M}.

\begin{acknowledgements}
 REG and LM  thank INAF for the support (Large Grant 2023 EPOCH) and for the MiniGrant 2022 Checs. They acknowledge financial support under the National Recovery and Resilience Plan (NRRP), Mission 4, Component 2, Investment 1.1, Call for tender No. 104 published on 2.2.2022 by the Italian Ministry of University and Research (MUR), funded by the European Union – NextGenerationEU– Project ‘Cosmic POT’ (PI: L. Magrini) Grant Assignment Decree No. 2022X4TM3H  by the Italian Ministry of Ministry of University and Research (MUR). 
 DR and MR research is supported by INAF (Theory Grant 2022, {\it ``An in-depth theoretical study of CNO element evolution in galaxies''}, Fu.~Ob.~1.05.12.06.08).
 DM research is funded by the European Union –NextGenerationEU" RRF M4C2 1.1  n: 2022HY2NSX. "CHRONOS: adjusting the clock(s) to unveil the CHRONO-chemo-dynamical Structure of the Galaxy” (PI: S. Cassisi)
AMA gratefully acknowledges support from the Swedish Research Council (VR 2020-03940) and from the Crafoord Foundation via the Royal Swedish Academy of Sciences (CR 2024-0015). 
 Use was made of the Simbad database, operated at the CDS, Strasbourg, France, and of NASA’s Astrophysics Data System Bibliographic Services. 
 This publication makes use of data products from the Two Micron
 All Sky Survey, which is a joint project of the University of
 Massachusetts and the Infrared Processing and Analysis
 Center/California Institute of Technology, funded by the National Aeronautics and Space Administration and the National Science Foundation.
 This research used Astropy (\url{http://www.astropy.org}) a community-developed core Python package for Astronomy \citep{astropy:2018}.
 This work presents results from the European Space Agency (ESA)
 space mission Gaia. Gaia data are processed by the Gaia Data Processing and Analysis Consortium (DPAC). Funding for the DPAC is provided by national institutions, in particular the institutions participating in the Gaia MultiLateral Agreement (MLA). The Gaia mission website is \url{https://www.cosmos.esa.int/gaia}. The Gaia archive website is \url{https://archives.esac.esa.int/gaia}.
\end{acknowledgements}

\bibliographystyle{aa.bst}

\bibliography{Faint2}

\begin{thebibliography}{188}
\expandafter\ifx\csname natexlab\endcsname\relax\def\natexlab#1{#1}\fi

\bibitem[{{Abdurro'uf} {et~al.}(2022){Abdurro'uf}, {Accetta}, {Aerts}, {Silva Aguirre}, {Ahumada}, {Ajgaonkar}, {Filiz Ak}, {Alam}, {Allende Prieto}, {Almeida}, {Anders}, {Anderson}, {Andrews}, {Anguiano}, {Aquino-Ort{\'\i}z}, {Arag{\'o}n-Salamanca}, {Argudo-Fern{\'a}ndez}, {Ata}, {Aubert}, {Avila-Reese}, {Badenes}, {Barb{\'a}}, {Barger}, {Barrera-Ballesteros}, {Beaton}, {Beers}, {Belfiore}, {Bender}, {Bernardi}, {Bershady}, {Beutler}, {Bidin}, {Bird}, {Bizyaev}, {Blanc}, {Blanton}, {Boardman}, {Bolton}, {Boquien}, {Borissova}, {Bovy}, {Brandt}, {Brown}, {Brownstein}, {Brusa}, {Buchner}, {Bundy}, {Burchett}, {Bureau}, {Burgasser}, {Cabang}, {Campbell}, {Cappellari}, {Carlberg}, {Wanderley}, {Carrera}, {Cash}, {Chen}, {Chen}, {Cherinka}, {Chiappini}, {Choi}, {Chojnowski}, {Chung}, {Clerc}, {Cohen}, {Comerford}, {Comparat}, {da Costa}, {Covey}, {Crane}, {Cruz-Gonzalez}, {Culhane}, {Cunha}, {Dai}, {Damke}, {Darling}, {Davidson}, {Davies}, {Dawson}, {De Lee}, {Diamond-Stanic}, {Cano-D{\'\i}az}, {S{\'a}nchez},
  {Donor}, {Duckworth}, {Dwelly}, {Eisenstein}, {Elsworth}, {Emsellem}, {Eracleous}, {Escoffier}, {Fan}, {Farr}, {Feng}, {Fern{\'a}ndez-Trincado}, {Feuillet}, {Filipp}, {Fillingham}, {Frinchaboy}, {Fromenteau}, {Galbany}, {Garc{\'\i}a}, {Garc{\'\i}a-Hern{\'a}ndez}, {Ge}, {Geisler}, {Gelfand}, {G{\'e}ron}, {Gibson}, {Goddy}, {Godoy-Rivera}, {Grabowski}, {Green}, {Greener}, {Grier}, {Griffith}, {Guo}, {Guy}, {Hadjara}, {Harding}, {Hasselquist}, {Hayes}, {Hearty}, {Hern{\'a}ndez}, {Hill}, {Hogg}, {Holtzman}, {Horta}, {Hsieh}, {Hsu}, {Hsu}, {Huber}, {Huertas-Company}, {Hutchinson}, {Hwang}, {Ibarra-Medel}, {Chitham}, {Ilha}, {Imig}, {Jaekle}, {Jayasinghe}, {Ji}, {Johnson}, {Jones}, {J{\"o}nsson}, {Katkov}, {Khalatyan}, {Kinemuchi}, {Kisku}, {Knapen}, {Kneib}, {Kollmeier}, {Kong}, {Kounkel}, {Kreckel}, {Krishnarao}, {Lacerna}, {Lane}, {Langgin}, {Lavender}, {Law}, {Lazarz}, {Leung}, {Leung}, {Lewis}, {Li}, {Li}, {Lian}, {Liang}, {Lin}, {Lin}, {Lin}, {Lintott}, {Long}, {Longa-Pe{\~n}a}, {L{\'o}pez-Cob{\'a}}, {Lu},
  {Lundgren}, {Luo}, {Mackereth}, {de la Macorra}, {Mahadevan}, {Majewski}, {Manchado}, {Mandeville}, {Maraston}, {Margalef-Bentabol}, {Masseron}, {Masters}, {Mathur}, {McDermid}, {Mckay}, {Merloni}, {Merrifield}, {Meszaros}, {Miglio}, {Di Mille}, {Minniti}, {Minsley}, {Monachesi}, {Moon}, {Mosser}, {Mulchaey}, {Muna}, {Mu{\~n}oz}, {Myers}, {Myers}, {Nadathur}, {Nair}, {Nandra}, {Neumann}, {Newman}, {Nidever}, {Nikakhtar}, {Nitschelm}, {O'Connell}, {Garma-Oehmichen}, {Luan Souza de Oliveira}, {Olney}, {Oravetz}, {Ortigoza-Urdaneta}, {Osorio}, {Otter}, {Pace}, {Padilla}, {Pan}, {Pan}, {Parikh}, {Parker}, {Peirani}, {Pe{\~n}a Ram{\'\i}rez}, {Penny}, {Percival}, {Perez-Fournon}, {Pinsonneault}, {Poidevin}, {Poovelil}, {Price-Whelan}, {B{\'a}rbara de Andrade Queiroz}, {Raddick}, {Ray}, {Rembold}, {Riddle}, {Riffel}, {Riffel}, {Rix}, {Robin}, {Rodr{\'\i}guez-Puebla}, {Roman-Lopes}, {Rom{\'a}n-Z{\'u}{\~n}iga}, {Rose}, {Ross}, {Rossi}, {Rubin}, {Salvato}, {S{\'a}nchez}, {S{\'a}nchez-Gallego}, {Sanderson}, {Santana
  Rojas}, {Sarceno}, {Sarmiento}, {Sayres}, {Sazonova}, {Schaefer}, {Schiavon}, {Schlegel}, {Schneider}, {Schultheis}, {Schwope}, {Serenelli}, {Serna}, {Shao}, {Shapiro}, {Sharma}, {Shen}, {Shetrone}, {Shu}, {Simon}, {Skrutskie}, {Smethurst}, {Smith}, {Sobeck}, {Spoo}, {Sprague}, {Stark}, {Stassun}, {Steinmetz}, {Stello}, {Stone-Martinez}, {Storchi-Bergmann}, {Stringfellow}, {Stutz}, {Su}, {Taghizadeh-Popp}, {Talbot}, {Tayar}, {Telles}, {Teske}, {Thakar}, {Theissen}, {Tkachenko}, {Thomas}, {Tojeiro}, {Hernandez Toledo}, {Troup}, {Trump}, {Trussler}, {Turner}, {Tuttle}, {Unda-Sanzana}, {V{\'a}zquez-Mata}, {Valentini}, {Valenzuela}, {Vargas-Gonz{\'a}lez}, {Vargas-Maga{\~n}a}, {Alfaro}, {Villanova}, {Vincenzo}, {Wake}, {Warfield}, {Washington}, {Weaver}, {Weijmans}, {Weinberg}, {Weiss}, {Westfall}, {Wild}, {Wilde}, {Wilson}, {Wilson}, {Wilson}, {Wolf}, {Wood-Vasey}, {Yan}, {Zamora}, {Zasowski}, {Zhang}, {Zhao}, {Zheng}, {Zheng}, \& {Zhu}}]{APOGEE}
{Abdurro'uf}, {Accetta}, K., {Aerts}, C., {et~al.} 2022, \apjs, 259, 35

\bibitem[{{Aguado} {et~al.}(2019){Aguado}, {Gonz{\'a}lez Hern{\'a}ndez}, {Allende Prieto}, \& {Rebolo}}]{aguado2019ApJ...874L..21A}
{Aguado}, D.~S., {Gonz{\'a}lez Hern{\'a}ndez}, J.~I., {Allende Prieto}, C., \& {Rebolo}, R. 2019, \apjl, 874, L21

\bibitem[{{Aguado} {et~al.}(2022){Aguado}, {Molaro}, {Caffau}, {Gonz{\'a}lez Hern{\'a}ndez}, {Zapatero Osorio}, {Bonifacio}, {Allende Prieto}, {Rebolo}, {Damasso}, {Su{\'a}rez Mascare{\~n}o}, {Howell}, {Furlan}, {Cristiani}, {Cupani}, {Di Marcantonio}, {D'Odorico}, {Lovis}, {Martins}, {Milakovi}, {Murphy}, {Nunes}, {Pepe}, {Santos}, {Schmidt}, \& {Sozzetti}}]{aguado2022A&A...668A..86A}
{Aguado}, D.~S., {Molaro}, P., {Caffau}, E., {et~al.} 2022, \aap, 668, A86

\bibitem[{{Allard} {et~al.}(2008){Allard}, {Kielkopf}, {Cayrel}, \& {van't Veer-Menneret}}]{allard2008}
{Allard}, N.~F., {Kielkopf}, J.~F., {Cayrel}, R., \& {van't Veer-Menneret}, C. 2008, \aap, 480, 581

\bibitem[{{Amarante} {et~al.}(2022){Amarante}, {Debattista}, {Beraldo e Silva}, {Laporte}, \& {Deg}}]{Amarante2022ApJ...937...12A}
{Amarante}, J. A.~S., {Debattista}, V.~P., {Beraldo e Silva}, L., {Laporte}, C. F.~P., \& {Deg}, N. 2022, \apj, 937, 12

\bibitem[{{Amarsi} {et~al.}(2020){Amarsi}, {Lind}, {Osorio}, {Nordlander}, {Bergemann}, {Reggiani}, {Wang}, {Buder}, {Asplund}, {Barklem}, {Wehrhahn}, {Sk{\'u}lad{\'o}ttir}, {Kobayashi}, {Karakas}, {Gao}, {Bland-Hawthorn}, {de Silva}, {Kos}, {Lewis}, {Martell}, {Sharma}, {Simpson}, {Zucker}, {{\v{C}}otar}, {Horner}, \& {GALAH Collaboration}}]{amarsi2020A&A...642A..62A}
{Amarsi}, A.~M., {Lind}, K., {Osorio}, Y., {et~al.} 2020, \aap, 642, A62

\bibitem[{{Amarsi} {et~al.}(2019){Amarsi}, {Nissen}, {Asplund}, {Lind}, \& {Barklem}}]{amarsi2019}
{Amarsi}, A.~M., {Nissen}, P.~E., {Asplund}, M., {Lind}, K., \& {Barklem}, P.~S. 2019, \aap, 622, L4

\bibitem[{{Amarsi} {et~al.}(2018){Amarsi}, {Nordlander}, {Barklem}, {Asplund}, {Collet}, \& {Lind}}]{amarsi2018}
{Amarsi}, A.~M., {Nordlander}, T., {Barklem}, P.~S., {et~al.} 2018, A\&A, 615, A139

\bibitem[{{Andrievsky} {et~al.}(2010){Andrievsky}, {Spite}, {Korotin}, {Spite}, {Bonifacio}, {Cayrel}, {Fran{\c{c}}ois}, \& {Hill}}]{andrievsky2010A&A...509A..88A}
{Andrievsky}, S.~M., {Spite}, M., {Korotin}, S.~A., {et~al.} 2010, \aap, 509, A88

\bibitem[{{Aoki} {et~al.}(2013){Aoki}, {Beers}, {Lee}, {Honda}, {Ito}, {Takada-Hidai}, {Frebel}, {Suda}, {Fujimoto}, {Carollo}, \& {Sivarani}}]{aoki2013AJ....145...13A}
{Aoki}, W., {Beers}, T.~C., {Lee}, Y.~S., {et~al.} 2013, \aj, 145, 13

\bibitem[{{Aoki} {et~al.}(2006){Aoki}, {Frebel}, {Christlieb}, {Norris}, {Beers}, {Minezaki}, {Barklem}, {Honda}, {Takada-Hidai}, {Asplund}, {Ryan}, {Tsangarides}, {Eriksson}, {Steinhauer}, {Deliyannis}, {Nomoto}, {Fujimoto}, {Ando}, {Yoshii}, \& {Kajino}}]{aoki2006ApJ...639..897A}
{Aoki}, W., {Frebel}, A., {Christlieb}, N., {et~al.} 2006, \apj, 639, 897

\bibitem[{{Arentsen} {et~al.}(2023){Arentsen}, {Aguado}, {Sestito}, {Gonz{\'a}lez Hern{\'a}ndez}, {Martin}, {Starkenburg}, {Jablonka}, \& {Yuan}}]{arentsen2023MNRAS.519.5554A}
{Arentsen}, A., {Aguado}, D.~S., {Sestito}, F., {et~al.} 2023, \mnras, 519, 5554

\bibitem[{{Arnone} {et~al.}(2005){Arnone}, {Ryan}, {Argast}, {Norris}, \& {Beers}}]{arnone2005A&A...430..507A}
{Arnone}, E., {Ryan}, S.~G., {Argast}, D., {Norris}, J.~E., \& {Beers}, T.~C. 2005, \aap, 430, 507

\bibitem[{{Asplund} {et~al.}(2021){Asplund}, {Amarsi}, \& {Grevesse}}]{asplund2021A&A...653A.141A}
{Asplund}, M., {Amarsi}, A.~M., \& {Grevesse}, N. 2021, \aap, 653, A141

\bibitem[{{Asplund} {et~al.}(2009){Asplund}, {Grevesse}, {Sauval}, \& {Scott}}]{2009ARA&A..47..481A}
{Asplund}, M., {Grevesse}, N., {Sauval}, A.~J., \& {Scott}, P. 2009, \araa, 47, 481

\bibitem[{{Astropy Collaboration} {et~al.}(2018){Astropy Collaboration}, {Price-Whelan}, {Sip{\H{o}}cz}, {G{\"u}nther}, {Lim}, {Crawford}, {Conseil}, {Shupe}, {Craig}, {Dencheva}, {Ginsburg}, {VanderPlas}, {Bradley}, {P{\'e}rez-Su{\'a}rez}, {de Val-Borro}, {Aldcroft}, {Cruz}, {Robitaille}, {Tollerud}, {Ardelean}, {Babej}, {Bach}, {Bachetti}, {Bakanov}, {Bamford}, {Barentsen}, {Barmby}, {Baumbach}, {Berry}, {Biscani}, {Boquien}, {Bostroem}, {Bouma}, {Brammer}, {Bray}, {Breytenbach}, {Buddelmeijer}, {Burke}, {Calderone}, {Cano Rodr{\'\i}guez}, {Cara}, {Cardoso}, {Cheedella}, {Copin}, {Corrales}, {Crichton}, {D'Avella}, {Deil}, {Depagne}, {Dietrich}, {Donath}, {Droettboom}, {Earl}, {Erben}, {Fabbro}, {Ferreira}, {Finethy}, {Fox}, {Garrison}, {Gibbons}, {Goldstein}, {Gommers}, {Greco}, {Greenfield}, {Groener}, {Grollier}, {Hagen}, {Hirst}, {Homeier}, {Horton}, {Hosseinzadeh}, {Hu}, {Hunkeler}, {Ivezi{\'c}}, {Jain}, {Jenness}, {Kanarek}, {Kendrew}, {Kern}, {Kerzendorf}, {Khvalko}, {King}, {Kirkby}, {Kulkarni},
  {Kumar}, {Lee}, {Lenz}, {Littlefair}, {Ma}, {Macleod}, {Mastropietro}, {McCully}, {Montagnac}, {Morris}, {Mueller}, {Mumford}, {Muna}, {Murphy}, {Nelson}, {Nguyen}, {Ninan}, {N{\"o}the}, {Ogaz}, {Oh}, {Parejko}, {Parley}, {Pascual}, {Patil}, {Patil}, {Plunkett}, {Prochaska}, {Rastogi}, {Reddy Janga}, {Sabater}, {Sakurikar}, {Seifert}, {Sherbert}, {Sherwood-Taylor}, {Shih}, {Sick}, {Silbiger}, {Singanamalla}, {Singer}, {Sladen}, {Sooley}, {Sornarajah}, {Streicher}, {Teuben}, {Thomas}, {Tremblay}, {Turner}, {Terr{\'o}n}, {van Kerkwijk}, {de la Vega}, {Watkins}, {Weaver}, {Whitmore}, {Woillez}, {Zabalza}, \& {Astropy Contributors}}]{astropy:2018}
{Astropy Collaboration}, {Price-Whelan}, A.~M., {Sip{\H{o}}cz}, B.~M., {et~al.} 2018, AJ, 156, 123

\bibitem[{{Bailer-Jones}(2023)}]{bailer-jones2023AJ....166..269B}
{Bailer-Jones}, C.~A.~L. 2023, \aj, 166, 269

\bibitem[{{Baratella} {et~al.}(2020){Baratella}, {D'Orazi}, {Carraro}, {Desidera}, {Randich}, {Magrini}, {Adibekyan}, {Smiljanic}, {Spina}, {Tsantaki}, {Tautvai{\v{s}}ien{\.{e}}}, {Sousa}, {Jofr{\'e}}, {Jim{\'e}nez-Esteban}, {Delgado-Mena}, {Martell}, {Van der Swaelmen}, {Roccatagliata}, {Gilmore}, {Alfaro}, {Bayo}, {Bensby}, {Bragaglia}, {Franciosini}, {Gonneau}, {Heiter}, {Hourihane}, {Jeffries}, {Koposov}, {Morbidelli}, {Prisinzano}, {Sacco}, {Sbordone}, {Worley}, {Zaggia}, \& {Lewis}}]{Baratella2020A&A...634A..34B}
{Baratella}, M., {D'Orazi}, V., {Carraro}, G., {et~al.} 2020, \aap, 634, A34

\bibitem[{{Baratella} {et~al.}(2021){Baratella}, {D'Orazi}, {Sheminova}, {Spina}, {Carraro}, {Gratton}, {Magrini}, {Randich}, {Lugaro}, {Pignatari}, {Romano}, {Biazzo}, {Bragaglia}, {Casali}, {Desidera}, {Frasca}, {de Silva}, {Melo}, {Van der Swaelmen}, {Tautvai{\v{s}}ien{\.{e}}}, {Jim{\'e}nez-Esteban}, {Gilmore}, {Bensby}, {Smiljanic}, {Bayo}, {Franciosini}, {Gonneau}, {Hourihane}, {Jofr{\'e}}, {Monaco}, {Morbidelli}, {Sacco}, {Sbordone}, {Worley}, \& {Zaggia}}]{Baratella2021A&A...653A..67B}
{Baratella}, M., {D'Orazi}, V., {Sheminova}, V., {et~al.} 2021, \aap, 653, A67

\bibitem[{{Barb{\'a}} {et~al.}(2019){Barb{\'a}}, {Minniti}, {Geisler}, {Alonso-Garc{\'\i}a}, {Hempel}, {Monachesi}, {Arias}, \& {G{\'o}mez}}]{barba2019ApJ...870L..24B}
{Barb{\'a}}, R.~H., {Minniti}, D., {Geisler}, D., {et~al.} 2019, \apjl, 870, L24

\bibitem[{{Barklem}(2007)}]{barklem2007}
{Barklem}, P.~S. 2007, \aap, 466, 327

\bibitem[{{Barklem} \& {O'Mara}(1997)}]{1997MNRAS.290..102B}
{Barklem}, P.~S. \& {O'Mara}, B.~J. 1997, \mnras, 290, 102

\bibitem[{{Barklem} {et~al.}(2000){Barklem}, {Piskunov}, \& {O'Mara}}]{BPO2000}
{Barklem}, P.~S., {Piskunov}, N., \& {O'Mara}, B.~J. 2000, \aap, 363, 1091

\bibitem[{{Barklem} {et~al.}(2002){Barklem}, {Stempels}, {Allende Prieto}, {Kochukhov}, {Piskunov}, \& {O'Mara}}]{BPO2002}
{Barklem}, P.~S., {Stempels}, H.~C., {Allende Prieto}, C., {et~al.} 2002, \aap, 385, 951

\bibitem[{{Belokurov} {et~al.}(2018){Belokurov}, {Erkal}, {Evans}, {Koposov}, \& {Deason}}]{Belokurov2018}
{Belokurov}, V., {Erkal}, D., {Evans}, N.~W., {Koposov}, S.~E., \& {Deason}, A.~J. 2018, \mnras, 478, 611

\bibitem[{{Belokurov} \& {Kravtsov}(2022)}]{belokurov22}
{Belokurov}, V. \& {Kravtsov}, A. 2022, \mnras, 514, 689

\bibitem[{{Belokurov} {et~al.}(2020){Belokurov}, {Sanders}, {Fattahi}, {Smith}, {Deason}, {Evans}, \& {Grand}}]{belokurov2020MNRAS.494.3880B}
{Belokurov}, V., {Sanders}, J.~L., {Fattahi}, A., {et~al.} 2020, \mnras, 494, 3880

\bibitem[{{Bergemann} {et~al.}(2017){Bergemann}, {Collet}, {Amarsi}, {Kovalev}, {Ruchti}, \& {Magic}}]{bergemann2017ApJ...847...15B}
{Bergemann}, M., {Collet}, R., {Amarsi}, A.~M., {et~al.} 2017, \apj, 847, 15

\bibitem[{{Bergemann} {et~al.}(2012){Bergemann}, {Lind}, {Collet}, {Magic}, \& {Asplund}}]{bergemann2012}
{Bergemann}, M., {Lind}, K., {Collet}, R., {Magic}, Z., \& {Asplund}, M. 2012, \mnras, 427, 27

\bibitem[{{Berni}(2024)}]{Berni2024arXiv240911429B}
{Berni}, L. 2024, arXiv e-prints, arXiv:2409.11429

\bibitem[{{Bonaca} {et~al.}(2020){Bonaca}, {Conroy}, {Cargile}, {Naidu}, {Johnson}, {Zaritsky}, {Ting}, {Caldwell}, {Han}, \& {van Dokkum}}]{bonaca2020ApJ...897L..18B}
{Bonaca}, A., {Conroy}, C., {Cargile}, P.~A., {et~al.} 2020, \apjl, 897, L18

\bibitem[{{Borre} {et~al.}(2022){Borre}, {Aguirre B{\o}rsen-Koch}, {Helmi}, {Koppelman}, {Nielsen}, {R{\o}rsted}, {Stello}, {Stokholm}, {Winther}, {Davies}, {Hon}, {Kruijssen}, {Laporte}, {Reyes}, \& {Yu}}]{borre2022MNRAS.514.2527B}
{Borre}, C.~C., {Aguirre B{\o}rsen-Koch}, V., {Helmi}, A., {et~al.} 2022, \mnras, 514, 2527

\bibitem[{Bromm {et~al.}(2001)Bromm, Ferrara, Coppi, \& Larson}]{bromm01}
Bromm, V., Ferrara, A., Coppi, P., \& Larson, R. 2001, Monthly Notices of the Royal Astronomical Society, 328, 969

\bibitem[{{Brook} {et~al.}(2003){Brook}, {Kawata}, {Gibson}, \& {Flynn}}]{brook2003ApJ...585L.125B}
{Brook}, C.~B., {Kawata}, D., {Gibson}, B.~K., \& {Flynn}, C. 2003, \apjl, 585, L125

\bibitem[{Brown {et~al.}(2014)Brown, Tumlinson, Geha, Simon, Vargas, VandenBerg, Kirby, Kalirai, Avila, Gennaro, {et~al.}}]{Brown14}
Brown, T.~M., Tumlinson, J., Geha, M., {et~al.} 2014, The Astrophysical Journal, 796, 91

\bibitem[{{Buder} {et~al.}(2021){Buder}, {Sharma}, {Kos}, {Amarsi}, {Nordlander}, {Lind}, {Martell}, {Asplund}, {Bland-Hawthorn}, {Casey}, {de Silva}, {D'Orazi}, {Freeman}, {Hayden}, {Lewis}, {Lin}, {Schlesinger}, {Simpson}, {Stello}, {Zucker}, {Zwitter}, {Beeson}, {Buck}, {Casagrande}, {Clark}, {{\v{C}}otar}, {da Costa}, {de Grijs}, {Feuillet}, {Horner}, {Kafle}, {Khanna}, {Kobayashi}, {Liu}, {Montet}, {Nandakumar}, {Nataf}, {Ness}, {Spina}, {Tepper-Garc{\'\i}a}, {Ting}, {Traven}, {Vogrin{\v{c}}i{\v{c}}}, {Wittenmyer}, {Wyse}, {{\v{Z}}erjal}, \& {Galah Collaboration}}]{buder2021MNRAS.506..150B}
{Buder}, S., {Sharma}, S., {Kos}, J., {et~al.} 2021, \mnras, 506, 150

\bibitem[{{Caffau} {et~al.}(2012){Caffau}, {Bonifacio}, {Fran{\c{c}}ois}, {Spite}, {Spite}, {Zaggia}, {Ludwig}, {Steffen}, {Mashonkina}, {Monaco}, {Sbordone}, {Molaro}, {Cayrel}, {Plez}, {Hill}, {Hammer}, \& {Randich}}]{caffau2012A&A...542A..51C}
{Caffau}, E., {Bonifacio}, P., {Fran{\c{c}}ois}, P., {et~al.} 2012, \aap, 542, A51

\bibitem[{{Casagrande} {et~al.}(2021){Casagrande}, {Lin}, {Rains}, {Liu}, {Buder}, {Horner}, {Asplund}, {Lewis}, {Martell}, {Nordlander}, {Stello}, {Ting}, {Wittenmyer}, {Bland-Hawthorn}, {Casey}, {De Silva}, {D'Orazi}, {Freeman}, {Hayden}, {Kos}, {Lind}, {Schlesinger}, {Sharma}, {Simpson}, {Zucker}, \& {Zwitter}}]{casagrande2021}
{Casagrande}, L., {Lin}, J., {Rains}, A.~D., {et~al.} 2021, \mnras, 507, 2684

\bibitem[{{Casagrande} {et~al.}(2014){Casagrande}, {Portinari}, {Glass}, {Laney}, {Silva Aguirre}, {Datson}, {Andersen}, {Nordstr{\"o}m}, {Holmberg}, {Flynn}, \& {Asplund}}]{casagrande2014MNRAS.439.2060C}
{Casagrande}, L., {Portinari}, L., {Glass}, I.~S., {et~al.} 2014, \mnras, 439, 2060

\bibitem[{{Casagrande} {et~al.}(2010){Casagrande}, {Ram{\'{\i}}rez}, {Mel{\'e}ndez}, {Bessell}, \& {Asplund}}]{Casagrande2010}
{Casagrande}, L., {Ram{\'{\i}}rez}, I., {Mel{\'e}ndez}, J., {Bessell}, M., \& {Asplund}, M. 2010, \aap, 512, A54

\bibitem[{{Cayrel} {et~al.}(2004){Cayrel}, {Depagne}, {Spite}, {Hill}, {Spite}, {Fran{\c{c}}ois}, {Plez}, {Beers}, {Primas}, {Andersen}, {Barbuy}, {Bonifacio}, {Molaro}, \& {Nordstr{\"o}m}}]{cayrel2004A&A...416.1117C}
{Cayrel}, R., {Depagne}, E., {Spite}, M., {et~al.} 2004, \aap, 416, 1117

\bibitem[{{Ceccarelli} {et~al.}(2024){Ceccarelli}, {Massari}, {Mucciarelli}, {Bellazzini}, {Nunnari}, {Cusano}, {Lardo}, {Romano}, {Ilyin}, \& {Stokholm}}]{ceccarelli24}
{Ceccarelli}, E., {Massari}, D., {Mucciarelli}, A., {et~al.} 2024, \aap, 684, A37

\bibitem[{{Cescutti} {et~al.}(2022){Cescutti}, {Bonifacio}, {Caffau}, {Monaco}, {Franchini}, {Lombardo}, {Matas Pinto}, {Lucertini}, {Fran{\c{c}}ois}, {Spitoni}, {Lallement}, {Sbordone}, {Mucciarelli}, {Spite}, {Hansen}, {Di Marcantonio}, {Ku{\v{c}}inskas}, {Dobrovolskas}, {Korn}, {Valentini}, {Magrini}, {Cristallo}, \& {Matteucci}}]{cescutti2022A&A...668A.168C}
{Cescutti}, G., {Bonifacio}, P., {Caffau}, E., {et~al.} 2022, \aap, 668, A168

\bibitem[{{Chang} \& {Tang}(1990)}]{1990JQSRT..43..207C}
{Chang}, T.~N. \& {Tang}, X. 1990, \jqsrt, 43, 207

\bibitem[{{Chiappini} {et~al.}(2001){Chiappini}, {Matteucci}, \& {Romano}}]{Chiappini01}
{Chiappini}, C., {Matteucci}, F., \& {Romano}, D. 2001, \apj, 554, 1044

\bibitem[{{Christlieb} {et~al.}(2002){Christlieb}, {Bessell}, {Beers}, {Gustafsson}, {Korn}, {Barklem}, {Karlsson}, {Mizuno-Wiedner}, \& {Rossi}}]{christlieb2002Natur.419..904C}
{Christlieb}, N., {Bessell}, M.~S., {Beers}, T.~C., {et~al.} 2002, \nat, 419, 904

\bibitem[{{Cohen} {et~al.}(2013){Cohen}, {Christlieb}, {Thompson}, {McWilliam}, {Shectman}, {Reimers}, {Wisotzki}, \& {Kirby}}]{cohen2013ApJ...778...56C}
{Cohen}, J.~G., {Christlieb}, N., {Thompson}, I., {et~al.} 2013, \apj, 778, 56

\bibitem[{{Cristallo} {et~al.}(2011){Cristallo}, {Piersanti}, {Straniero}, {Gallino}, {Dom{\'\i}nguez}, {Abia}, {Di Rico}, {Quintini}, \& {Bisterzo}}]{Cristallo2011}
{Cristallo}, S., {Piersanti}, L., {Straniero}, O., {et~al.} 2011, \apjs, 197, 17

\bibitem[{{Cristallo} {et~al.}(2009){Cristallo}, {Straniero}, {Gallino}, {Piersanti}, {Dom{\'\i}nguez}, \& {Lederer}}]{Cristallo2009}
{Cristallo}, S., {Straniero}, O., {Gallino}, R., {et~al.} 2009, \apj, 696, 797

\bibitem[{{Cristallo} {et~al.}(2015){Cristallo}, {Straniero}, {Piersanti}, \& {Gobrecht}}]{Cristallo2015}
{Cristallo}, S., {Straniero}, O., {Piersanti}, L., \& {Gobrecht}, D. 2015, \apjs, 219, 40

\bibitem[{{da Silva} \& {Smiljanic}(2023)}]{da_silva2023A&A...677A..74D}
{da Silva}, A.~R. \& {Smiljanic}, R. 2023, \aap, 677, A74

\bibitem[{{De Silva} {et~al.}(2015){De Silva}, {Freeman}, {Bland-Hawthorn}, {Martell}, {de Boer}, {Asplund}, {Keller}, {Sharma}, {Zucker}, {Zwitter}, {Anguiano}, {Bacigalupo}, {Bayliss}, {Beavis}, {Bergemann}, {Campbell}, {Cannon}, {Carollo}, {Casagrande}, {Casey}, {Da Costa}, {D'Orazi}, {Dotter}, {Duong}, {Heger}, {Ireland}, {Kafle}, {Kos}, {Lattanzio}, {Lewis}, {Lin}, {Lind}, {Munari}, {Nataf}, {O'Toole}, {Parker}, {Reid}, {Schlesinger}, {Sheinis}, {Simpson}, {Stello}, {Ting}, {Traven}, {Watson}, {Wittenmyer}, {Yong}, \& {{\v{Z}}erjal}}]{GALAH}
{De Silva}, G.~M., {Freeman}, K.~C., {Bland-Hawthorn}, J., {et~al.} 2015, \mnras, 449, 2604

\bibitem[{{Dekker} {et~al.}(2000){Dekker}, {D'Odorico}, {Kaufer}, {Delabre}, \& {Kotzlowski}}]{2000SPIE.4008..534D}
{Dekker}, H., {D'Odorico}, S., {Kaufer}, A., {Delabre}, B., \& {Kotzlowski}, H. 2000, in Society of Photo-Optical Instrumentation Engineers (SPIE) Conference Series, Vol. 4008, Optical and IR Telescope Instrumentation and Detectors, ed. M.~{Iye} \& A.~F. {Moorwood}, 534--545

\bibitem[{{dos Santos} {et~al.}(2016){dos Santos}, {Mel{\'e}ndez}, {do Nascimento}, {Bedell}, {Ram{\'\i}rez}, {Bean}, {Asplund}, {Spina}, {Dreizler}, {Alves-Brito}, \& {Casagrande}}]{2016A&A...592A.156D}
{dos Santos}, L.~A., {Mel{\'e}ndez}, J., {do Nascimento}, J.-D., {et~al.} 2016, \aap, 592, A156

\bibitem[{{Dutra-Ferreira} {et~al.}(2016){Dutra-Ferreira}, {Pasquini}, {Smiljanic}, {Porto de Mello}, \& {Steffen}}]{2016A&A...585A..75D}
{Dutra-Ferreira}, L., {Pasquini}, L., {Smiljanic}, R., {Porto de Mello}, G.~F., \& {Steffen}, M. 2016, \aap, 585, A75

\bibitem[{{Eggen} {et~al.}(1962){Eggen}, {Lynden-Bell}, \& {Sandage}}]{eggen1962ApJ...136..748E}
{Eggen}, O.~J., {Lynden-Bell}, D., \& {Sandage}, A.~R. 1962, \apj, 136, 748

\bibitem[{{Feuillet} {et~al.}(2021){Feuillet}, {Sahlholdt}, {Feltzing}, \& {Casagrande}}]{feuillet2021MNRAS.508.1489F}
{Feuillet}, D.~K., {Sahlholdt}, C.~L., {Feltzing}, S., \& {Casagrande}, L. 2021, \mnras, 508, 1489

\bibitem[{{Fran{\c{c}}ois} {et~al.}(2024){Fran{\c{c}}ois}, {Cescutti}, {Bonifacio}, {Caffau}, {Monaco}, {Steffen}, {Puschnig}, {Calura}, {Cristallo}, {Di Marcantonio}, {Dobrovolskas}, {Franchini}, {Gallagher}, {Hansen}, {Korn}, {Ku{\v{c}}inskas}, {Lallement}, {Lombardo}, {Lucertini}, {Magrini}, {Matas Pinto}, {Matteucci}, {Mucciarelli}, {Sbordone}, {Spite}, {Spitoni}, \& {Valentini}}]{francois2024A&A...686A.295F}
{Fran{\c{c}}ois}, P., {Cescutti}, G., {Bonifacio}, P., {et~al.} 2024, \aap, 686, A295

\bibitem[{{Fran{\c{c}}ois} {et~al.}(2007){Fran{\c{c}}ois}, {Depagne}, {Hill}, {Spite}, {Spite}, {Plez}, {Beers}, {Andersen}, {James}, {Barbuy}, {Cayrel}, {Bonifacio}, {Molaro}, {Nordstr{\"o}m}, \& {Primas}}]{francois2007A&A...476..935F}
{Fran{\c{c}}ois}, P., {Depagne}, E., {Hill}, V., {et~al.} 2007, \aap, 476, 935

\bibitem[{{Fran{\c{c}}ois} {et~al.}(2004){Fran{\c{c}}ois}, {Matteucci}, {Cayrel}, {Spite}, {Spite}, \& {Chiappini}}]{francois2004A&A...421..613F}
{Fran{\c{c}}ois}, P., {Matteucci}, F., {Cayrel}, R., {et~al.} 2004, \aap, 421, 613

\bibitem[{{Frebel} {et~al.}(2005){Frebel}, {Aoki}, {Christlieb}, {Ando}, {Asplund}, {Barklem}, {Beers}, {Eriksson}, {Fechner}, {Fujimoto}, {Honda}, {Kajino}, {Minezaki}, {Nomoto}, {Norris}, {Ryan}, {Takada-Hidai}, {Tsangarides}, \& {Yoshii}}]{frebel2005Natur.434..871F}
{Frebel}, A., {Aoki}, W., {Christlieb}, N., {et~al.} 2005, \nat, 434, 871

\bibitem[{{Gallart} {et~al.}(2019){Gallart}, {Bernard}, {Brook}, {Ruiz-Lara}, {Cassisi}, {Hill}, \& {Monelli}}]{gallart2019NatAs...3..932G}
{Gallart}, C., {Bernard}, E.~J., {Brook}, C.~B., {et~al.} 2019, Nature Astronomy, 3, 932

\bibitem[{{Gerber} {et~al.}(2023){Gerber}, {Magg}, {Plez}, {Bergemann}, {Heiter}, {Olander}, \& {Hoppe}}]{gerber2023}
{Gerber}, J.~M., {Magg}, E., {Plez}, B., {et~al.} 2023, \aap, 669, A43

\bibitem[{{Gilmore} {et~al.}(2022){Gilmore}, {Randich}, {Worley}, {Hourihane}, {Gonneau}, {Sacco}, {Lewis}, {Magrini}, {Fran{\c{c}}ois}, {Jeffries}, {Koposov}, {Bragaglia}, {Alfaro}, {Allende Prieto}, {Blomme}, {Korn}, {Lanzafame}, {Pancino}, {Recio-Blanco}, {Smiljanic}, {Van Eck}, {Zwitter}, {Bensby}, {Flaccomio}, {Irwin}, {Franciosini}, {Morbidelli}, {Damiani}, {Bonito}, {Friel}, {Vink}, {Prisinzano}, {Abbas}, {Hatzidimitriou}, {Held}, {Jordi}, {Paunzen}, {Spagna}, {Jackson}, {Ma{\'\i}z Apell{\'a}niz}, {Asplund}, {Bonifacio}, {Feltzing}, {Binney}, {Drew}, {Ferguson}, {Micela}, {Negueruela}, {Prusti}, {Rix}, {Vallenari}, {Bergemann}, {Casey}, {de Laverny}, {Frasca}, {Hill}, {Lind}, {Sbordone}, {Sousa}, {Adibekyan}, {Caffau}, {Daflon}, {Feuillet}, {Gebran}, {Gonzalez Hernandez}, {Guiglion}, {Herrero}, {Lobel}, {Merle}, {Mikolaitis}, {Montes}, {Morel}, {Ruchti}, {Soubiran}, {Tabernero}, {Tautvai{\v{s}}ien{\.{e}}}, {Traven}, {Valentini}, {Van der Swaelmen}, {Villanova}, {Viscasillas V{\'a}zquez}, {Bayo},
  {Biazzo}, {Carraro}, {Edvardsson}, {Heiter}, {Jofr{\'e}}, {Marconi}, {Martayan}, {Masseron}, {Monaco}, {Walton}, {Zaggia}, {Aguirre B{\o}rsen-Koch}, {Alves}, {Balaguer-Nunez}, {Barklem}, {Barrado}, {Bellazzini}, {Berlanas}, {Binks}, {Bressan}, {Capuzzo-Dolcetta}, {Casagrande}, {Casamiquela}, {Collins}, {D'Orazi}, {Dantas}, {Debattista}, {Delgado-Mena}, {Di Marcantonio}, {Drazdauskas}, {Evans}, {Famaey}, {Franchini}, {Fr{\'e}mat}, {Fu}, {Geisler}, {Gerhard}, {Gonz{\'a}lez Solares}, {Grebel}, {Guti{\'e}rrez Albarr{\'a}n}, {Jim{\'e}nez-Esteban}, {J{\"o}nsson}, {Khachaturyants}, {Kordopatis}, {Kos}, {Lagarde}, {Ludwig}, {Mahy}, {Mapelli}, {Marfil}, {Martell}, {Messina}, {Miglio}, {Minchev}, {Moitinho}, {Montalban}, {Monteiro}, {Morossi}, {Mowlavi}, {Mucciarelli}, {Murphy}, {Nardetto}, {Ortolani}, {Paletou}, {Palou{\v{s}}}, {Pickering}, {Quirrenbach}, {Re Fiorentin}, {Read}, {Romano}, {Ryde}, {Sanna}, {Santos}, {Seabroke}, {Spina}, {Steinmetz}, {Stonkut{\'e}}, {Sutorius}, {Th{\'e}venin}, {Tosi}, {Tsantaki},
  {Wright}, {Wyse}, {Zoccali}, {Zorec}, \& {Zucker}}]{Gilmore2022A&A...666A.120G}
{Gilmore}, G., {Randich}, S., {Worley}, C.~C., {et~al.} 2022, \aap, 666, A120

\bibitem[{{Giribaldi} {et~al.}(2021){Giribaldi}, {da Silva}, {Smiljanic}, \& {Cornejo Espinoza}}]{giribaldi2021A&A...650A.194G}
{Giribaldi}, R.~E., {da Silva}, A.~R., {Smiljanic}, R., \& {Cornejo Espinoza}, D. 2021, \aap, 650, A194

\bibitem[{{Giribaldi} \& {Smiljanic}(2023)}]{giribaldi2023A&A...673A..18G}
{Giribaldi}, R.~E. \& {Smiljanic}, R. 2023, \aap, 673, A18

\bibitem[{{Giribaldi} {et~al.}(2019){Giribaldi}, {Ubaldo-Melo}, {Porto de Mello}, {Pasquini}, {Ludwig}, {Ulmer-Moll}, \& {Lorenzo-Oliveira}}]{giribaldi2019A&A...624A..10G}
{Giribaldi}, R.~E., {Ubaldo-Melo}, M.~L., {Porto de Mello}, G.~F., {et~al.} 2019, \aap, 624, A10

\bibitem[{{Giribaldi} {et~al.}(2023){Giribaldi}, {Van Eck}, {Merle}, {Jorissen}, {Krynski}, {Planquart}, {Valentini}, {Chiappini}, \& {Van Winckel}}]{giribaldi2023A&A...679A.110G}
{Giribaldi}, R.~E., {Van Eck}, S., {Merle}, T., {et~al.} 2023, \aap, 679, A110

\bibitem[{{Gonzalez-Jara} {et~al.}(2025){Gonzalez-Jara}, {Tissera}, {Monachesi}, {Sillero}, {Pallero}, {Pedrosa}, {Tau}, {Tapia-Contreras}, \& {Bignone}}]{Gonzalez2025A&A...693A.282G}
{Gonzalez-Jara}, J., {Tissera}, P.~B., {Monachesi}, A., {et~al.} 2025, \aap, 693, A282

\bibitem[{{Grand} {et~al.}(2020){Grand}, {Kawata}, {Belokurov}, {Deason}, {Fattahi}, {Fragkoudi}, {G{\'o}mez}, {Marinacci}, \& {Pakmor}}]{grand2020MNRAS.497.1603G}
{Grand}, R. J.~J., {Kawata}, D., {Belokurov}, V., {et~al.} 2020, \mnras, 497, 1603

\bibitem[{{Gruyters} {et~al.}(2013){Gruyters}, {Korn}, {Richard}, {Grundahl}, {Collet}, {Mashonkina}, {Osorio}, \& {Barklem}}]{gruyters2013A&A...555A..31G}
{Gruyters}, P., {Korn}, A.~J., {Richard}, O., {et~al.} 2013, \aap, 555, A31

\bibitem[{{Gustafsson} {et~al.}(2008){Gustafsson}, {Edvardsson}, {Eriksson}, {J{\o}rgensen}, {Nordlund}, \& {Plez}}]{gustafson2008}
{Gustafsson}, B., {Edvardsson}, B., {Eriksson}, K., {et~al.} 2008, A\&A, 486, 951

\bibitem[{{Hansen} {et~al.}(2013){Hansen}, {Bergemann}, {Cescutti}, {Fran{\c{c}}ois}, {Arcones}, {Karakas}, {Lind}, \& {Chiappini}}]{hansen2013A&A...551A..57H}
{Hansen}, C.~J., {Bergemann}, M., {Cescutti}, G., {et~al.} 2013, \aap, 551, A57

\bibitem[{{Hayek} {et~al.}(2011){Hayek}, {Asplund}, {Collet}, \& {Nordlund}}]{2011A&A...529A.158H}
{Hayek}, W., {Asplund}, M., {Collet}, R., \& {Nordlund}, {\r{A}}. 2011, \aap, 529, A158

\bibitem[{{Heger} \& {Woosley}(2002)}]{heger02}
{Heger}, A. \& {Woosley}, S.~E. 2002, \apj, 567, 532

\bibitem[{{Heger} \& {Woosley}(2010)}]{Heger+woosley10}
{Heger}, A. \& {Woosley}, S.~E. 2010, \apj, 724, 341

\bibitem[{{Heiter} {et~al.}(2021){Heiter}, {Lind}, {Bergemann}, {Asplund}, {Mikolaitis}, {Barklem}, {Masseron}, {de Laverny}, {Magrini}, {Edvardsson}, {J{\"o}nsson}, {Pickering}, {Ryde}, {Bayo Ar{\'a}n}, {Bensby}, {Casey}, {Feltzing}, {Jofr{\'e}}, {Korn}, {Pancino}, {Damiani}, {Lanzafame}, {Lardo}, {Monaco}, {Morbidelli}, {Smiljanic}, {Worley}, {Zaggia}, {Randich}, \& {Gilmore}}]{heiter2021A&A...645A.106H}
{Heiter}, U., {Lind}, K., {Bergemann}, M., {et~al.} 2021, \aap, 645, A106

\bibitem[{{Helmi} {et~al.}(2018){Helmi}, {Babusiaux}, {Koppelman}, {Massari}, {Veljanoski}, \& {Brown}}]{helmi2018}
{Helmi}, A., {Babusiaux}, C., {Koppelman}, H.~H., {et~al.} 2018, \nat, 563, 85

\bibitem[{{Helmi} \& {de Zeeuw}(2000)}]{Helmi2000MNRAS.319..657H}
{Helmi}, A. \& {de Zeeuw}, P.~T. 2000, \mnras, 319, 657

\bibitem[{{Horta} {et~al.}(2023){Horta}, {Cunningham}, {Sanderson}, {Johnston}, {Panithanpaisal}, {Arora}, {Necib}, {Wetzel}, {Bailin}, \& {Faucher-Gigu{\`e}re}}]{Horta2023ApJ...943..158H}
{Horta}, D., {Cunningham}, E.~C., {Sanderson}, R.~E., {et~al.} 2023, \apj, 943, 158

\bibitem[{{Horta} {et~al.}(2021){Horta}, {Schiavon}, {Mackereth}, {Pfeffer}, {Mason}, {Kisku}, {Fragkoudi}, {Allende Prieto}, {Cunha}, {Hasselquist}, {Holtzman}, {Majewski}, {Nataf}, {O'Connell}, {Schultheis}, \& {Smith}}]{Horta21}
{Horta}, D., {Schiavon}, R.~P., {Mackereth}, J.~T., {et~al.} 2021, \mnras, 500, 1385

\bibitem[{{Hourihane} {et~al.}(2023){Hourihane}, {Fran{\c{c}}ois}, {Worley}, {Magrini}, {Gonneau}, {Casey}, {Gilmore}, {Randich}, {Sacco}, {Recio-Blanco}, {Korn}, {Allende Prieto}, {Smiljanic}, {Blomme}, {Bragaglia}, {Walton}, {Van Eck}, {Bensby}, {Lanzafame}, {Frasca}, {Franciosini}, {Damiani}, {Lind}, {Bergemann}, {Bonifacio}, {Hill}, {Lobel}, {Montes}, {Feuillet}, {Tautvai{\v{s}}ien{\.{e}}}, {Guiglion}, {Tabernero}, {Gonz{\'a}lez Hern{\'a}ndez}, {Gebran}, {Van der Swaelmen}, {Mikolaitis}, {Daflon}, {Merle}, {Morel}, {Lewis}, {Gonz{\'a}lez Solares}, {Murphy}, {Jeffries}, {Jackson}, {Feltzing}, {Prusti}, {Carraro}, {Biazzo}, {Prisinzano}, {Jofr{\'e}}, {Zaggia}, {Drazdauskas}, {Stonkut{\'e}}, {Marfil}, {Jim{\'e}nez-Esteban}, {Mahy}, {Guti{\'e}rrez Albarr{\'a}n}, {Berlanas}, {Santos}, {Morbidelli}, {Spina}, \& {Minkevi{\v{c}}i{\={u}}t{\.{e}}}}]{Hourihane2023A&A...676A.129H}
{Hourihane}, A., {Fran{\c{c}}ois}, P., {Worley}, C.~C., {et~al.} 2023, \aap, 676, A129

\bibitem[{{Iwamoto} {et~al.}(1999){Iwamoto}, {Brachwitz}, {Nomoto}, {Kishimoto}, {Umeda}, {Hix}, \& {Thielemann}}]{Iwamoto99}
{Iwamoto}, K., {Brachwitz}, F., {Nomoto}, K., {et~al.} 1999, \apjs, 125, 439

\bibitem[{{Jacobson} {et~al.}(2015){Jacobson}, {Keller}, {Frebel}, {Casey}, {Asplund}, {Bessell}, {Da Costa}, {Lind}, {Marino}, {Norris}, {Pe{\~n}a}, {Schmidt}, {Tisserand}, {Walsh}, {Yong}, \& {Yu}}]{jacobson2015ApJ...807..171J}
{Jacobson}, H.~R., {Keller}, S., {Frebel}, A., {et~al.} 2015, \apj, 807, 171

\bibitem[{{Jean-Baptiste} {et~al.}(2017){Jean-Baptiste}, {Di Matteo}, {Haywood}, {G{\'o}mez}, {Montuori}, {Combes}, \& {Semelin}}]{JeanBaptiste2017}
{Jean-Baptiste}, I., {Di Matteo}, P., {Haywood}, M., {et~al.} 2017, \aap, 604, A106

\bibitem[{{Jenkins} {et~al.}(2021){Jenkins}, {Li}, {Pace}, {Ji}, {Koposov}, \& {Mutlu-Pakdil}}]{Jenkins2021}
{Jenkins}, S.~A., {Li}, T.~S., {Pace}, A.~B., {et~al.} 2021, \apj, 920, 92

\bibitem[{{Jofr{\'e}} {et~al.}(2014){Jofr{\'e}}, {Heiter}, {Soubiran}, {Blanco-Cuaresma}, {Worley}, {Pancino}, {Cantat-Gaudin}, {Magrini}, {Bergemann}, {Gonz{\'a}lez Hern{\'a}ndez}, {Hill}, {Lardo}, {de Laverny}, {Lind}, {Masseron}, {Montes}, {Mucciarelli}, {Nordlander}, {Recio Blanco}, {Sobeck}, {Sordo}, {Sousa}, {Tabernero}, {Vallenari}, \& {Van Eck}}]{jofre2014A&A...564A.133J}
{Jofr{\'e}}, P., {Heiter}, U., {Soubiran}, C., {et~al.} 2014, \aap, 564, A133

\bibitem[{{Karovicova} {et~al.}(2020){Karovicova}, {White}, {Nordlander}, {Casagrande}, {Ireland}, {Huber}, \& {Jofr{\'e}}}]{karovicova2020A&A...640A..25K}
{Karovicova}, I., {White}, T.~R., {Nordlander}, T., {et~al.} 2020, \aap, 640, A25

\bibitem[{{Keller} {et~al.}(2014){Keller}, {Bessell}, {Frebel}, {Casey}, {Asplund}, {Jacobson}, {Lind}, {Norris}, {Yong}, {Heger}, {Magic}, {da Costa}, {Schmidt}, \& {Tisserand}}]{keller2014Natur.506..463K}
{Keller}, S.~C., {Bessell}, M.~S., {Frebel}, A., {et~al.} 2014, \nat, 506, 463

\bibitem[{{Kennicutt}(1998)}]{Kennicutt98}
{Kennicutt}, Robert~C., J. 1998, \araa, 36, 189

\bibitem[{{Khoperskov} {et~al.}(2023){Khoperskov}, {Minchev}, {Libeskind}, {Haywood}, {Di Matteo}, {Belokurov}, {Steinmetz}, {Gomez}, {Grand}, {Hoffman}, {Knebe}, {Sorce}, {Spaare}, {Tempel}, \& {Vogelsberger}}]{Khoperskov2023A&A...677A..90K}
{Khoperskov}, S., {Minchev}, I., {Libeskind}, N., {et~al.} 2023, \aap, 677, A90

\bibitem[{{Kim} {et~al.}(2002){Kim}, {Demarque}, {Yi}, \& {Alexander}}]{kim2002}
{Kim}, Y.-C., {Demarque}, P., {Yi}, S.~K., \& {Alexander}, D.~R. 2002, \apjs, 143, 499

\bibitem[{{Kobayashi} {et~al.}(2020){Kobayashi}, {Karakas}, \& {Lugaro}}]{kobayashi20}
{Kobayashi}, C., {Karakas}, A.~I., \& {Lugaro}, M. 2020, \apj, 900, 179

\bibitem[{{Kobayashi} {et~al.}(2006){Kobayashi}, {Umeda}, {Nomoto}, {Tominaga}, \& {Ohkubo}}]{kobayashi06}
{Kobayashi}, C., {Umeda}, H., {Nomoto}, K., {Tominaga}, N., \& {Ohkubo}, T. 2006, \apj, 653, 1145

\bibitem[{{Koch} {et~al.}(2009){Koch}, {Wilkinson}, {Kleyna}, {Irwin}, {Zucker}, {Belokurov}, {Gilmore}, {Fellhauer}, \& {Evans}}]{Koch2009ApJ...690..453K}
{Koch}, A., {Wilkinson}, M.~I., {Kleyna}, J.~T., {et~al.} 2009, \apj, 690, 453

\bibitem[{{Koppelman} {et~al.}(2020){Koppelman}, {Bos}, \& {Helmi}}]{koppelman20}
{Koppelman}, H.~H., {Bos}, R. O.~Y., \& {Helmi}, A. 2020, \aap, 642, L18

\bibitem[{{Koppelman} {et~al.}(2019){Koppelman}, {Helmi}, {Massari}, {Price-Whelan}, \& {Starkenburg}}]{koppelman19}
{Koppelman}, H.~H., {Helmi}, A., {Massari}, D., {Price-Whelan}, A.~M., \& {Starkenburg}, T.~K. 2019, \aap, 631, L9

\bibitem[{{Korn} {et~al.}(2007){Korn}, {Grundahl}, {Richard}, {Mashonkina}, {Barklem}, {Collet}, {Gustafsson}, \& {Piskunov}}]{korn2007ApJ...671..402K}
{Korn}, A.~J., {Grundahl}, F., {Richard}, O., {et~al.} 2007, \apj, 671, 402

\bibitem[{{Kroupa} {et~al.}(1993){Kroupa}, {Tout}, \& {Gilmore}}]{Kroupa1993}
{Kroupa}, P., {Tout}, C.~A., \& {Gilmore}, G. 1993, \mnras, 262, 545

\bibitem[{{Kruijssen} {et~al.}(2020){Kruijssen}, {Pfeffer}, {Chevance}, {Bonaca}, {Trujillo-Gomez}, {Bastian}, {Reina-Campos}, {Crain}, \& {Hughes}}]{kruijssen20}
{Kruijssen}, J.~M.~D., {Pfeffer}, J.~L., {Chevance}, M., {et~al.} 2020, \mnras, 498, 2472

\bibitem[{{Lai} {et~al.}(2011){Lai}, {Lee}, {Bolte}, {Lucatello}, {Beers}, {Johnson}, {Sivarani}, \& {Rockosi}}]{Lai2011ApJ...738...51L}
{Lai}, D.~K., {Lee}, Y.~S., {Bolte}, M., {et~al.} 2011, \apj, 738, 51

\bibitem[{{Larson}(1981)}]{Larson1981MNRAS.194..809L}
{Larson}, R.~B. 1981, \mnras, 194, 809

\bibitem[{{Leenaarts} \& {Carlsson}(2009)}]{2009ASPC..415...87L}
{Leenaarts}, J. \& {Carlsson}, M. 2009, in Astronomical Society of the Pacific Conference Series, Vol. 415, The Second Hinode Science Meeting: Beyond Discovery-Toward Understanding, ed. B.~{Lites}, M.~{Cheung}, T.~{Magara}, J.~{Mariska}, \& K.~{Reeves}, 87

\bibitem[{{Limongi} \& {Chieffi}(2018)}]{Limongi+18}
{Limongi}, M. \& {Chieffi}, A. 2018, VizieR Online Data Catalog, J/ApJS/237/13

\bibitem[{LOWESS(2021-03-29)}]{LOWESS}
LOWESS. 2021-03-29, {}, last accessed 16 September 2017

\bibitem[{{Ludwig} {et~al.}(2009){Ludwig}, {Behara}, {Steffen}, \& {Bonifacio}}]{ludwig2009}
{Ludwig}, H.-G., {Behara}, N.~T., {Steffen}, M., \& {Bonifacio}, P. 2009, \aap, 502, L1

\bibitem[{{Magic} {et~al.}(2013){Magic}, {Collet}, {Asplund}, {Trampedach}, {Hayek}, {Chiavassa}, {Stein}, \& {Nordlund}}]{2013A&A...557A..26M}
{Magic}, Z., {Collet}, R., {Asplund}, M., {et~al.} 2013, \aap, 557, A26

\bibitem[{{Magrini} {et~al.}(2023){Magrini}, {Bensby}, {Brucalassi}, {Randich}, {Jeffries}, {de Silva}, {Skuladottir}, {Smiljanic}, {Gonzalez}, {Hill}, {Lagarde}, {Tolstoy}, {Arroyo-Polonio}, {Baratella}, {Barnes}, {Battaglia}, {Baumgardt}, {Bellazzini}, {Biazzo}, {Bragaglia}, {Carter}, {Casali}, {Cescutti}, {Danielski}, {Delgado Mena}, {Drazdauskas}, {Gieles}, {Giribaldi}, {Hawkins}, {Hoeijmakers}, {Jablonka}, {Kamath}, {Louth}, {Fabiola Marino}, {Martell}, {Merle}, {Montet}, {Murphy}, {Nisini}, {Nordlander}, {D'Orazi}, {Pino}, {Romano}, {Sacco}, {Sandford}, {Sollima}, {Spina}, {Tautvaisiene}, {Ting}, {Tozzi}, {Van der Swaelmen}, {Van Eck}, {Watson}, {Worley}, \& {Zocchi}}]{Magrini2023arXiv231208270M}
{Magrini}, L., {Bensby}, T., {Brucalassi}, A., {et~al.} 2023, arXiv e-prints, arXiv:2312.08270

\bibitem[{{Mainieri} {et~al.}(2024){Mainieri}, {Anderson}, {Brinchmann}, {Cimatti}, {Ellis}, {Hill}, {Kneib}, {McLeod}, {Opitom}, {Roth}, {Sanchez-Saez}, {Smiljanic}, {Tolstoy}, {Bacon}, {Randich}, {Adamo}, {Annibali}, {Arevalo}, {Audard}, {Barsanti}, {Battaglia}, {Bayo Aran}, {Belfiore}, {Bellazzini}, {Bellini}, {Beltran}, {Berni}, {Bianchi}, {Biazzo}, {Bisero}, {Bisogni}, {Bland-Hawthorn}, {Blondin}, {Bodensteiner}, {Boffin}, {Bonito}, {Bono}, {Bouche}, {Bowman}, {Braga}, {Bragaglia}, {Branchesi}, {Brucalassi}, {Bryant}, {Bryson}, {Busa}, {Camera}, {Carbone}, {Casali}, {Casali}, {Casasola}, {Castro}, {Catelan}, {Cavallo}, {Chiappini}, {Cioni}, {Colless}, {Colzi}, {Contarini}, {Couch}, {D'Ammando}, {d'Assignies D.}, {D'Orazi}, {da Silva}, {Dainotti}, {Damiani}, {Danielski}, {De Cia}, {de Jong}, {Dhawan}, {Dierickx}, {Driver}, {Dupletsa}, {Escoffier}, {Escorza}, {Fabrizio}, {Fiorentino}, {Fontana}, {Fontani}, {Forero Sanchez}, {Franois}, {Galindo-Guil}, {Gallazzi}, {Galli}, {Garcia}, {Garcia-Rojas},
  {Garilli}, {Grand}, {Guarcello}, {Hazra}, {Helmi}, {Herrero}, {Iglesias}, {Ilic}, {Irsic}, {Ivanov}, {Izzo}, {Jablonka}, {Joachimi}, {Kakkad}, {Kamann}, {Koposov}, {Kordopatis}, {Kovacevic}, {Kraljic}, {Kuncarayakti}, {Kwon}, {La Forgia}, {Lahav}, {Laigle}, {Lazzarin}, {Leaman}, {Leclercq}, {Lee}, {Lee}, {Lehnert}, {Lira}, {Loffredo}, {Lucatello}, {Magrini}, {Maguire}, {Mahler}, {Zahra Majidi}, {Malavasi}, {Mannucci}, {Marconi}, {Martin}, {Marulli}, {Massari}, {Matsuno}, {Mattheee}, {McGee}, {Merc}, {Merle}, {Miglio}, {Migliorini}, {Minchev}, {Minniti}, {Miret-Roig}, {Monreal Ibero}, {Montano}, {Montet}, {Moresco}, {Moretti}, {Moscardini}, {Moya}, {Mueller}, {Nanayakkara}, {Nicholl}, {Nordlander}, {Onori}, {Padovani}, {Pala}, {Panda}, {Pandey-Pommier}, {Pasquini}, {Pawlak}, {Pessi}, {Pisani}, {Popovic}, {Prisinzano}, {Raddi}, {Rainer}, {Rebassa-Mansergas}, {Richard}, {Rigault}, {Rocher}, {Romano}, {Rosati}, {Sacco}, {Sanchez-Janssen}, {Sander}, {Sanders}, {Sargent}, {Sarpa}, {Schimd}, {Schipani},
  {Sefusatti}, {Smith}, {Spina}, {Steinmetz}, {Tacchella}, {Tautvaisiene}, {Theissen}, {Thomas}, {Ting}, {Travouillon}, {Tresse}, {Trivedi}, {Tsantaki}, {Tsedrik}, {Urrutia}, {Valenti}, {Van der Swaelmen}, {Van Eck}, {Verdiani}, {Verdier}, {Vergani}, {Verhamme}, {Vernet}, {Verza}, {Viel}, {Vielzeuf}, {Vietri}, {Vink}, {Viscasillas Vazquez}, {Wang}, {Weilbacher}, {Wendt}, {Wright}, {Ye}, {Yeche}, {Yu}, {Zafar}, {Zibetti}, {Ziegler}, \& {Zinchenko}}]{Mainieri2024arXiv240305398M}
{Mainieri}, V., {Anderson}, R.~I., {Brinchmann}, J., {et~al.} 2024, arXiv e-prints, arXiv:2403.05398

\bibitem[{{Malhan} {et~al.}(2022){Malhan}, {Ibata}, {Sharma}, {Famaey}, {Bellazzini}, {Carlberg}, {D'Souza}, {Yuan}, {Martin}, \& {Thomas}}]{Malhan2022ApJ...926..107M}
{Malhan}, K., {Ibata}, R.~A., {Sharma}, S., {et~al.} 2022, \apj, 926, 107

\bibitem[{{Malhan} \& {Rix}(2024)}]{malhan2024ApJ...964..104M}
{Malhan}, K. \& {Rix}, H.-W. 2024, \apj, 964, 104

\bibitem[{{Mannucci} {et~al.}(2006){Mannucci}, {Della Valle}, \& {Panagia}}]{Mannucci06}
{Mannucci}, F., {Della Valle}, M., \& {Panagia}, N. 2006, \mnras, 370, 773

\bibitem[{{Mashonkina}(2013)}]{mashonkina2013A&A...550A..28M}
{Mashonkina}, L. 2013, \aap, 550, A28

\bibitem[{{Massari} {et~al.}(2019){Massari}, {Koppelman}, \& {Helmi}}]{massari19}
{Massari}, D., {Koppelman}, H.~H., \& {Helmi}, A. 2019, \aap, 630, L4

\bibitem[{{Matsuno} {et~al.}(2024){Matsuno}, {Amarsi}, {Carlos}, \& {Nissen}}]{matsuno2024A&A...688A..72M}
{Matsuno}, T., {Amarsi}, A.~M., {Carlos}, M., \& {Nissen}, P.~E. 2024, \aap, 688, A72

\bibitem[{{Matsuno} {et~al.}(2019){Matsuno}, {Aoki}, \& {Suda}}]{matsuno2019ApJ...874L..35M}
{Matsuno}, T., {Aoki}, W., \& {Suda}, T. 2019, \apjl, 874, L35

\bibitem[{Matteucci {et~al.}(2006)Matteucci, Panagia, Pipino, Mannucci, Recchi, \& Della~Valle}]{Matteucci06}
Matteucci, F., Panagia, N., Pipino, A., {et~al.} 2006, Monthly Notices of the Royal Astronomical Society, 372, 265

\bibitem[{{Mayor} {et~al.}(2003){Mayor}, {Pepe}, {Queloz}, {Bouchy}, {Rupprecht}, {Lo Curto}, {Avila}, {Benz}, {Bertaux}, {Bonfils}, {Dall}, {Dekker}, {Delabre}, {Eckert}, {Fleury}, {Gilliotte}, {Gojak}, {Guzman}, {Kohler}, {Lizon}, {Longinotti}, {Lovis}, {Megevand}, {Pasquini}, {Reyes}, {Sivan}, {Sosnowska}, {Soto}, {Udry}, {van Kesteren}, {Weber}, \& {Weilenmann}}]{mayor2003Msngr.114...20M}
{Mayor}, M., {Pepe}, F., {Queloz}, D., {et~al.} 2003, The Messenger, 114, 20

\bibitem[{{McMillan}(2017)}]{McMillan2017MNRAS.465...76M}
{McMillan}, P.~J. 2017, \mnras, 465, 76

\bibitem[{{Mel{\'e}ndez} \& {Barbuy}(2009)}]{melendez2009A&A...497..611M}
{Mel{\'e}ndez}, J. \& {Barbuy}, B. 2009, \aap, 497, 611

\bibitem[{{Mel{\'e}ndez} {et~al.}(2012){Mel{\'e}ndez}, {Bergemann}, {Cohen}, {Endl}, {Karakas}, {Ram{\'\i}rez}, {Cochran}, {Yong}, {MacQueen}, {Kobayashi}, \& {Asplund}}]{melendez2012A&A...543A..29M}
{Mel{\'e}ndez}, J., {Bergemann}, M., {Cohen}, J.~G., {et~al.} 2012, \aap, 543, A29

\bibitem[{{Mel{\'e}ndez} {et~al.}(2010){Mel{\'e}ndez}, {Casagrande}, {Ram{\'\i}rez}, {Asplund}, \& {Schuster}}]{melendez2010A&A...515L...3M}
{Mel{\'e}ndez}, J., {Casagrande}, L., {Ram{\'\i}rez}, I., {Asplund}, M., \& {Schuster}, W.~J. 2010, \aap, 515, L3

\bibitem[{{Meynet} \& {Maeder}(2002)}]{Meynet+02}
{Meynet}, G. \& {Maeder}, A. 2002, \aap, 390, 561

\bibitem[{{Molero} {et~al.}(2023){Molero}, {Magrini}, {Matteucci}, {Romano}, {Palla}, {Cescutti}, {Viscasillas V{\'a}zquez}, \& {Spitoni}}]{Molero2023MNRAS.523.2974M}
{Molero}, M., {Magrini}, L., {Matteucci}, F., {et~al.} 2023, \mnras, 523, 2974

\bibitem[{{Montalb{\'a}n} {et~al.}(2021){Montalb{\'a}n}, {Mackereth}, {Miglio}, {Vincenzo}, {Chiappini}, {Buldgen}, {Mosser}, {Noels}, {Scuflaire}, {Vrard}, {Willett}, {Davies}, {Hall}, {Nielsen}, {Khan}, {Rendle}, {van Rossem}, {Ferguson}, \& {Chaplin}}]{montalban2021NatAs...5..640M}
{Montalb{\'a}n}, J., {Mackereth}, J.~T., {Miglio}, A., {et~al.} 2021, Nature Astronomy, 5, 640

\bibitem[{{Monty} {et~al.}(2020){Monty}, {Venn}, {Lane}, {Lokhorst}, \& {Yong}}]{Monty2020MNRAS.497.1236M}
{Monty}, S., {Venn}, K.~A., {Lane}, J. M.~M., {Lokhorst}, D., \& {Yong}, D. 2020, \mnras, 497, 1236

\bibitem[{{Mori} {et~al.}(2024){Mori}, {Di Matteo}, {Salvadori}, {Khoperskov}, {Pagnini}, \& {Haywood}}]{Mori2024A&A...690A.136M}
{Mori}, A., {Di Matteo}, P., {Salvadori}, S., {et~al.} 2024, \aap, 690, A136

\bibitem[{{Mucciarelli} {et~al.}(2021){Mucciarelli}, {Massari}, {Minelli}, {Romano}, {Bellazzini}, {Ferraro}, {Matteucci}, \& {Origlia}}]{Mucciarelli21}
{Mucciarelli}, A., {Massari}, D., {Minelli}, A., {et~al.} 2021, Nature Astronomy, 5, 1247

\bibitem[{{Mucciarelli} {et~al.}(2022){Mucciarelli}, {Monaco}, {Bonifacio}, {Salaris}, {Deal}, {Spite}, {Richard}, \& {Lallement}}]{Mucciarelli22}
{Mucciarelli}, A., {Monaco}, L., {Bonifacio}, P., {et~al.} 2022, \aap, 661, A153

\bibitem[{{Myeong} {et~al.}(2019){Myeong}, {Vasiliev}, {Iorio}, {Evans}, \& {Belokurov}}]{myeong2019MNRAS.488.1235M}
{Myeong}, G.~C., {Vasiliev}, E., {Iorio}, G., {Evans}, N.~W., \& {Belokurov}, V. 2019, \mnras, 488, 1235

\bibitem[{{Nissen} \& {Schuster}(2010)}]{nissen2010A&A...511L..10N}
{Nissen}, P.~E. \& {Schuster}, W.~J. 2010, \aap, 511, L10

\bibitem[{{Nordlander} {et~al.}(2024){Nordlander}, {Gruyters}, {Richard}, \& {Korn}}]{nordlander2024MNRAS.52712120N}
{Nordlander}, T., {Gruyters}, P., {Richard}, O., \& {Korn}, A.~J. 2024, \mnras, 527, 12120

\bibitem[{{Norris} {et~al.}(2007){Norris}, {Christlieb}, {Korn}, {Eriksson}, {Bessell}, {Beers}, {Wisotzki}, \& {Reimers}}]{norris2007ApJ...670..774N}
{Norris}, J.~E., {Christlieb}, N., {Korn}, A.~J., {et~al.} 2007, \apj, 670, 774

\bibitem[{{Norris} {et~al.}(2013){Norris}, {Yong}, {Bessell}, {Christlieb}, {Asplund}, {Gilmore}, {Wyse}, {Beers}, {Barklem}, {Frebel}, \& {Ryan}}]{norris2013ApJ...762...28N}
{Norris}, J.~E., {Yong}, D., {Bessell}, M.~S., {et~al.} 2013, \apj, 762, 28

\bibitem[{{Omukai} {et~al.}(2005){Omukai}, {Tsuribe}, {Schneider}, \& {Ferrara}}]{Omukai05}
{Omukai}, K., {Tsuribe}, T., {Schneider}, R., \& {Ferrara}, A. 2005, The Astrophysical Journal, 626, 627

\bibitem[{{Palla} {et~al.}(2020){Palla}, {Matteucci}, {Spitoni}, {Vincenzo}, \& {Grisoni}}]{Palla20}
{Palla}, M., {Matteucci}, F., {Spitoni}, E., {Vincenzo}, F., \& {Grisoni}, V. 2020, \mnras, 498, 1710

\bibitem[{{Pehlivan Rhodin} {et~al.}(2017){Pehlivan Rhodin}, {Hartman}, {Nilsson}, \& {J{\"o}nsson}}]{2017A&A...598A.102P}
{Pehlivan Rhodin}, A., {Hartman}, H., {Nilsson}, H., \& {J{\"o}nsson}, P. 2017, \aap, 598, A102

\bibitem[{{Pepe} {et~al.}(2021){Pepe}, {Cristiani}, {Rebolo}, {Santos}, {Dekker}, {Cabral}, {Di Marcantonio}, {Figueira}, {Lo Curto}, {Lovis}, {Mayor}, {M{\'e}gevand}, {Molaro}, {Riva}, {Zapatero Osorio}, {Amate}, {Manescau}, {Pasquini}, {Zerbi}, {Adibekyan}, {Abreu}, {Affolter}, {Alibert}, {Aliverti}, {Allart}, {Allende Prieto}, {{\'A}lvarez}, {Alves}, {Avila}, {Baldini}, {Bandy}, {Barros}, {Benz}, {Bianco}, {Borsa}, {Bourrier}, {Bouchy}, {Broeg}, {Calderone}, {Cirami}, {Coelho}, {Conconi}, {Coretti}, {Cumani}, {Cupani}, {D'Odorico}, {Damasso}, {Deiries}, {Delabre}, {Demangeon}, {Dumusque}, {Ehrenreich}, {Faria}, {Fragoso}, {Genolet}, {Genoni}, {G{\'e}nova Santos}, {Gonz{\'a}lez Hern{\'a}ndez}, {Hughes}, {Iwert}, {Kerber}, {Knudstrup}, {Landoni}, {Lavie}, {Lillo-Box}, {Lizon}, {Maire}, {Martins}, {Mehner}, {Micela}, {Modigliani}, {Monteiro}, {Monteiro}, {Moschetti}, {Murphy}, {Nunes}, {Oggioni}, {Oliveira}, {Oshagh}, {Pall{\'e}}, {Pariani}, {Poretti}, {Rasilla}, {Rebord{\~a}o}, {Redaelli}, {Santana Tschudi},
  {Santin}, {Santos}, {S{\'e}gransan}, {Schmidt}, {Segovia}, {Sosnowska}, {Sozzetti}, {Sousa}, {Span{\`o}}, {Su{\'a}rez Mascare{\~n}o}, {Tabernero}, {Tenegi}, {Udry}, \& {Zanutta}}]{pepe2021A&A...645A..96P}
{Pepe}, F., {Cristiani}, S., {Rebolo}, R., {et~al.} 2021, \aap, 645, A96

\bibitem[{{Pereira} {et~al.}(2013){Pereira}, {Asplund}, {Collet}, {Thaler}, {Trampedach}, \& {Leenaarts}}]{pereira2013}
{Pereira}, T.~M.~D., {Asplund}, M., {Collet}, R., {et~al.} 2013, \aap, 554, A118

\bibitem[{{Prantzos} {et~al.}(2018){Prantzos}, {Abia}, {Limongi}, {Chieffi}, \& {Cristallo}}]{prantzos18}
{Prantzos}, N., {Abia}, C., {Limongi}, M., {Chieffi}, A., \& {Cristallo}, S. 2018, \mnras, 476, 3432

\bibitem[{{Pr{\v{s}}a} {et~al.}(2016){Pr{\v{s}}a}, {Harmanec}, {Torres}, {Mamajek}, {Asplund}, {Capitaine}, {Christensen-Dalsgaard}, {Depagne}, {Haberreiter}, {Hekker}, {Hilton}, {Kopp}, {Kostov}, {Kurtz}, {Laskar}, {Mason}, {Milone}, {Montgomery}, {Richards}, {Schmutz}, {Schou}, \& {Stewart}}]{2016AJ....152...41P}
{Pr{\v{s}}a}, A., {Harmanec}, P., {Torres}, G., {et~al.} 2016, \aj, 152, 41

\bibitem[{{Ram{\'\i}rez} {et~al.}(2013){Ram{\'\i}rez}, {Allende Prieto}, \& {Lambert}}]{ramirez2013ApJ...764...78R}
{Ram{\'\i}rez}, I., {Allende Prieto}, C., \& {Lambert}, D.~L. 2013, \apj, 764, 78

\bibitem[{{Randich} {et~al.}(2022){Randich}, {Gilmore}, {Magrini}, {Sacco}, {Jackson}, {Jeffries}, {Worley}, {Hourihane}, {Gonneau}, {Viscasillas Vazquez}, {Franciosini}, {Lewis}, {Alfaro}, {Allende Prieto}, {Bensby}, {Blomme}, {Bragaglia}, {Flaccomio}, {Fran{\c{c}}ois}, {Irwin}, {Koposov}, {Korn}, {Lanzafame}, {Pancino}, {Recio-Blanco}, {Smiljanic}, {Van Eck}, {Zwitter}, {Asplund}, {Bonifacio}, {Feltzing}, {Binney}, {Drew}, {Ferguson}, {Micela}, {Negueruela}, {Prusti}, {Rix}, {Vallenari}, {Bayo}, {Bergemann}, {Biazzo}, {Carraro}, {Casey}, {Damiani}, {Frasca}, {Heiter}, {Hill}, {Jofr{\'e}}, {de Laverny}, {Lind}, {Marconi}, {Martayan}, {Masseron}, {Monaco}, {Morbidelli}, {Prisinzano}, {Sbordone}, {Sousa}, {Zaggia}, {Adibekyan}, {Bonito}, {Caffau}, {Daflon}, {Feuillet}, {Gebran}, {Gonzalez Hernandez}, {Guiglion}, {Herrero}, {Lobel}, {Maiz Apellaniz}, {Merle}, {Mikolaitis}, {Montes}, {Morel}, {Soubiran}, {Spina}, {Tabernero}, {Tautvai{\v{s}}iene}, {Traven}, {Valentini}, {Van der Swaelmen}, {Villanova}, {Wright},
  {Abbas}, {Aguirre B{\o}rsen-Koch}, {Alves}, {Balaguer-Nunez}, {Barklem}, {Barrado}, {Berlanas}, {Binks}, {Bressan}, {Capuzzo-Dolcetta}, {Casagrande}, {Casamiquela}, {Collins}, {D'Orazi}, {Dantas}, {Debattista}, {Delgado-Mena}, {Di Marcantonio}, {Drazdauskas}, {Evans}, {Famaey}, {Franchini}, {Fr{\'e}mat}, {Friel}, {Fu}, {Geisler}, {Gerhard}, {Gonzalez Solares}, {Grebel}, {Gutierrez Albarran}, {Hatzidimitriou}, {Held}, {Jim{\'e}nez-Esteban}, {J{\"o}nsson}, {Jordi}, {Khachaturyants}, {Kordopatis}, {Kos}, {Lagarde}, {Mahy}, {Mapelli}, {Marfil}, {Martell}, {Messina}, {Miglio}, {Minchev}, {Moitinho}, {Montalban}, {Monteiro}, {Morossi}, {Mowlavi}, {Mucciarelli}, {Murphy}, {Nardetto}, {Ortolani}, {Paletou}, {Palou{\v{s}}}, {Paunzen}, {Pickering}, {Quirrenbach}, {Re Fiorentin}, {Read}, {Romano}, {Ryde}, {Sanna}, {Santos}, {Seabroke}, {Spagna}, {Steinmetz}, {Stonkut{\'e}}, {Sutorius}, {Th{\'e}venin}, {Tosi}, {Tsantaki}, {Vink}, {Wright}, {Wyse}, {Zoccali}, {Zorec}, {Zucker}, \& {Walton}}]{Randich2022A&A...666A.121R}
{Randich}, S., {Gilmore}, G., {Magrini}, L., {et~al.} 2022, \aap, 666, A121

\bibitem[{{Reggiani} \& {Mel{\'e}ndez}(2018)}]{reggiani2018MNRAS.475.3502R}
{Reggiani}, H. \& {Mel{\'e}ndez}, J. 2018, \mnras, 475, 3502

\bibitem[{{Reggiani} {et~al.}(2016){Reggiani}, {Mel{\'e}ndez}, {Yong}, {Ram{\'\i}rez}, \& {Asplund}}]{reggiani2016A&A...586A..67R}
{Reggiani}, H., {Mel{\'e}ndez}, J., {Yong}, D., {Ram{\'\i}rez}, I., \& {Asplund}, M. 2016, \aap, 586, A67

\bibitem[{Rey {et~al.}(2023)Rey, Agertz, Starkenburg, Renaud, Joshi, Pontzen, Martin, Feuillet, \& Read}]{rey_10.1093/mnras/stad513}
Rey, M.~P., Agertz, O., Starkenburg, T.~K., {et~al.} 2023, Monthly Notices of the Royal Astronomical Society, 521, 995

\bibitem[{{Rodr{\'\i}guez D{\'\i}az} {et~al.}(2024){Rodr{\'\i}guez D{\'\i}az}, {Lagae}, {Amarsi}, {Bigot}, {Zhou}, {Aguirre B{\o}rsen-Koch}, {Lind}, {Trampedach}, \& {Collet}}]{2024A&A...688A.212R}
{Rodr{\'\i}guez D{\'\i}az}, L.~F., {Lagae}, C., {Amarsi}, A.~M., {et~al.} 2024, \aap, 688, A212

\bibitem[{{Romano} {et~al.}(2020){Romano}, {Franchini}, {Grisoni}, {Spitoni}, {Matteucci}, \& {Morossi}}]{Romano2020}
{Romano}, D., {Franchini}, M., {Grisoni}, V., {et~al.} 2020, \aap, 639, A37

\bibitem[{{Romano} {et~al.}(2010){Romano}, {Karakas}, {Tosi}, \& {Matteucci}}]{Romano10}
{Romano}, D., {Karakas}, A.~I., {Tosi}, M., \& {Matteucci}, F. 2010, \aap, 522, A32

\bibitem[{{Romano} {et~al.}(1999){Romano}, {Matteucci}, {Molaro}, \& {Bonifacio}}]{Romano99}
{Romano}, D., {Matteucci}, F., {Molaro}, P., \& {Bonifacio}, P. 1999, \aap, 352, 117

\bibitem[{{Romano} {et~al.}(2000){Romano}, {Matteucci}, {Salucci}, \& {Chiappini}}]{Romano20}
{Romano}, D., {Matteucci}, F., {Salucci}, P., \& {Chiappini}, C. 2000, \apj, 539, 235

\bibitem[{{Romano} {et~al.}(2019){Romano}, {Matteucci}, {Zhang}, {Ivison}, \& {Ventura}}]{Romano19}
{Romano}, D., {Matteucci}, F., {Zhang}, Z.-Y., {Ivison}, R.~J., \& {Ventura}, P. 2019, \mnras, 490, 2838

\bibitem[{{Romano} {et~al.}(2017){Romano}, {Matteucci}, {Zhang}, {Papadopoulos}, \& {Ivison}}]{Romano17}
{Romano}, D., {Matteucci}, F., {Zhang}, Z.~Y., {Papadopoulos}, P.~P., \& {Ivison}, R.~J. 2017, \mnras, 470, 401

\bibitem[{{Rossi} {et~al.}(2024{\natexlab{a}}){Rossi}, {Romano}, {Mucciarelli}, {Ceccarelli}, {Massari}, \& {Zamorani}}]{Rossi24CNO}
{Rossi}, M., {Romano}, D., {Mucciarelli}, A., {et~al.} 2024{\natexlab{a}}, arXiv e-prints, arXiv:2406.14615

\bibitem[{{Rossi} {et~al.}(2021){Rossi}, {Salvadori}, \& {Sk{\'u}lad{\'o}ttir}}]{Rossi+21}
{Rossi}, M., {Salvadori}, S., \& {Sk{\'u}lad{\'o}ttir}, {\'A}. 2021, \mnras, 503, 6026

\bibitem[{{Rossi} {et~al.}(2023){Rossi}, {Salvadori}, {Sk{\'u}lad{\'o}ttir}, \& {Vanni}}]{Rossi+23}
{Rossi}, M., {Salvadori}, S., {Sk{\'u}lad{\'o}ttir}, {\'A}., \& {Vanni}, I. 2023, \mnras, 522, L1

\bibitem[{{Rossi} {et~al.}(2024{\natexlab{b}}){Rossi}, {Salvadori}, {Sk{\'u}lad{\'o}ttir}, {Vanni}, \& {Koutsouridou}}]{Rossi2024arXiv}
{Rossi}, M., {Salvadori}, S., {Sk{\'u}lad{\'o}ttir}, {\'A}., {Vanni}, I., \& {Koutsouridou}, I. 2024{\natexlab{b}}, arXiv e-prints, arXiv:2406.12960

\bibitem[{Salvadori \& Ferrara(2009)}]{SS09}
Salvadori, S. \& Ferrara, A. 2009, Monthly Notices of the Royal Astronomical Society: Letters, 395, L6

\bibitem[{Salvadori {et~al.}(2015)Salvadori, Skuladottir, \& Tolstoy}]{SS15}
Salvadori, S., Skuladottir, A., \& Tolstoy, E. 2015, Monthly Notices of the Royal Astronomical Society, 454, 1320

\bibitem[{{Schlafly} \& {Finkbeiner}(2011)}]{Schlafly2011ApJ...737..103S}
{Schlafly}, E.~F. \& {Finkbeiner}, D.~P. 2011, \apj, 737, 103

\bibitem[{{Schlegel} {et~al.}(1998){Schlegel}, {Finkbeiner}, \& {Davis}}]{schlegel1998ApJ...500..525S}
{Schlegel}, D.~J., {Finkbeiner}, D.~P., \& {Davis}, M. 1998, \apj, 500, 525

\bibitem[{{Schneider} {et~al.}(2003){Schneider}, {Ferrara}, {Salvaterra}, {Omukai}, \& {Bromm}}]{Schneider2003}
{Schneider}, R., {Ferrara}, A., {Salvaterra}, R., {Omukai}, K., \& {Bromm}, V. 2003, \nat, 422, 869

\bibitem[{{Searle} \& {Zinn}(1978)}]{searle1978ApJ...225..357S}
{Searle}, L. \& {Zinn}, R. 1978, \apj, 225, 357

\bibitem[{{Semenova} {et~al.}(2020){Semenova}, {Bergemann}, {Deal}, {Serenelli}, {Hansen}, {Gallagher}, {Bayo}, {Bensby}, {Bragaglia}, {Carraro}, {Morbidelli}, {Pancino}, \& {Smiljanic}}]{semenova2020}
{Semenova}, E., {Bergemann}, M., {Deal}, M., {et~al.} 2020, \aap, 643, A164

\bibitem[{{Sestito} {et~al.}(2023){Sestito}, {Venn}, {Arentsen}, {Aguado}, {Kielty}, {Lardo}, {Martin}, {Navarro}, {Starkenburg}, {Waller}, {Carlberg}, {Fran{\c{c}}ois}, {Gonz{\'a}lez Hern{\'a}ndez}, {Kordopatis}, {Vitali}, \& {Yuan}}]{sestito2023MNRAS.518.4557S}
{Sestito}, F., {Venn}, K.~A., {Arentsen}, A., {et~al.} 2023, \mnras, 518, 4557

\bibitem[{Simon(2019)}]{Simon19}
Simon, J.~D. 2019, Monthly Notices of the Royal Astronomical Society

\bibitem[{{Sitnova} {et~al.}(2015){Sitnova}, {Zhao}, {Mashonkina}, {Chen}, {Liu}, {Pakhomov}, {Tan}, {Bolte}, {Alexeeva}, {Grupp}, {Shi}, \& {Zhang}}]{sitnova2015ApJ...808..148S}
{Sitnova}, T., {Zhao}, G., {Mashonkina}, L., {et~al.} 2015, \apj, 808, 148

\bibitem[{{Sitnova} {et~al.}(2019){Sitnova}, {Mashonkina}, {Ezzeddine}, \& {Frebel}}]{Sitnova2019MNRAS.485.3527S}
{Sitnova}, T.~M., {Mashonkina}, L.~I., {Ezzeddine}, R., \& {Frebel}, A. 2019, \mnras, 485, 3527

\bibitem[{{Sitnova} {et~al.}(2024){Sitnova}, {Yuan}, {Matsuno}, {Mashonkina}, {Alexeeva}, {Holmbeck}, {Sestito}, {Lombardo}, {Banerjee}, {Martin}, \& {Jiang}}]{sitnova2024A&A...690A.331S}
{Sitnova}, T.~M., {Yuan}, Z., {Matsuno}, T., {et~al.} 2024, \aap, 690, A331

\bibitem[{{Smiljanic} {et~al.}(2009){Smiljanic}, {Pasquini}, {Bonifacio}, {Galli}, {Gratton}, {Randich}, \& {Wolff}}]{smiljanic2009A&A...499..103S}
{Smiljanic}, R., {Pasquini}, L., {Bonifacio}, P., {et~al.} 2009, \aap, 499, 103

\bibitem[{{Souto} {et~al.}(2019){Souto}, {Allende Prieto}, {Cunha}, {Pinsonneault}, {Smith}, {Garcia-Dias}, {Bovy}, {Garc{\'\i}a-Hern{\'a}ndez}, {Holtzman}, {Johnson}, {J{\"o}nsson}, {Majewski}, {Shetrone}, {Sobeck}, {Zamora}, {Pan}, \& {Nitschelm}}]{souto2019ApJ...874...97S}
{Souto}, D., {Allende Prieto}, C., {Cunha}, K., {et~al.} 2019, \apj, 874, 97

\bibitem[{{Souto} {et~al.}(2018){Souto}, {Cunha}, {Smith}, {Allende Prieto}, {Garc{\'\i}a-Hern{\'a}ndez}, {Pinsonneault}, {Holzer}, {Frinchaboy}, {Holtzman}, {Johnson}, {J{\"o}nsson}, {Majewski}, {Shetrone}, {Sobeck}, {Stringfellow}, {Teske}, {Zamora}, {Zasowski}, {Carrera}, {Stassun}, {Fernandez-Trincado}, {Villanova}, {Minniti}, \& {Santana}}]{souto2018ApJ...857...14S}
{Souto}, D., {Cunha}, K., {Smith}, V.~V., {et~al.} 2018, \apj, 857, 14

\bibitem[{{Spite} {et~al.}(2013){Spite}, {Caffau}, {Bonifacio}, {Spite}, {Ludwig}, {Plez}, \& {Christlieb}}]{spite2013A&A...552A.107S}
{Spite}, M., {Caffau}, E., {Bonifacio}, P., {et~al.} 2013, \aap, 552, A107

\bibitem[{{Spitoni} {et~al.}(2019){Spitoni}, {Silva Aguirre}, {Matteucci}, {Calura}, \& {Grisoni}}]{Spitoni2019}
{Spitoni}, E., {Silva Aguirre}, V., {Matteucci}, F., {Calura}, F., \& {Grisoni}, V. 2019, \aap, 623, A60

\bibitem[{{Spitoni} {et~al.}(2021){Spitoni}, {Verma}, {Silva Aguirre}, {Vincenzo}, {Matteucci}, {Vai{\v{c}}ekauskait{\.{e}}}, {Palla}, {Grisoni}, \& {Calura}}]{Spitoni2021}
{Spitoni}, E., {Verma}, K., {Silva Aguirre}, V., {et~al.} 2021, \aap, 647, A73

\bibitem[{{Starkenburg} {et~al.}(2017){Starkenburg}, {Martin}, {Youakim}, {Aguado}, {Allende Prieto}, {Arentsen}, {Bernard}, {Bonifacio}, {Caffau}, {Carlberg}, {C{\^o}t{\'e}}, {Fouesneau}, {Fran{\c{c}}ois}, {Franke}, {Gonz{\'a}lez Hern{\'a}ndez}, {Gwyn}, {Hill}, {Ibata}, {Jablonka}, {Longeard}, {McConnachie}, {Navarro}, {S{\'a}nchez-Janssen}, {Tolstoy}, \& {Venn}}]{Starkenburg2017MNRAS.471.2587S}
{Starkenburg}, E., {Martin}, N., {Youakim}, K., {et~al.} 2017, \mnras, 471, 2587

\bibitem[{{Stehl{\'e}} \& {Hutcheon}(1999)}]{stehle1999}
{Stehl{\'e}}, C. \& {Hutcheon}, R. 1999, \aaps, 140, 93

\bibitem[{{Stonkut{\.{e}}} {et~al.}(2016){Stonkut{\.{e}}}, {Koposov}, {Howes}, {Feltzing}, {Worley}, {Gilmore}, {Ruchti}, {Kordopatis}, {Randich}, {Zwitter}, {Bensby}, {Bragaglia}, {Smiljanic}, {Costado}, {Tautvai{\v{s}}ien{\.{e}}}, {Casey}, {Korn}, {Lanzafame}, {Pancino}, {Franciosini}, {Hourihane}, {Jofr{\'e}}, {Lardo}, {Lewis}, {Magrini}, {Monaco}, {Morbidelli}, {Sacco}, \& {Sbordone}}]{Stonkut2016MNRAS.460.1131S}
{Stonkut{\.{e}}}, E., {Koposov}, S.~E., {Howes}, L.~M., {et~al.} 2016, \mnras, 460, 1131

\bibitem[{{Timmes} {et~al.}(1995){Timmes}, {Woosley}, \& {Weaver}}]{timmes95}
{Timmes}, F.~X., {Woosley}, S.~E., \& {Weaver}, T.~A. 1995, \apjs, 98, 617

\bibitem[{{van den Hoek} \& {Groenewegen}(1997)}]{VanDenHoek+97}
{van den Hoek}, L.~B. \& {Groenewegen}, M.~A.~T. 1997, \aaps, 123, 305

\bibitem[{{Van der Swaelmen} {et~al.}(2024){Van der Swaelmen}, {Viscasillas Vazquez}, {Magrini}, {Recio-Blanco}, {Palicio}, {Worley}, {Vallenari}, {Spina}, {Fran{\c{c}}ois}, {Tautvai{\v{s}}iene}, {Sacco}, {Randich}, \& {de Laverny}}]{Van2024arXiv240704204V}
{Van der Swaelmen}, M., {Viscasillas Vazquez}, C., {Magrini}, L., {et~al.} 2024, arXiv e-prints, arXiv:2407.04204

\bibitem[{{Vidal} {et~al.}(1970){Vidal}, {Cooper}, \& {Smith}}]{vidal1970}
{Vidal}, C.~R., {Cooper}, J., \& {Smith}, E.~W. 1970, \jqsrt, 10, 1011

\bibitem[{Vincenzo {et~al.}(2019)Vincenzo, Spitoni, Calura, Matteucci, Silva Aguirre, Miglio, \& Cescutti}]{vincenzo10.1093/mnrasl/slz070}
Vincenzo, F., Spitoni, E., Calura, F., {et~al.} 2019, Monthly Notices of the Royal Astronomical Society: Letters, 487, L47

\bibitem[{{{\v{S}}koda} \& {{\v{S}}lechta}(2004)}]{skoda2004ASPC..310..571S}
{{\v{S}}koda}, P. \& {{\v{S}}lechta}, M. 2004, in Astronomical Society of the Pacific Conference Series, Vol. 310, IAU Colloq. 193: Variable Stars in the Local Group, ed. D.~W. {Kurtz} \& K.~R. {Pollard}, 571

\bibitem[{{{\v{S}}koda} {et~al.}(2008){{\v{S}}koda}, {{\v{S}}urlan}, \& {Tomi{\'c}}}]{skoda2008SPIE.7014E..5XS}
{{\v{S}}koda}, P., {{\v{S}}urlan}, B., \& {Tomi{\'c}}, S. 2008, in Society of Photo-Optical Instrumentation Engineers (SPIE) Conference Series, Vol. 7014, Ground-based and Airborne Instrumentation for Astronomy II, ed. I.~S. {McLean} \& M.~M. {Casali}, 70145X

\bibitem[{{Witten} {et~al.}(2024){Witten}, {Laporte}, {Martin-Alvarez}, {Sijacki}, {Yuan}, {Haehnelt}, {Baker}, {Dunlop}, {Ellis}, {Grogin}, {Illingworth}, {Katz}, {Koekemoer}, {Magee}, {Maiolino}, {McClymont}, {P{\'e}rez-Gonz{\'a}lez}, {Pusk{\'a}s}, {Roberts-Borsani}, {Santini}, \& {Simmonds}}]{Witten2024NatAs...8..384W}
{Witten}, C., {Laporte}, N., {Martin-Alvarez}, S., {et~al.} 2024, Nature Astronomy, 8, 384

\bibitem[{{Yi} {et~al.}(2003){Yi}, {Kim}, \& {Demarque}}]{yi2003}
{Yi}, S.~K., {Kim}, Y.-C., \& {Demarque}, P. 2003, \apjs, 144, 259

\bibitem[{{Yong} {et~al.}(2013){Yong}, {Norris}, {Bessell}, {Christlieb}, {Asplund}, {Beers}, {Barklem}, {Frebel}, \& {Ryan}}]{yong2013ApJ...762...26Y}
{Yong}, D., {Norris}, J.~E., {Bessell}, M.~S., {et~al.} 2013, \apj, 762, 26

\end{thebibliography}

\begin{appendix} 


\newpage
\section{Additional Figures}

\begin{figure*}
    \centering
    \includegraphics[width=0.85\linewidth]{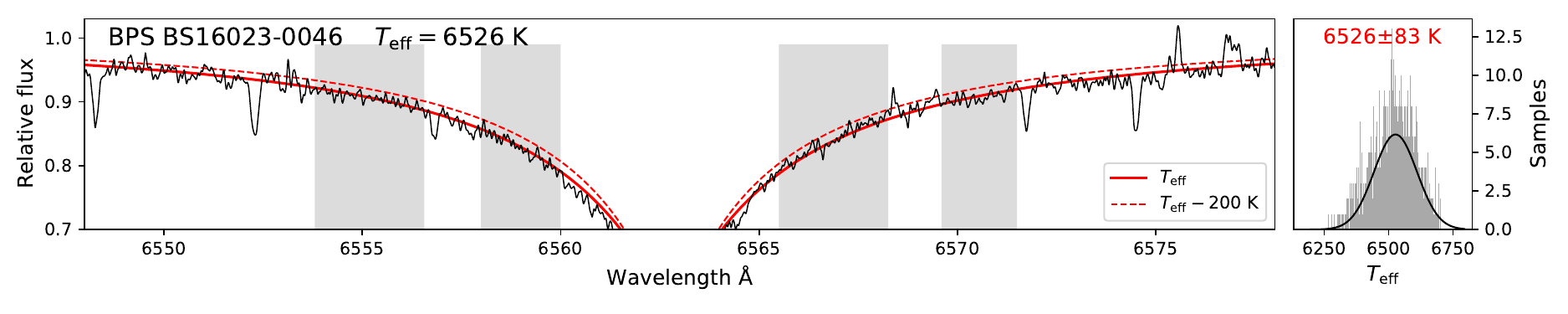}
    \includegraphics[width=0.85\linewidth]{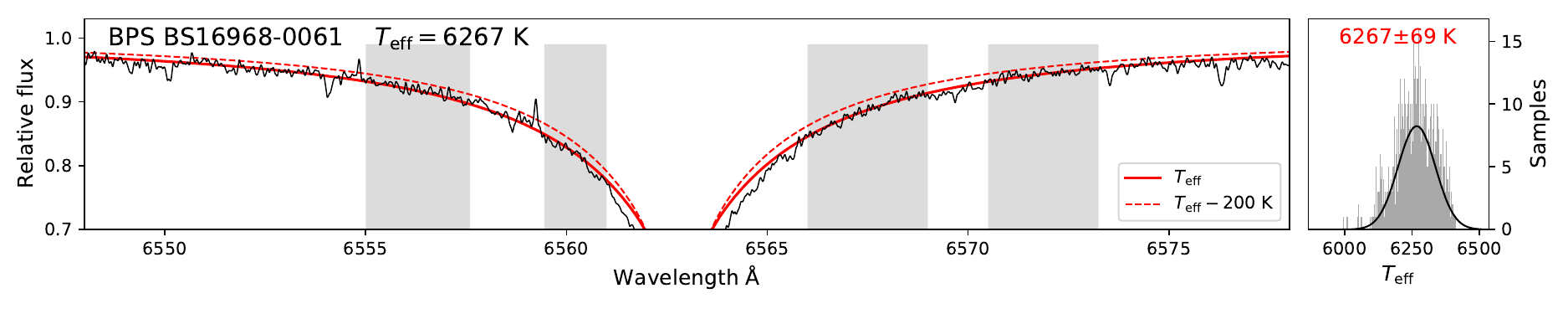}
    \includegraphics[width=0.85\linewidth]{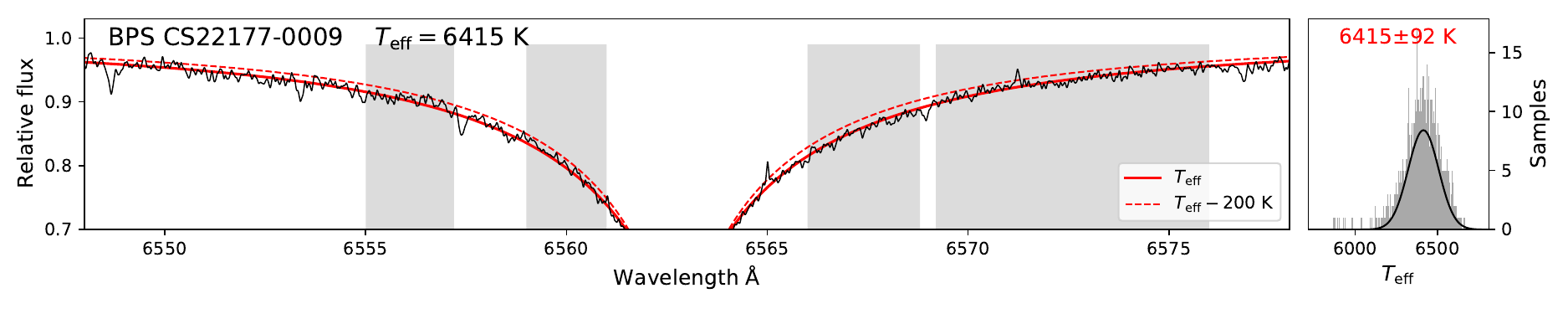}
    \includegraphics[width=0.85\linewidth]{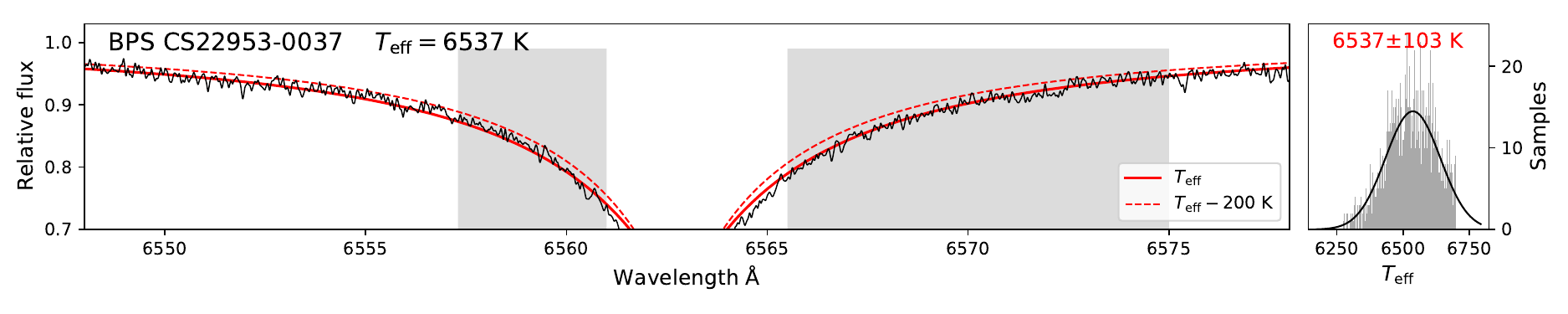}
    \includegraphics[width=0.85\linewidth]{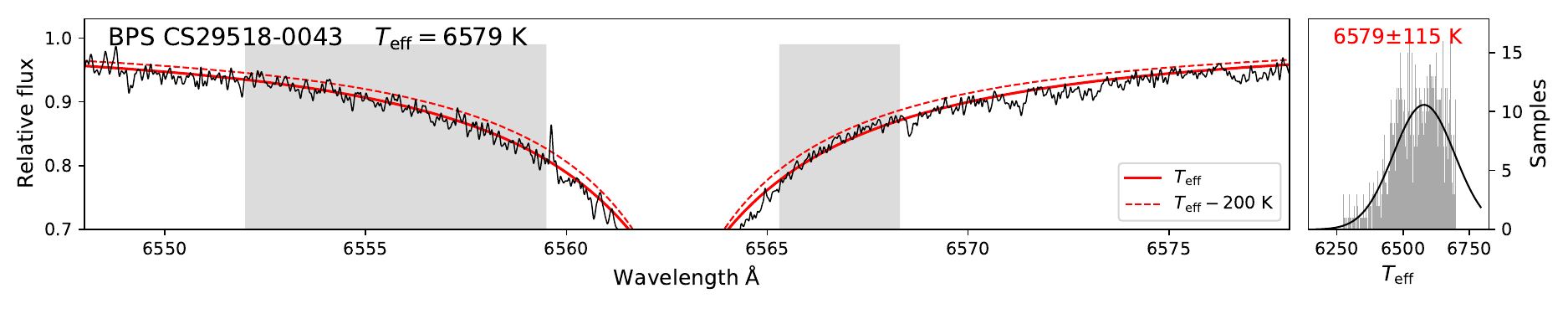}
    \caption{\tiny H$\alpha$ profile fits of five dwarf stars from \cite{melendez2010A&A...515L...3M}. 
    In left panels, observational spectra is shown in black. The fitted 3D NLTE synthetic H$\alpha$ line profile is represented by the red continuous line, and a synthetic profile related to a \teff $-200$~K  is represented by the dashed red line. Shades represent the widows of fits employed. Right panels show histograms of the temperature values related every pixel within the windows of fits. The histograms are fitted by Gaussian distributions (black lines), whose medians and 1$\sigma$ dispersions, are the most likely \teff\ values and their corresponding errors.}
    \label{fig:Teffa}
\end{figure*}

\begin{figure*}[h]
    \centering
    \includegraphics[width=0.3\linewidth]{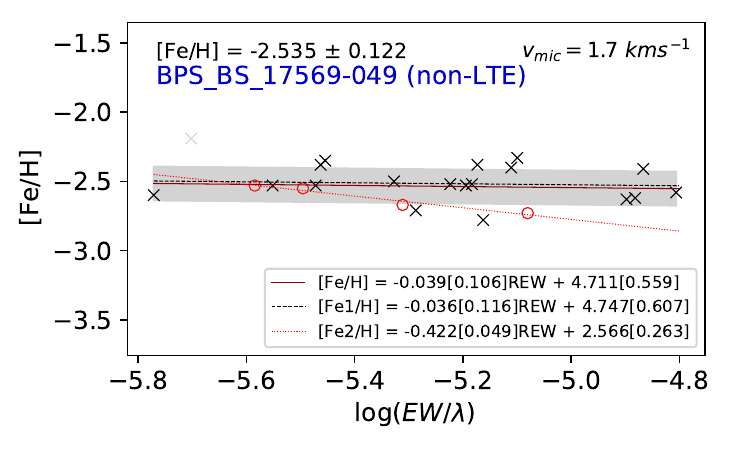}
    \includegraphics[width=0.3\linewidth]{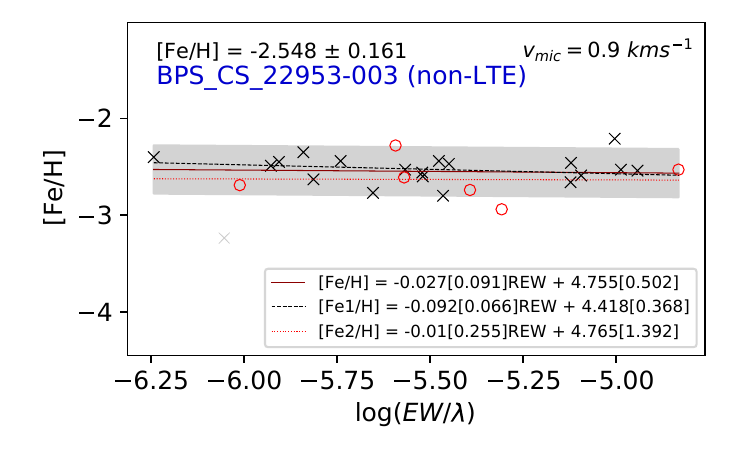}
    \includegraphics[width=0.3\linewidth]{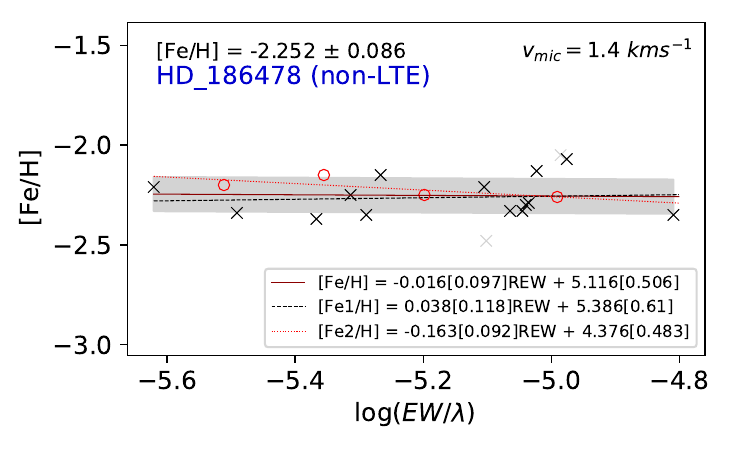}
    \includegraphics[width=0.3\linewidth]{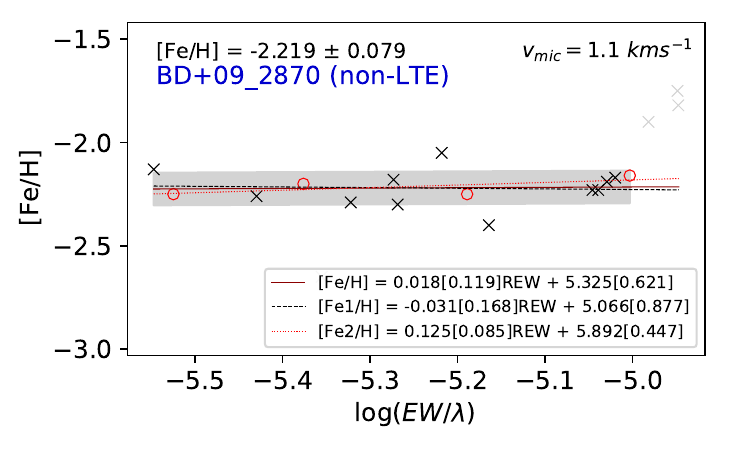}
    \includegraphics[width=0.3\linewidth]{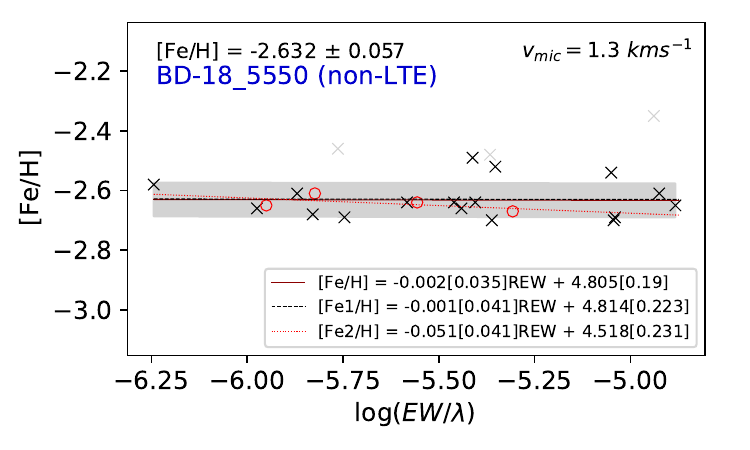}
    \includegraphics[width=0.3\linewidth]{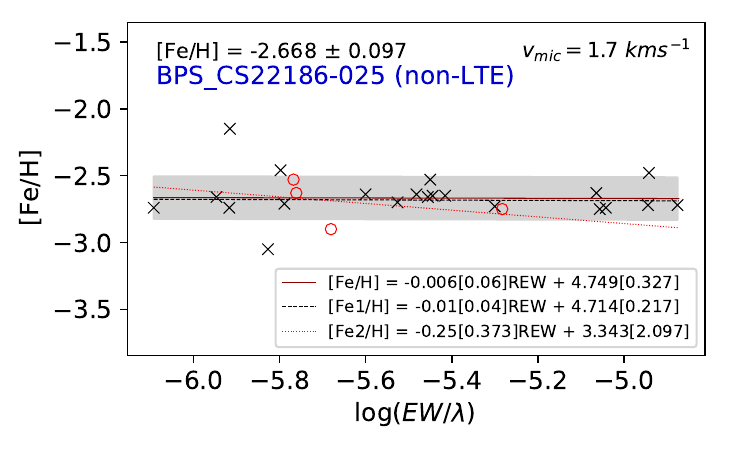}
    \includegraphics[width=0.3\linewidth]{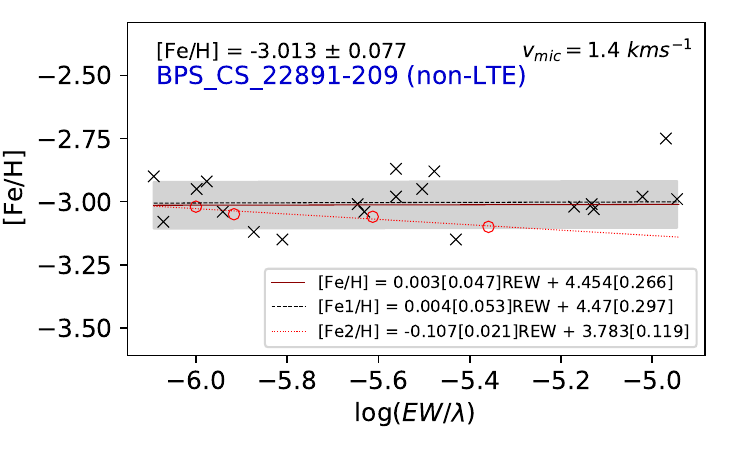}
    \includegraphics[width=0.3\linewidth]{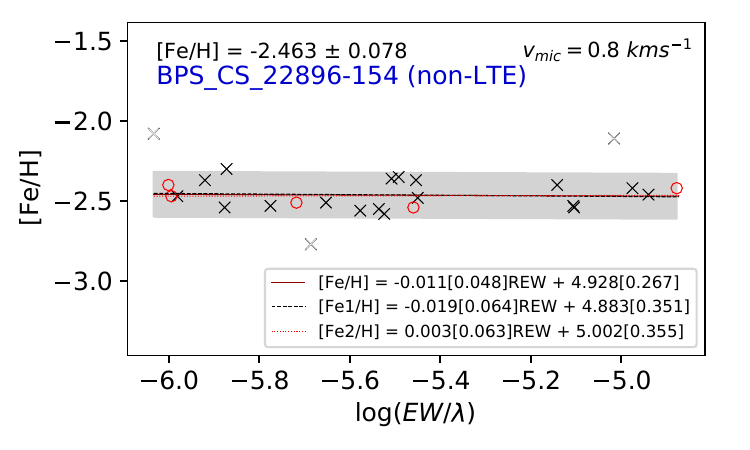}
    \includegraphics[width=0.3\linewidth]{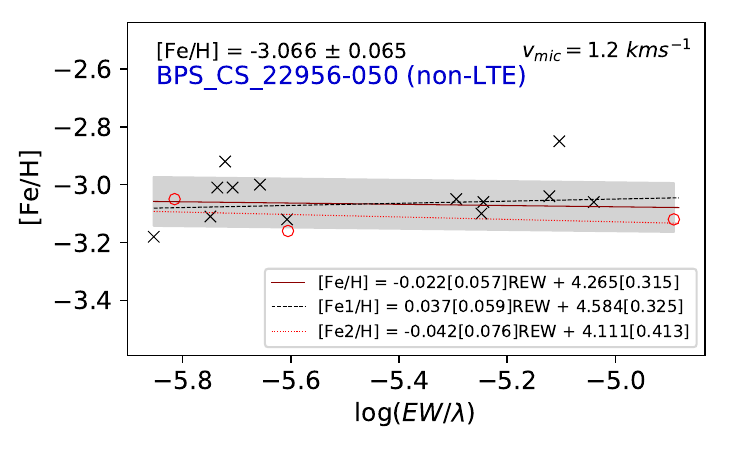}
    \includegraphics[width=0.3\linewidth]{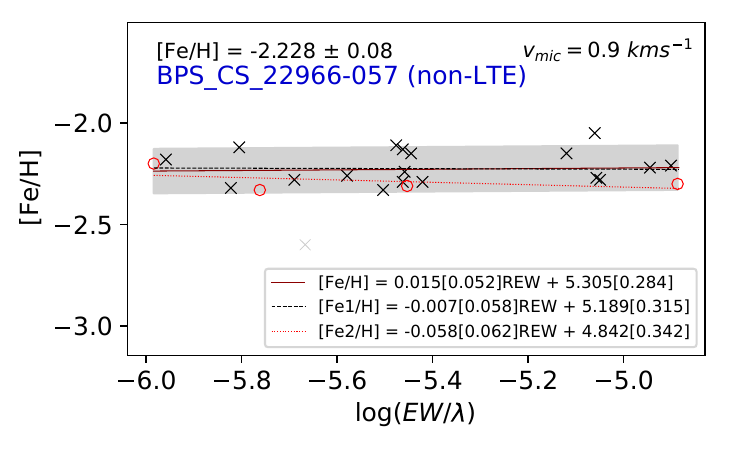}
    \includegraphics[width=0.3\linewidth]{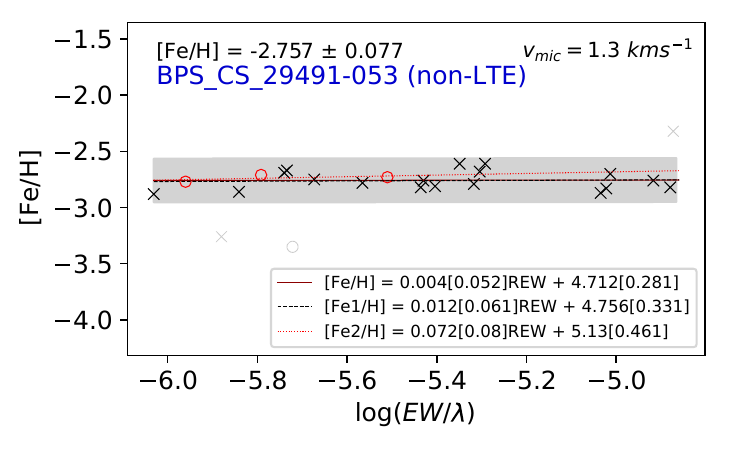}
    \includegraphics[width=0.3\linewidth]{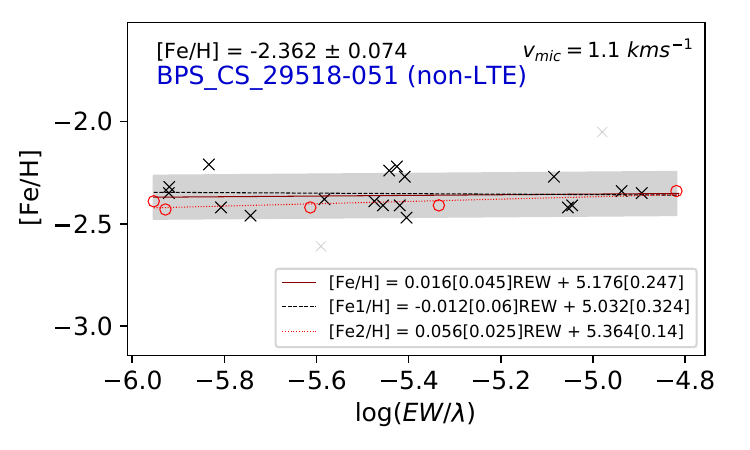}
    \includegraphics[width=0.3\linewidth]{Fe_HD122563_R38513_vmic1.2_vmac6.96_REW-4.8_.pdf}
    \caption{\tiny Determination of metallicity and microturbulence of stars of Precision Sample II. Abundances of \ion{Fe}{I} lines are represented by cross symbols, whereas those of \ion{Fe}{II} are represented by red circles. Regressions of each species are represented by black dotted and red dotted lines, respectively.
    The red line represents the regression of both \ion{Fe}{I} and \ion{Fe}{II} species together. 
    The shade represents 1$\sigma$ dispersion. 
    Regressions are computed considering 2$\sigma$ clipping.
    }
    \label{fig:FeH}
\end{figure*}

\begin{figure*}
    \centering
    \includegraphics[width=0.85\linewidth]{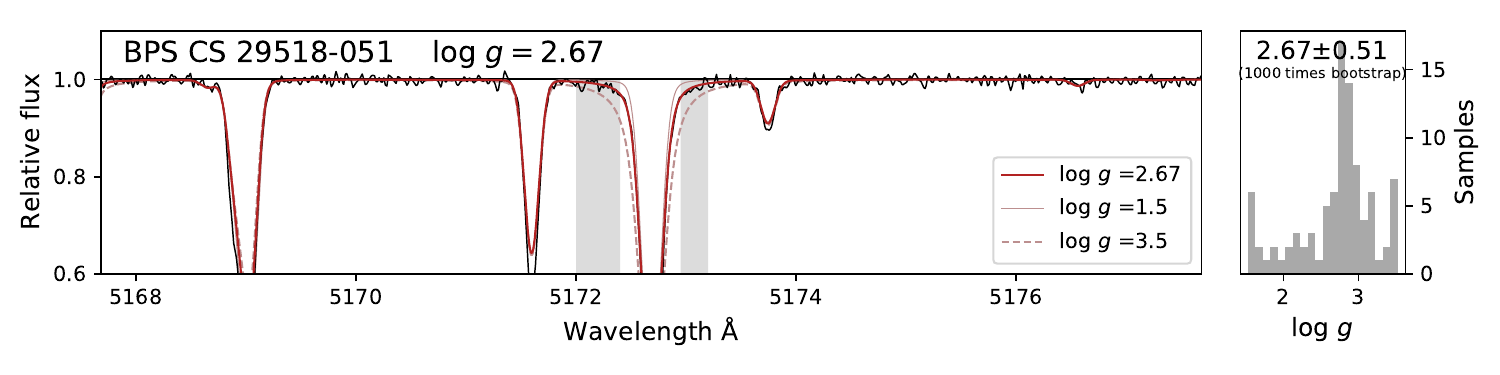}
    \includegraphics[width=0.85\linewidth]{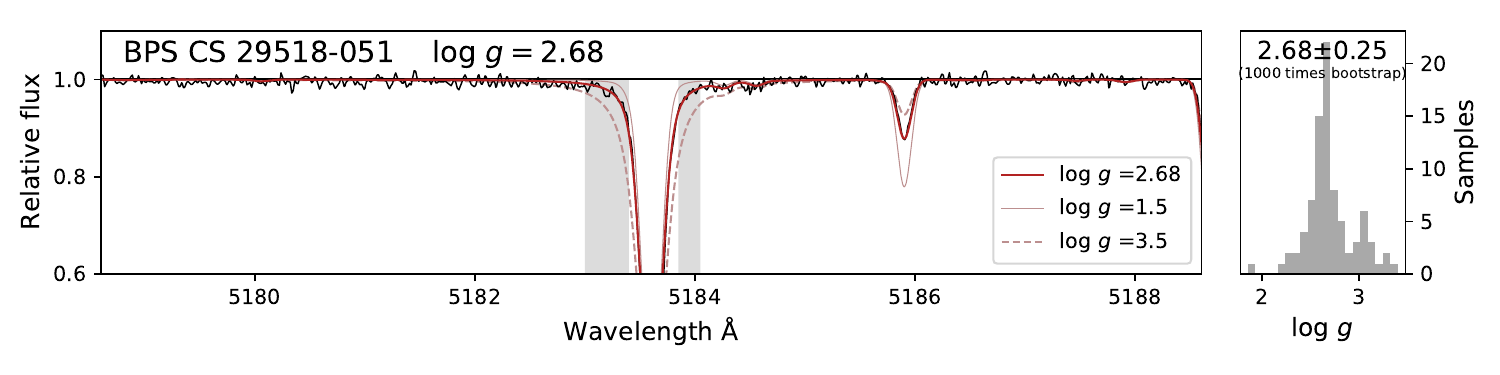}
    
    \caption{\tiny Magnesium profile fits of the star BPS CS 29518$-051$. The observational profile is represented in black. The Red lines represent synthetic profiles. The thick line corresponds to the lost likely \logg. The  right panels show histograms of the \logg\ values associated with all the pixels within the shaded windows. The most probable \logg\ and its error are obtained by bootstrapping.}
    \label{fig:logg_BPS_CS_29518-051}
\end{figure*}

\begin{figure*}
    \centering
    \includegraphics[width=0.85\linewidth]{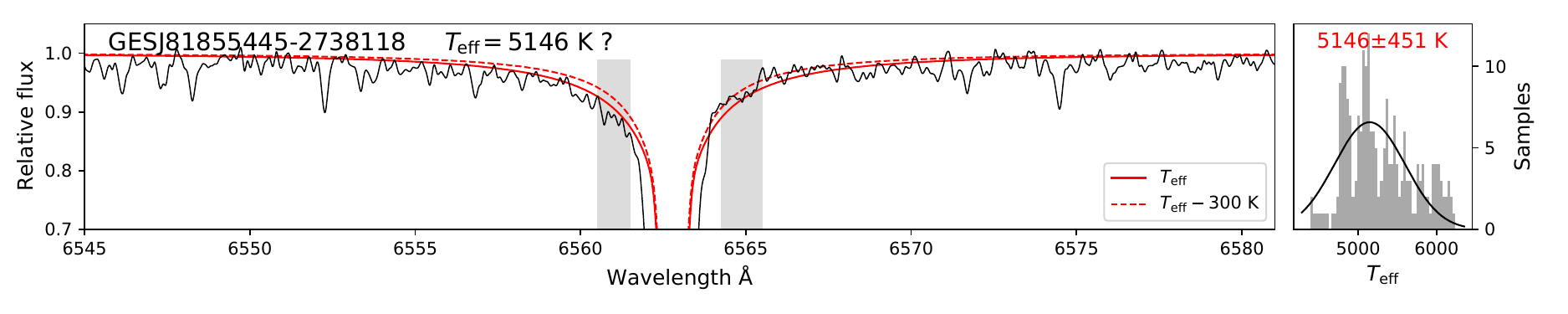}
    
    \caption{\tiny Probable flux emission at the red wing of the H$\alpha$ profile of the star GESJ81855445-2738118. The elements of the plot are the same as those in Fig.~\ref{fig:Teffa}.
    The red wing is more compatible with a cool profile (dashed red line), whereas the blue wing is deeper, therefore hotter, than the hot profile (continuum red line).}
    \label{fig:Ha_emission}
\end{figure*}

\begin{table*}
\caption{Effective temperature determinations from diverse methods, spectral resolution and signal to noise.}
\label{tab:new_teff}
\centering
\scriptsize
\begin{threeparttable}
\begin{tabular}{lccccccccc}
\hline\hline
Star & \teff($B - RP$) & \teff($G - BP$) & \teffP\ / $T_{\mathrm{eff}}^{\mathrm{int}}$ / $T_{\mathrm{eff}}^{\mathrm{IRFM}}$ & \teff(H$\alpha$) & $\langle$\teff$\rangle$ & $\delta T_{\mathrm{eff}}^{E(B-V)}$ & 
comments & R & S/N\\
\hline
BD+24 1676 & --- & --- & --- / --- / $6387 \pm 92$ & $6440 \pm 33$ & $6433 \pm 23$ & --- & --- & 56990 & 489 \\
BD$-10$ 388 & --- & --- & --- / --- / $6260 \pm 61$ & $6293 \pm 43$ & $6285 \pm 19$ & --- & --- &51690 & 528\\
BPS CS22166-030 & --- & --- & --- / --- / $6039 \pm 59$ & $6255 \pm 39$ & $6203 \pm 130$ & --- & --- & 51690& 544 \\
CD-35 14849 & ---  & --- & --- / --- / $6396 \pm 65$ & $6377 \pm 34$ &  $6380 \pm 10$ & --- & --- & 51690 & 526\\
BD+03 740 & --- & --- & --- / --- / $6419 \pm 83$ & $6385 \pm 48$ & $6388 \pm 13$ & --- & --- & 51690 & 655\\
BD$-13$ 3442 & --- & --- & --- / --- / $6434 \pm 80$ & $6446 \pm 34$ & $6444 \pm 6$ & --- & --- & 56990 & 778\\
BD+26 2621 & ---  & --- & --- / --- / $6336 \pm 72$ & $6457 \pm 31$ & $6436 \pm 65$ & --- & --- & 51690 & 381\\
CD$-33$ 1173 &---  & --- & --- / --- / $6685 \pm 68$ & $6602 \pm 32$ & $6617 \pm 46$ & --- & --- & 51690 & 564\\ 
CD$-71$ 1234 & --- & --- & --- / --- / $6408 \pm 69 $ & $6432 \pm 31$ & $6427 \pm 13$ & --- & --- & 51690 & 547 \\
HE 0926$-0508$ & $6645 \pm 31$ & $6481 \pm 69$ & --- / --- / $6451 \pm 79$ & $6458 \pm 32$ & $6457 \pm 3$ & 77 & --- & 42310 & 178\\
LP 815$-43$ & ---  & --- & --- / --- / $6535 \pm 68$ & $6539 \pm 41$ & $6538 \pm 2$ & --- & --- & 107200 & 665\\
LP 831$-70$ & --- & --- & --- / --- / $6408 \pm 67$ & $6379 \pm 35$ & $6384 \pm 16$ & --- & --- & 51690 & 315\\
UCAC2 20056019 & $6390 \pm 27$ & $6412 \pm 68$ & --- / --- / $6393 \pm 14$ & $6443 \pm 33$ & $6401 \pm 25$ & 9 & --- & 42310& 124 \\
Wolf 1492 & $6513 \pm 32$& $6577 \pm 75$ & --- / --- / $6523 \pm 52$ & $6592 \pm 32$ & $6572 \pm 44$ & 41 & ---  & 42310 & 226 \\
HD 140283 & --- & --- & --- / $5792 \pm 55$ / --- & $5810 \pm 32$ & $5805 \pm  11$  & --- & --- & 14000 & 955 \\
BPS BS16023$-046$ & $6497 \pm 30$ & $6524 \pm 73$ & --- / --- / $6501 \pm 7$ & $6526 \pm 83$ & $6502 \pm 7$ & 13 & --- & 42310 & 46 \\ 
BPS BS16968$-061$ & $6241 \pm 32$ & $6276 \pm 72$ & --- / --- / $6247 \pm 37$ & $6267 \pm 69$ & $6252 \pm 12$ & $ 32$ & --- & 42310 & 169 \\ 
BPS CS22177$-009$  & $6368 \pm 34$ & $6398 \pm 76$ & --- / --- / $6373 \pm 35$& $6415 \pm 92$ & $6353 \pm 74$ & 31 & --- & 42310 & 140\\ 
BPS CS22953$-037$ & $6505 \pm 30$ & $6529 \pm 73$ & --- / --- / $6508 \pm 23$ &  $6537 \pm 103$ & $6510 \pm 9$ & 20 & --- & 42310 & 97\\ 
BPS CS29518$-043$ & $6513 \pm 30$ & $6543 \pm 74$ & --- / --- / $6517 \pm 20$ & $6579 \pm  115$ & $6519 \pm 15$ & 14 & --- &42310 & 98 \\
\hline
BPS BS 17569$-049$ & $4798 \pm 20$ & $4800 \pm 33$ & $4798 \pm 61$ / --- / --- & $4790 \pm 55$ & $4794 \pm 6$ & 12 & --- & 42310 & 275\\
BPS CS 22953$-003$ & $5171 \pm 19$ & $5189 \pm 40$ & $5174 \pm 72$ / --- / --- & $5155 \pm 47$ & $5160 \pm 12$ & 39 & --- & 42310 & 171\\
HD 186478 & $4769 \pm 34$& $4770 \pm 45$ & $4770 \pm 111$ / --- / --- & $4804 \pm 75$ & $4793 \pm 22$ & 93 & SB & 42310 & 341\\
BD+09~2870 & $4721 \pm 18$ & $4731 \pm 33$ & $4723 \pm  61$ / --- / --- & $4725 \pm 80$ & $4724 \pm 1$  & 9 &  --- & 42310 & 169 \\
BD$-18$~5550 &  $4909 \pm 51$ & $4900 \pm 63$  & $4906 \pm 95$ / --- / --- & $5030 \pm 76$ &  $4981 \pm 85$ & 74 & ---  & 42310 & 520\\
BPS~CS 22186$-025$   & $5056 \pm 20$ & $5065 \pm 39$ & $5057 \pm 60$ / --- / --- & $5036 \pm 41$ & $5042 \pm 14$ & 11 & --- & 42310 & 113 \\
BPS CS 22891$-209$ & $4826 \pm 27$& $4808 \pm 41$ & $4820 \pm 66$ / --- / --- & $4896 \pm 132$  & $4835 \pm 43$ & 25 & --- & 42310 & 250 \\
BPS CS 22896$-154$ & $5309 \pm 26$ & $5312 \pm 46$& $5310 \pm 65$ / --- / --- & $5274 \pm 53$ & $5288 \pm 25$ & 24 & --- & 42310 & 179 \\
BPS CS 22956$-050$  & $5039 \pm 21$& $5028 \pm 40$ & $5036 \pm 62$ / --- / --- & $4960 \pm 187$ & $5028 \pm 32$ & 12 & --- & 42310 & 150\\
BPS CS 22966$-057$ & $5549 \pm 23$ & $5542 \pm 52$ & $5548 \pm 61$ / --- / --- & $5469 \pm 146$ & $5536 \pm 40$ & $9$ & --- & 42310 & 160\\
BPS CS 29491$-053$ & $4850 \pm 20$ & $4834 \pm 35$ & $4846 \pm 61$ / --- / --- & $4775 \pm 135$ & $4834 \pm 38$ & 7 & --- & 42310 & 145\\
BPS CS 29518$-051$ & $5339 \pm 20$ & $5341 \pm 45$ & $5339 \pm60$ / --- / --- & $5340 \pm 91$ & $5339 \pm 6$ & 7 & --- & 42310 & 159\\
HD 122563 & --- & --- & --- / $4635 \pm 34$ / --- & $4616 \pm 54$ & $4627 \pm 13$ & --- & --- & 19000 & 660\\
\hline
GESJ17543319-4110487 & $5105 \pm 70$ & $5087 \pm 92$ & $5098 \pm 113$ / --- / --- & $5129 \pm 160$ & $5108 \pm 21$ & 95 & --- & 47000 & 40
\\
GESJ17545552-3803393 & $4772 \pm 84$ & $4725 \pm 101$ & $4754 \pm 124$ / --- / --- & $4717 \pm 269$ & $4748 \pm 20$ & 104 & --- & 47000 & 28\\
GESJ17574658-3847500$^{\filledstar}$ & $5088 \pm 81$ & $5542 \pm 112$ & $5251 \pm 306$ / --- / --- & --- & ---& 119 & H$\alpha$(em)/1.47 & 47000 & 27\\
GESJ18185545-2738118$^{\filledstar}$ & $4859 \pm 98$ & $5015 \pm 125$ & $4921 \pm 197$ / --- / --- & --- & --- & 154 & H$\alpha$(em)/1.21 & 47000 & 31\\
GESJ18215385-3410188 & $4919 \pm 44$ & $4945 \pm 62$ & $4928 \pm 77$ / --- / --- & $4904 \pm 237$ & $4925 \pm 10$ & 46 & --- & 47000 & 72\\
GESJ18261627-3154272 & $4965 \pm 68$ & $5194 \pm 90$ & $5014\pm149$ / --- / --- & $4827 \pm 156$ & $4924 \pm 132$ & 103 & --- & 47000 & 31\\
GESJ18265160-3159404$^{\filledstar}$ & $4603 \pm 74$ & $4561 \pm 88$ & $4586 \pm 102$ / --- / --- & --- & --- & 76 & H$\alpha$(em) & 47000 & 43\\
GESJ18361733-2700053 & $4908 \pm 83$ & $4864 \pm 103$ & $4890 \pm 135$ / --- / --- & $4789 \pm 86$ & $4818 \pm 65$ & 117 & --- & 47000 & 36\\
GESJ18370372-2806385 & $4831 \pm 64$ & $4836 \pm 79$ & $4833 \pm 110$ / --- / --- & $5044 \pm 211$ & $4878 \pm 122$ & 91 & --- & 47000 & 62\\
GESJ18374490-2808311 & $4589 \pm 69$ & $4572 \pm 82$ & $4582 \pm 97$ / --- / --- & $4727 \pm 150$ & $4627 \pm 99$ & 75 & --- & 47000 & 23\\
GESJ01293652-5020327$^\ddagger$ & $6278 \pm 43$ & $6331 \pm 69$ & $6293 \pm 69$ / --- / --- & $6343 \pm 71$ & $6368 \pm 35$ & 8 & H$\alpha$(em)/SB? & 47000 & 44\\
GESJ03372526-2724127$^{\filledstar}$ & $5814 \pm 29$ & $5824 \pm 56$ & $5816 \pm 61$ / --- / --- & --- & --- & 6 & H$\alpha$(art) & 47000 & 55\\
GESJ09475504-1027247$^{\filledstar}$ & $5403 \pm 26$ & $5402 \pm 57$ & $5403 \pm 64$ / --- / --- & --- & --- & 22 & H$\alpha$(art) & 47000 & 112\\
GESJ10091577-4127155$^{\filledstar}$ & $5368 \pm 56$ & $5348 \pm 79$ & $5361 \pm 100$ / --- / --- & --- & --- & 79 & H$\alpha$(art) & 47000 & 70\\
GESJ10142785-4052503$^{\filledstar}$ & $5206 \pm 60$ & $5191 \pm 80$ & $5201 \pm 102$ / --- / --- & --- & --- & 83 & H$\alpha$(art) & 47000 & 141\\
GESJ10580638-1542390 & $4828 \pm 35$ & $4819 \pm 57$ & $4826 \pm 64$ / --- / --- & $4955 \pm 201$ & $4838 \pm 53$ & 20 & --- & 47000 & 53\\
GESJ12555146-4507342$^{\filledstar}$ & $5053 \pm 38$ & $5066 \pm 63$ & $5057 \pm 73$ / --- / --- & --- & --- & 40 & H$\alpha$(em) & 47000 & 53\\
GESJ14214860-4408399$^{\filledstar}$ & $4678 \pm 48$& $4666 \pm 64$ & $4674 \pm 69$ / --- / --- & --- & --- & 34 & H$\alpha$(em) & 47000 & 44\\
GESJ14410090-4007413 & $4907 \pm 39$ & $4912 \pm 59$ & $4908 \pm 68$ / --- / --- & $4976 \pm 110$ & $4926 \pm 43$ & 31 & --- & 47000 & 99\\
GESJ15300154-2005148 & $5644 \pm 48$ & $5704 \pm 63$ & $5666 \pm 98 $ / --- / --- &  $5826 \pm 31$ & $5811 \pm 65$ & 66 & --- & 47000 & 80\\
GESJ22125747-4539203 & $5524 \pm 23$ & $5522 \pm 56$ & $5524 \pm61$ / --- / --- & $5417 \pm 116$ & $5500 \pm 62$ & 9 & --- & 47000 & 53\\
GESJ22494718-5001048 & $6301 \pm 41$ & $6322 \pm 69$ & $6307 \pm 62$ / --- / --- & $6235 \pm 230$ & $6302 \pm25$ & 5 & --- & 47000 & 30 \\

\hline
\end{tabular}
\begin{tablenotes}
\item{} \textbf{Notes.} {Second and third columns list temperatures derived from the Gaia colours using the relations of \cite{casagrande2021}; the errors accompanying the values include those induced by \logg\ and [Fe/H] errors.
Fourth column lists either \teffP, $T_{\mathrm{eff}}^{\mathrm{int}}$, or $T_{\mathrm{eff}}^{\mathrm{IRFM}}$; where \teffP\ is the weighted average of the quantities in the first two columns. The corresponding errors are internal and include the influence of the interstellar extinction uncertainty in the budget.
$T_{\mathrm{eff}}^{\mathrm{int}}$ and $T_{\mathrm{eff}}^{\mathrm{IRFM}}$ (direct measurements, not from calibrated relations) are fundamental determinations. Stars with them do not include \teffP\ to avoid potential biases. $T_{\mathrm{eff}}^{\mathrm{IRFM}}$ and $T_{\mathrm{eff}}^{\mathrm{int}}$ are provided by \cite{Casagrande2010} and  \cite{karovicova2020A&A...640A..25K}, respectively.
Fifth column displays \teff(H$\alpha$) along with its internal uncertainty, according to the expansion in \cite{giribaldi2021A&A...650A.194G} and \cite{giribaldi2023A&A...679A.110G}.
Sixth column lists the weighted average of \teffa\ and the temperature in preceding column.
The total uncertainty may be obtained by adding the accuracy of the standard scale in quadrature \cite[50~K][]{giribaldi2021A&A...650A.194G,giribaldi2023A&A...679A.110G}, however we note that this estimate is dominated by the precision of the interferometic measurements \citep{casagrande2014MNRAS.439.2060C,giribaldi2021A&A...650A.194G,giribaldi2023A&A...679A.110G}.
Seventh column lists the error in temperature induced by the uncertainty in $E(B-V)$.
{\bf Eighth} column lists the comments according to the coding: Spectroscopic binary (SB), emission around the H$\alpha$ core (H$\alpha$(em)), artefact at the H$\alpha$ profile (H$\alpha$(art)), Gaia ruwe value is given when $> 1$.
The two last columns list the nominal spectral resolution and S/N around the Mg line at 5528~\AA.
Stars indicated with the symbol ($\filledstar$) have \teff\ determined only with Gaia colour calibrations because of {\it distorted} H$\alpha$ profiles. 
The star accompanied by the symbol ($\ddagger$) has unreliable atmospheric parameters, see main text in Sect~\ref{sec:parameters}.
} 
\end{tablenotes}
\end{threeparttable}
\end{table*}

\begin{table}
\caption{Iron line list.}
\label{tab:Fe_linelist}
\centering
\tiny
\begin{tabular}{lcccc}
\hline\hline
Element & Wavelength (\AA) & evolutionary stage \\
\hline
Fe I & 4202.03 & d/g \\
Fe I & 4210.35 & d/g \\
Fe I & 4466.55 & d/g\\
Fe II & 4508.28 & d/g \\
Fe I & 4918.99 & d/g\\
Fe II & 4923.92 & d/g\\
Fe I & 4966.09 & g \\
Fe I & 4982.50 & g\\
Fe II & 5018.44 & d \\
Fe I & 5049.82 & d/g\\
Fe II & 5169.00 & d\\
Fe I & 5197.58 & d/g\\
Fe I & 5232.94 & d/g\\
Fe II & 5234.62 & g\\
Fe I & 5266.55 & d/g\\
Fe I & 5302.30 & d/g\\
Fe I & 5324.18 & d/g\\
Fe II & 5362.86 & g\\
Fe I & 5397.10 & d\\
Fe I & 5434.52 & d\\
Fe II & 5534.84 & g\\
Fe I & 5569.62 & g\\
Fe I & 5624.54 & d/g\\
Fe I & 6065.48 & g\\
Fe I & 6200.30 & g\\
Fe I & 6213.40 & g\\
Fe I & 6246.32 & g\\
Fe I & 6252.56 & g\\
Fe I & 6265.10 & g\\
Fe I & 6336.82 & g\\
Fe I & 6411.65 & g\\
Fe I & 6430.85 & g\\
Fe II & 6456.38 & g\\
Fe I & 6750.10 & g\\
\hline
\end{tabular}
\begin{tablenotes}
\item{First and second columns are self-explanatory. Third column \\ indicated the stars in which the lines majorly appear, where "d"\\ stands for dwarf and "g" stands for giant.} 
\end{tablenotes}
\end{table}

 \begin{figure*}
    \centering
    \includegraphics[width=0.85\linewidth]{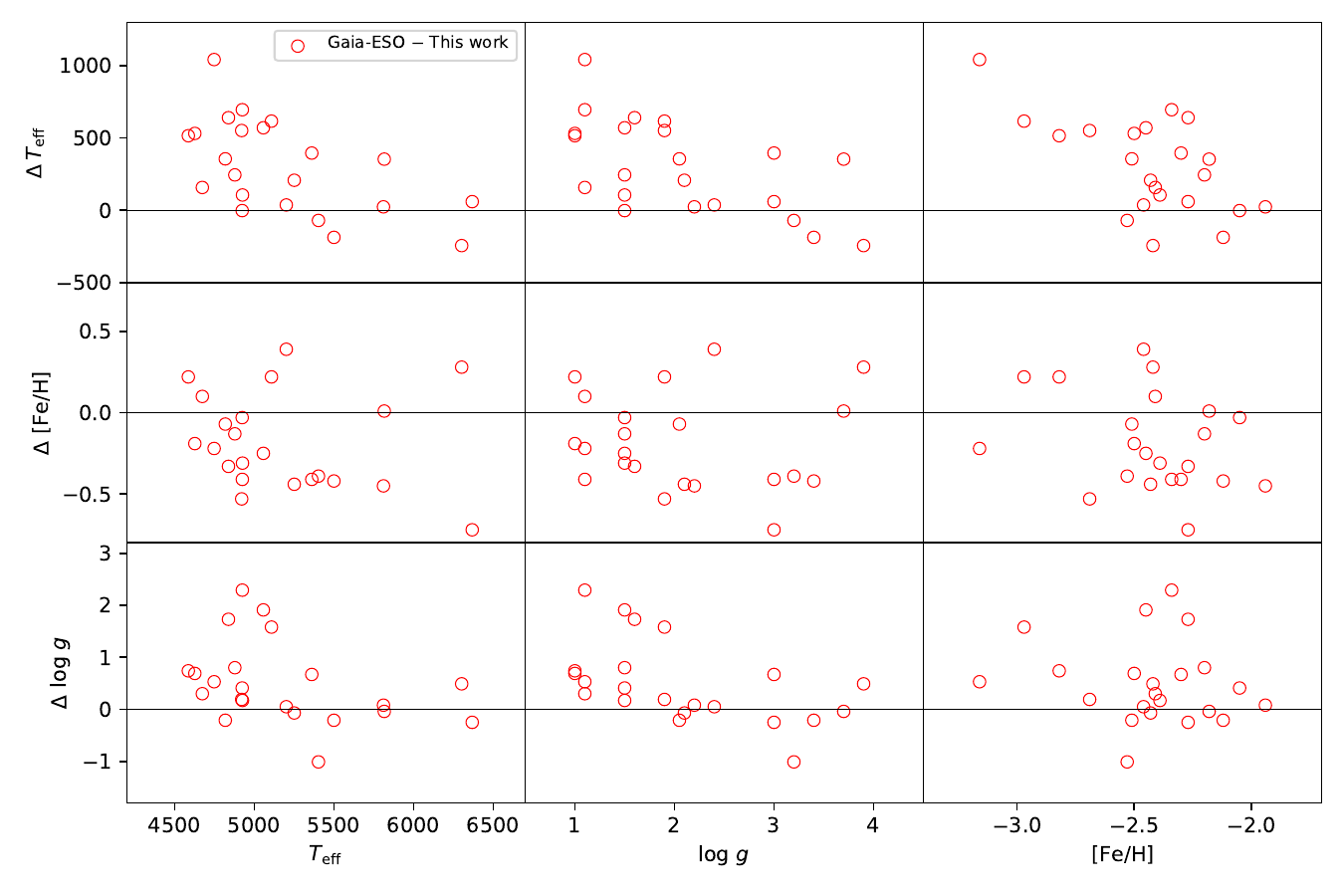}
    \caption{\tiny {\bf Comparison of atmospheric parameters.} Horizontal axis display parameters determined in this work.}
    \label{fig:GESO_comparison}
\end{figure*}

 \begin{figure}
    \centering
    \includegraphics[width=0.48\linewidth]{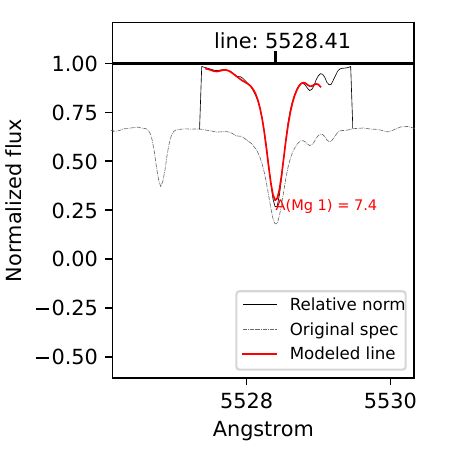}
    \includegraphics[width=0.48\linewidth]{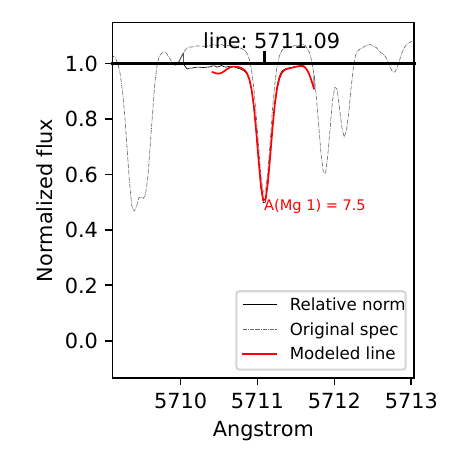}
    \caption{\tiny Mg line fits in the solar spectrum { reflected in Ganymede}. 
    {The dashed line represents the observed spectrum normalised by our automatic pipeline. It shows obvious flux mislocation with respect to the unity.
    The continuous black line displays the  spectrum re-normalised on the fly by our line-fitting algorithm, which uses the pseudo-continuum of the synthetic spectrum (red line) as the flux reference.
    Abundances recovered by fitting are noted in the plots.}
    } 
    \label{fig:Mg_fits}
\end{figure}

 \begin{figure}
    \centering
    \includegraphics[width=0.99\linewidth]{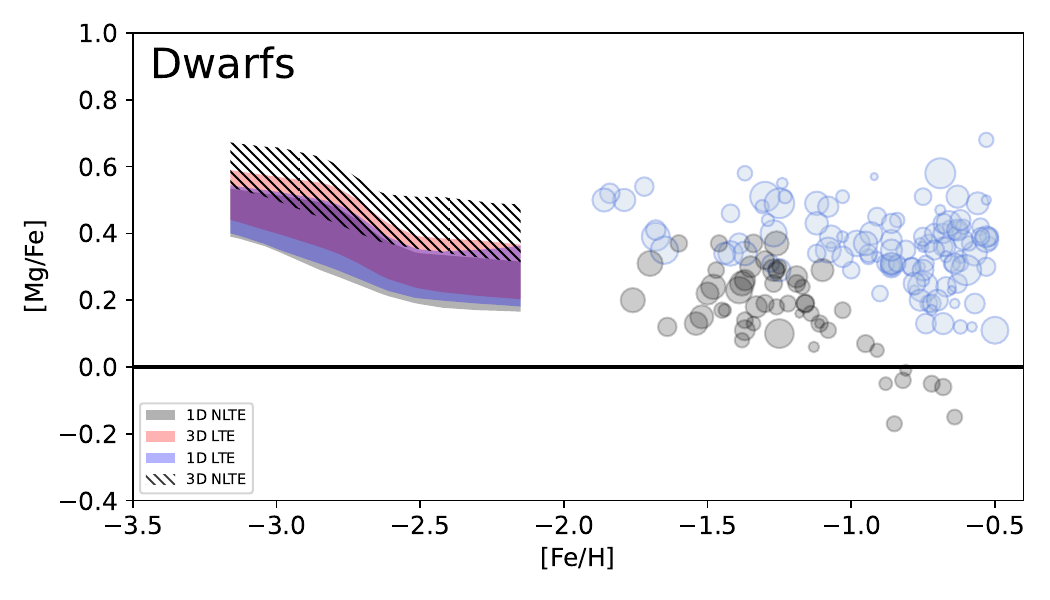}
    \caption{\tiny {\bf LOWESS dispersions of [Mg/Fe] of dwarfs.} 
    Shades of different colours display 1D LTE, 3D LTE, 1D NLTE, and 3D NLTE assumptions according to the legends. Same way as in Fig. \ref{fig:pop},
    Milky Way and Enceladus populations are plotted in blue and gray symbols as reference. 
    }
    \label{fig:dwarfs_dispersions}
\end{figure}

\begin{figure}
    \centering
     \includegraphics[width=0.99\linewidth]{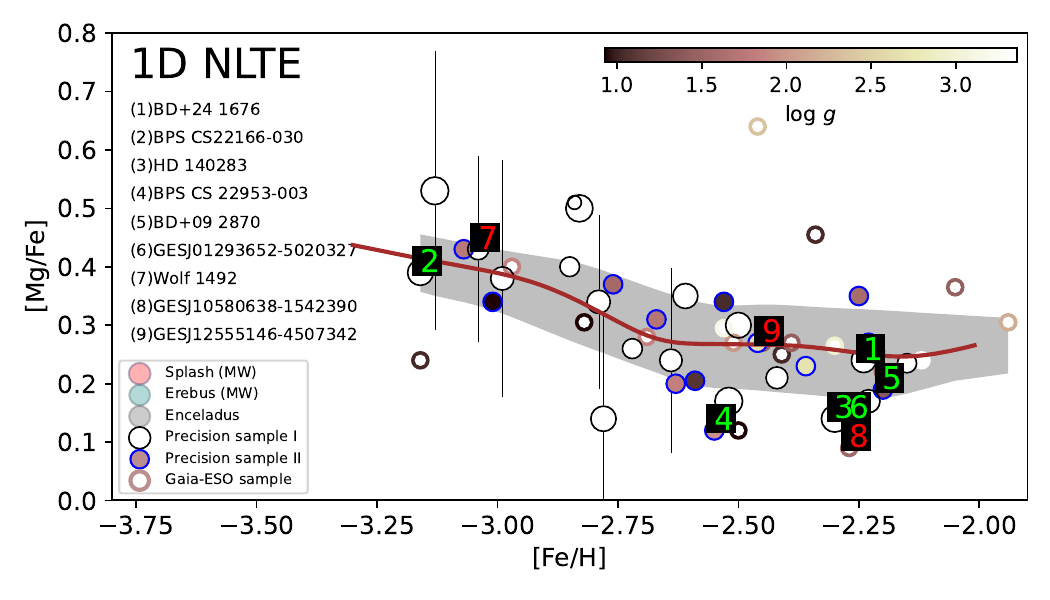}
     \includegraphics[width=0.99\linewidth]{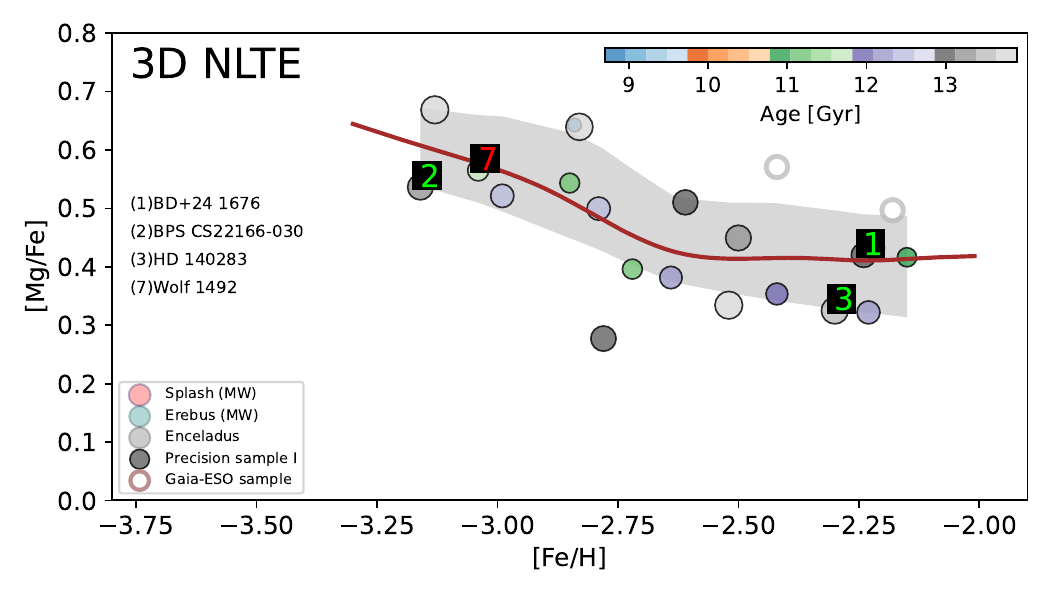}
    \caption{\tiny  Close up of the 1D NLTE and 3D NLTE [Mg/Fe] vs. [Fe/H] diagram. The plots display the same as right column panels in Fig.~\ref{fig:pop}, but remarking the stars that in Fig.~\ref{fig:lindblad} show Enceladus-like orbits (green numbers) and Sequoia-like orbits (red numbers).
    }
    \label{fig:GE_like}
\end{figure}

\section{Curve of growth of the Mg line at 5528~\AA}
\label{sec:growth}

\begin{figure}
    \centering
    \includegraphics[width=0.95\linewidth]{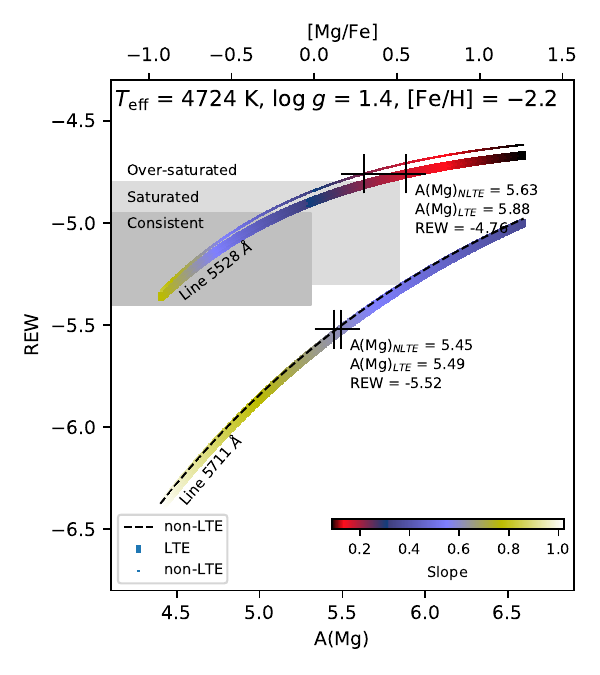}\\
    
    \caption{\tiny 
    Theoretical curves of growth of the Mg lines at 5528 and 5711~\AA. The curves correspond to the star BD+09~2870, the parameters of which are given in the plot. Thick and thin lines represent NLTE and LTE, respectively. The NLTE curve of the line at 5711~\AA\ is marked by the dashed line.
    Cross symbols indicate REW and corresponding Mg abundances in LTE and NLTE for each line; values are noted in the plot.
    }
    \label{fig:curve_growth}
\end{figure}

Figure~\ref{fig:curve_growth} shows the theoretical curve of growth of both Mg lines for BD+09~2870, a cool (\teff\ = 4724 K) and relatively low gravity (\logg\ = 1.4) star in the over-saturated region of the plot.
Thick lines represent 1D LTE and the thin lines represent 1D NLTE. Both lines are colour-coded by the slope $\delta$REW/$\delta$A(Mg).
The NLTE curve of the line 5528~\AA\ bifurcates from that of LTE as A(Mg) and REW increase, predicting progressively lower A(Mg) determinations.  
The difference between 1D LTE and 1D NLTE is larger in  the saturated and over-saturated regions of the curve because  the slope approximates to zero.
For instance, the REW $ = -4.76$ of the 5528~\AA\ line of the star corresponds to [Mg/Fe] values of 0.30 and 0.65~dex in the 1D NLTE and 1D LTE curves, respectively (see crosses in the plot).
On the other hand, the curves of growth of the line at 5711~\AA\ do not get as saturated as those of 5528~\AA, even for extremely high abundances such as [Mg/Fe] $\gtrsim 0.8$~dex.
Additionally, LTE and NLTE curves are almost over-imposed, with a constant separation of REW+0.02 or [Mg/Fe]$-0.04$ of the latter with respect to the former.

\end{appendix}

\end{document}